\documentclass[11pt]{article}
\usepackage{slashed,verbatim,subfigure}
\usepackage[numbers,sort&compress]{natbib}
\usepackage{amsmath}
\usepackage{amssymb}
\usepackage{appendix}
\usepackage{graphicx}
\usepackage{dcolumn}
\usepackage{bm}
\usepackage[colorlinks=true,linktocpage=true,
linkcolor=blue,citecolor=blue]{hyperref}
\usepackage[a4paper]{geometry}
\catcode`@11
\def\seceqaa{\@addtoreset{equation}{section}
\def\theequation{A\arabic{equation}}}
\def\seceqbb{\@addtoreset{equation}{section}
\def\theequation{B\arabic{equation}}}
\def\seceqcc{\@addtoreset{equation}{section}
\def\theequation{C\arabic{equation}}}
\def\seceqdd{\@addtoreset{equation}{section}
\def\theequation{D\arabic{equation}}}
\def\seceqee{\@addtoreset{equation}{section}
\def\theequation{E\arabic{equation}}}
\def\seceqff{\@addtoreset{equation}{section}
\def\theequation{F\arabic{equation}}}
\def\seceqgg{\@addtoreset{equation}{section}
\def\theequation{G\arabic{equation}}}
\def\seceqhh{\@addtoreset{equation}{section}
\def\theequation{H\arabic{equation}}}
\catcode`@11
\date{\today}

\begin{document}

\large
\title{\bf{(Phenomenology/Lattice-Compatible) $SU(3)$ M$\chi$PT HD up to ${\cal O}(p^4)$ and the  ${\cal O}\left(R^4\right)$-Large-$N$ Connection}}
\author{Vikas Yadav\footnote{email- vyadav@ph.iitr.ac.in}, ~ Gopal Yadav\footnote{email- gyadav@ph.iitr.ac.in}~~and~~Aalok Misra\footnote{email- aalok.misra@ph.iitr.ac.in}\vspace{0.1in}\\
Department of Physics,\\
Indian Institute of Technology Roorkee, Roorkee 247667, India}
\date{}
\maketitle
\begin{abstract}
Obtaining the values of the coupling constants of the  low energy effective theory corresponding to QCD, compatible with experimental data, even in the (vector) mesonic sector from (the  ${\cal M}$-theory  uplift of) a UV-complete string theory dual, has thus far been missing in the literature. We take the first step in this direction by obtaining the values of the coupling constants of the ${\cal O}(p^4)$ $\chi$PT Lagrangian in the chiral limit involving the NGBs and $\rho$ meson (and its flavor partners) from the  ${\cal M}$-theory /type IIA dual of large-$N$ thermal QCD, inclusive of the ${\cal O}(R^4)$ corrections. We observe that ensuring compatibility with phenomenological/lattice results (the values ) as given in \cite{Ecker-2015}, requires a relationship relating the ${\cal O}(R^4)$ corrections and large-$N$ suppression. In other words, QCD demands that the higher derivative corrections and the large-$N$ suppressed corrections  in its M/string theory dual, are related. As a bonus, we explicitly show that the ${\cal O}(R^4)$ corrections in the UV to the  ${\cal M}$-theory  uplift of the type IIB dual of large-$N$ thermal QCD at low temperatures, can be consistently set to be vanishingly small.
\end{abstract}

\begin{center}
\footnotesize{\it Dedicated by one of the authors (AM) to the memory of his supervisor, the late Professor Daniel Koltun who introduced him to $\chi$PT 25 years ago - a subject the former has returned back to after 22 years - and instilled in the former the importance of uncompromising independence of thought and inquiry.}
\end{center}
\thispagestyle{empty}
\newpage
\tableofcontents

\newpage

\section{Introduction}

Chiral Perturbation Theory ($\chi$PT) is an effective field theory of Quantum Chromodynamics (QCD) which describes the low energy regime (IR) of Quantum Chromodynamics (QCD). In $\chi$PT, the degrees of freedom are hadrons (which include mesons and baryons and hyperons but we will study only mesonic $\chi$PT). $\chi$PT Lagrangian consists of various terms which are invariant under chiral symmetry $SU(N_f)_L\times SU(N_f)_R$, charge conjugation and parity symmetry of QCD. Chiral symmetry $SU(N_f)_L\times SU(N_f)_R$ is spontaneously broken to $SU(N_f)_V$ yielding $(N_f^2 -1)$ (pseudo-)Goldstone bosons where $N_f$ is the number of flavors. As an effective field theory, it is renormalizable  order-by-order in momentum.

The $\rho$ vector meson can also be incorporated by augmenting the chiral symmetry with the inclusion of a HLS (Hidden Local Symmetry) whose gauge group is $G_{global}\times H_{local}$, where $G_{global} = SU(N_f)_L\times SU(N_f)_R$ and $H = SU(N_f)_V$ is the HLS \cite{HARADA}. The $\rho$ meson and its flavor partners  are identified with gauge boson of HLS. Chiral symmetry breaking scale is $\Lambda_\chi \sim 4\pi F_\pi \sim 1.1 GeV$ \cite{MG} in the chiral limit\cite{GL}. Since $\Lambda_\chi$  is much larger than the $\rho$ mass scale and $M_\rho < \Lambda_\chi$ , therefore we can expand the generating functional of QCD in terms of $p/\Lambda_\chi$ or $m/\Lambda_\chi$. At scale  $\mu = \Lambda_\chi$, perturbative expansion in $\mu/\Lambda_\chi$ breaks down. One can construct the most general form of the Lagrangian order by order in the derivative expansion consistently with the chiral symmetry. The ``universal" leading order Lagrangian is constructed from the terms of $O(p^2)$ .
J. Gasser and H. Leutwyler worked out the $SU(2)$\cite{GLF}$/SU(3)$ \cite{GL}  chiral perturbation theory lagrangian up to $O(p^4)$ and the renormalization of the coupling constants at scale $\mu = m_\eta$ \cite{GL}. At NLO, i.e. $O(p^4)$, there are 12 coupling constants ``$(L_{i = 1,2,..10}, H_1, H_2)$". Some of the low energy constants $L_4, L_5, L_6, L_8$ have been calculated from Lattice simulation from the MILC collaboration \cite{MILC}. In \cite{Pich}, the ten low energy constants $L_{i = 1,2,..10}$ were evaluated at scale $\mu = M_\rho$; \cite{EJP} also contains the low energy constants at $\mu = M_\rho$.  Updated values are given in \cite{Ecker-2015}.

QCD is a non-abelian gauge theory coupled to quarks. This theory can also be explored using gauge/gravity duality which is a duality between strongly coupled gauge theory and weakly coupled gravitational theory. Originally, it started with the AdS/CFT correspondance \cite{Maldacena} which is a duality between strongly coupled ${\cal N}=4\ SU(N)$ supersymmetric Yang Mills theory in large $N$ limit and type IIB string theory on  $AdS_5\times S^5$ background. ${\cal N}=4$ SU(N) supersymmetric Yang Mills theory is a conformal field theory but QCD is non-conformal. This correspondence has been generalised to study QCD. There are two approaches to study holographic theories. One is the top-down approach in which one starts with ten dimensional type IIB/IIA string theory or eleven dimensional  ${\cal M}$-theory  and then compactifies it to four dimensions with desired properties; this will be the approach followed in this work. The other is the bottom-up approach, in which one has to first know  what properties we are looking for and then construct the gravitational dual, which usually involves an $AdS_5$.

As regards a top-down approach, Kruczenski et al \cite{KMMW} considered an interesting probe $D6$-branes in a $D4$-brane type IIA supergravity background, which they used to explore various aspects of low energy phenomena in QCD. An important ingredient which was still missing from their model, however, were the massless pions as Nambu-Goldstone bosons associated with the spontaneous breaking of the $U(N_f)_L \times U(N_f)_R$ chiral symmetry in QCD. Following \cite{KMMW}, Sakai-Sugimoto (SS) in \cite{SS1,SS2} considered a nice modification by looking at a $D4/D8/\overline{D8}$ system in type IIA supergravity background with  anti-periodic boundary condition for fermions along a circle to break supersymmetry. This model exhibited chiral symmetry breaking  with $D8-\overline{D8}$ pairs merging into $D8$-branes. This model also yields massless pions which are identified with Nambu-Goldstone bosons associated with chiral symmetry breaking, and the lightest vector meson($\rho$ meson). The Sakai Sugimoto model is closely related to the HLS formalism which produces Kawarabayashi-Suzuki-Riazuddin-Fayyazuddin-type relation among the couplings. Chern-Simons term on the probe brane leads to the Wess-Zumino-Witten term. In \cite{SS2}, Sakai Sugimoto   obtained few low energy constants ($L_1, L_2, L_3$) of $SU(3)$ chiral perturbation theory at ${\cal O}(p^4)$ which were close to the values given in \cite{Pich} for suitable choice of a parameter $\kappa$. Although Sakai Sugimoto model reproduces various physical quantities of low energy QCD but this model deviates from realistic QCD above the energy scale  of the vector mesons because they obtain a four-dimensional theory by compactifying $D4$-branes on a circle of radius $\tilde{M}_{KK}^{-1}$ with an infinite tower of Kaluza-Klein modes of mass scale $\tilde{M}_{KK}$ arises. These Kaluza-Klein modes do not appear in realistic QCD. Further, the SS model caters  to the IR and is not UV complete. This was taken care of by the (only) UV-complete $D3, D5/\overline{D5}, D7/\overline{D7}$ holographic dual of large-$N$ thermal QCD of the McGill group \cite{metrics}, its type IIA SYZ mirror and the  ${\cal M}$-theory  uplift of the same (in particular in the `MQGP limit' (\ref{MQGP_limit})) as constructed in \cite{MQGP} (with one of the co-authors [AM]).

In\cite{HARADA}, the  authors considered the Sakai Sugimoto model \cite{SS1} as holographic QCD model and  proposed a method to integrate out infinite number of higher KK modes appearing in the expansion of five dimensional gauge field which consists of infinite number of vector and axial vector fields including pion as Nambu Goldstone boson arising due to spontaneous chiral symmetry breaking. First, they truncated the spectrum at certain level so that number of fields ``integrated in" becomes finite in the theory and then integrated out all the KK modes except pion and lowest lying vector mode($\rho$ meson and the flavor partners). Using this method they obtained the effective lagrangian up to $O(p^4)$. However this Lagrangian is not same as the $SU(3)$ chiral perturbation theory Lagrangian \cite{GL}. The authors in \cite{HMM} derived relations between the $SU(3)$ low energy coupling constants of \cite{GL} and $O(p^4)$ couplings in \cite{HARADA}.

The main takeaway from this paper is that QCD imposes a relationship between the higher derivative corrections and large-$N$ suppression when comparing our ${\cal M}$-theory/type IIA holographic computational results for the low energy coupling constants of $\chi$PT Lagrangian up to ${\cal O}(p^4)$ and corresponding experimental values of these coupling constants.

The rest of the paper is organized as follows. In section {\bf 2}, we give a quick review of the type IIB holographic dual of large-N thermal QCD as constructed in \cite{metrics}, its SYZ type IIA mirror (and its  ${\cal M}$-theory  uplift) as obtained in \cite{MQGP} and summarize results pertaining to the applications of the type IIA/ ${\cal M}$-theory  dual of holographic QCD phenomenology as obtained by the group to which the authors of this paper belong. In section {\bf 3}, we outline obtaining the chiral limit of the Chiral Perturbation Theory Lagrangian up to ${\cal O}(p^4)$ involving the $\pi$ and $\rho$ mesons as well as their flavor partners, using the HLS formalism/notation of \cite{HARADA}. In section {\bf 4} which forms the core of the paper with all the results, we obtain the values of the LECs of \cite{GL} up to ${\cal O}(p^4)$ as radial integrals using the type IIA SYZ mirror of \cite{metrics} inclusive of the ${\cal O}(R^4)$  ${\cal M}$-theory  corrections worked out in \cite{OR4-Yadav+Misra}. In the process of matching the holographic results with experimental values see that there is a deep connection between the large-$N$ suppression and the aforementioned $l_p^6$ (${\cal O}(R^4)$) corrections. Section {\bf 5} has a summary of the results and the lessons learnt. There are four supporting appendices. Appendix {\bf A} is about showing that the ${\cal O}(R^4)$ corrections to the  ${\cal M}$-theory  uplift of large-$N$ thermal QCD-like theories at low temperatures (i.e. below the deconfinement temperature) can consistently be made to be vanishingly small in the UV. Appendix {\bf B} has some details pertaining to the evaluation of the coupling constants appearing in the ``HLS" chiral Lagrangian as radial integrals, which is relevant to the computation of Section {\bf 4}. Appendix {\bf C} is a brief review of the HLS formalism and a derivation of the ${\cal O}(p^2)$ $SU(3)$ $\chi$PT Lagrangian of \cite{GL} as well as contributions to the low energy coupling constants (LECs) of ${\cal O}(p^4)$ $SU(3)$ $\chi$PT Lagrangian of \cite{GL} arising from integrating out the rho mesons from the theory. Appendix {\bf D} provides details of obtaining the DBI action for the type IIA flavor $D6$-branes incorporating the ${\cal O}(R^4)$ corrections to the ${\cal M}$-theory uplift of large-$N$ thermal QCD-like theories.

\section {Brief Review of the (UV Complete) Type IIB/IIA SYZ Mirror (and  ${\cal M}$-theory ) Holographic Dual of large-$N$ Thermal QCD at Intermediate Coupling, and Holographic QCD Phenomenology}

In the context of a UV-complete top-down holographic large-$N$ thermal QCD, the following summarizes the main features of the brane construct and the gravity dual of \cite{metrics}, its type IIA mirror and ${\cal M}$-theory uplift (up to ${\cal O}(R^4)$) as well as holographic phenomenological applications of the same.

\begin{itemize}
\item {\bf Brane construct of \cite{metrics}}:
The type IIB string dual of \cite{metrics} consists of $N$ $D3$-branes placed at the tip of six-dimensional conifold, with $M\ D5$-branes wrapping the vanishing $S^2$, referred to as fractional $D3$-branes, and $M\ \overline{D5}$-branes  distributed along the resolved $S^2$ placed at antipodal points relative to the $M$ $D5$-branes. Denoting the average $D5/\overline{D5}$ separation  by ${\cal R}_{D5/\overline{D5}}$, $r>{\cal R}_{D5/\overline{D5}}$, would be the UV.    The $N_f\ D7$-branes, holomorphically embedded via Ouyang embedding \cite{ouyang} in the resolved conifold geometry, ``smeared"/delocalized along the angular directions $\theta_{1,2}$, are present in the UV, the IR-UV interpolating region and dip into the (confining) IR (but do not touch the $D3$-branes; the shortest $D3-D7$ string corresponding to the lightest quark). In addition, $N_f\ \overline{D7}$-branes are present in the UV and the UV-IR interpolating region for the reason given below. The following table summarizes the aforementioned brane construct wherein $S^2(\theta_1,\phi_1)$ denotes the vanishing two-sphere and (NP/SP of) $S^2_a(\theta_2,\phi_2)$ is the (North Pole/South Pole of the) resolved/blown-up two-sphere - $a$ being the radius of the blown-up $S^2$ - and $r_{\rm UV}$ is the UV cut-off and $\frac{\epsilon}{\left({\cal R}_{D5/\overline{D5}} - |\mu_{\rm Ouyang}|^{\frac{2}{3}}\right)}<1$ in Table 1. Also, $\mu_{\rm Ouyang}$ is the Ouyang embedding parameter that is defined as:
\begin{equation}
\label{Ouyang-definition}
\left(r^6 + 9 a^2 r^4\right)^{\frac{1}{4}} e^{\frac{i}{2}\left(\psi-\phi_1-\phi_2\right)}\sin\left(\frac{\theta_1}{2}\right)\sin\left(\frac{\theta_2}{2}\right) = \mu_{\rm Ouyang},
\end{equation}
effected by (\ref{theta12-deloc}) for vanishingly small $|\mu_{\rm Ouyang}|$, while describing the embedding of the flavor $D7$-branes in the resolved conifold geometry.
\begin{table}[h]
\begin{center}
\begin{tabular}{|c|c|c|}\hline
&&\\
S. No. & Branes & World Volume \\
&&\\ \hline
&&\\
1. & $N\ D3$ & $\mathbb{R}^{1,3}(t,x^{1,2,3}) \times \{r=0\}$ \\
&&\\  \hline
&&\\
2. & $M\ D5$ & $\mathbb{R}^{1,3}(t,x^{1,2,3}) \times \{r=0\} \times S^2(\theta_1,\phi_1) \times {\rm NP}_{S^2_a(\theta_2,\phi_2)}$ \\
&&\\  \hline
&&\\
3. & $M\ \overline{D5}$ & $\mathbb{R}^{1,3}(t,x^{1,2,3}) \times \{r=0\}  \times S^2(\theta_1,\phi_1) \times {\rm SP}_{S^2_a(\theta_2,\phi_2)}$ \\
&&\\  \hline
&&\\
4. & $N_f\ D7$ & $\mathbb{R}^{1,3}(t,x^{1,2,3}) \times \mathbb{R}_+(r\in[|\mu_{\rm Ouyang}|^{\frac{2}{3}},r_{\rm UV}])  \times S^3(\theta_1,\phi_1,\psi) \times {\rm NP}_{S^2_a(\theta_2,\phi_2)}$ \\
&&\\  \hline
&&\\
5. & $N_f\ \overline{D7}$ & $\mathbb{R}^{1,3}(t,x^{1,2,3}) \times \mathbb{R}_+(r\in[{\cal R}_{D5/\overline{D5}}-\epsilon,r_{\rm UV}]) \times S^3(\theta_1,\phi_1,\psi) \times {\rm SP}_{S^2_a(\theta_2,\phi_2)}$ \\
&&\\  \hline
\end{tabular}
\end{center}
\caption{The Type IIB Brane Construct of \cite{metrics}}
\end{table}

\item
In the UV, one has $SU(N+M)\times SU(N+M)$ color gauge group and $SU(N_f)\times SU(N_f)$ flavor gauge group. There occurs a partial Higgsing of $SU(N+M)\times SU(N+M)$ to $SU(N+M)\times SU(N)$ as one goes from $r>{\cal R}_{D5/\overline{D5}}$  to $r<{\cal R}_{D5/\overline{D5}}$. This happens because at energies less than  ${\cal R}_{D5/\overline{D5}}$ (IR), the $\overline{D5}$-branes are integrated out resulting in the reduction of the rank of one of the product gauge groups (which is $SU(N + {\rm number\ of}\ D5-{\rm branes})\times SU(N + {\rm number\ of}\ \overline{D5}-{\rm branes})$). By the same token, the $\overline{D5}$-branes are ``integrated in" for energies more than  ${\cal R}_{D5/\overline{D5}}$ (UV), resulting in the conformal Klebanov-Witten-like $SU(M+N)\times SU(M+N)$ product color gauge group.

\item
The pair of gauge couplings, $g_{SU(N+M)}$ and $g_{SU(N)}$ can be shown to flow  oppositel. The flux of the NS-NS B through the vanishing $S^2$ is the reason for introduction of non-conformality which is why $M$ $\overline{D5}$-branes were included in \cite{metrics} to cancel the net $D5$-brane charge in the UV. Further, as the $N_f$ flavor $D7$-branes enter the RG flow of the gauge couplings via the dilaton, their contribution therefore needs to be canceled by $N_f\ \overline{D7}$-branes which is the reason for their inclusion in the UV.

Using UV-complete top-down type IIB holographic dual of lagre-$N$ thermal QCD at finite gauge/'t Hooft coupling\cite{metrics}, delocalized Strominger-Yau-Zaslow (SYZ) type IIA  mirror of \cite{metrics}  and its M theory uplift in the 'MQGP' limit as worked out in\cite{MQGP}, our group has made the following contributions to the top-down holographic dual of large N thermal QCD phenomenology at intermediate gauge/'t Hooft coupling.

\item
 In the IR, at the end of a Seiberg-like duality cascade,  the number of colors $N_c$ gets identified with $M$, which in the `MQGP limit' to be discussed below, can be tuned to equal 3 (See \cite{Misra+Gale}).

\item
{\bf Gravity dual of the brane construct of \cite{metrics}}: The finite temperature ($>T_c$) on the gauge/brane side is effected in the gravitational dual via a black hole in the latter. Turning on of the temperature (in addition to requiring a finite separation between the $M\ D5$-branes and $M\ \overline{D5}$-branes to provide a natural scale above which one is in the UV) corresponds in the gravitational dual to having a non-trivial resolution parameter of the conifold. IR confinement on the brane/gauge theory side corresponds to having a non-trivial deformation of the conifold geometry in the gravitational dual.  The gravity dual is hence given by a  resolved warped deformed conifold wherein the $D3$-branes and the $D5$-branes are replaced by fluxes in the IR, and the back-reactions are included in the warp factor and fluxes.

\item
{\bf Color-Flavor Enhancement of Length Scale in the IR}:  In the IR in the MQGP limit (\ref{MQGP_limit}) , with the inclusion of terms higher order in $g_s N_f$  in the RR and NS-NS three-form fluxes and the NLO terms in $N$ in the metric, there occurs an IR color-flavor enhancement of the length scale as compared to a Planckian length scale in Klebanov-Strassler(KS)'s model even for ${\cal O}(1)$ $M$, thereby ensuring that quantum corrections will be suppressed. This was discussed/summarized in \cite{NPB}/\cite{Misra+Gale}. Essentially, defining:
\begin{eqnarray}
\label{NeffMeffNfeff}
& & N_{\rm eff}(r) = N\left[ 1 + \frac{3 g_s M_{\rm eff}^2}{2\pi N}\left(\log r + \frac{3 g_s N_f^{\rm eff}}{2\pi}\left(\log r\right)^2\right)\right],\nonumber\\
& & M_{\rm eff}(r) = M + \frac{3g_s N_f M}{2\pi}\log r + \sum_{m\geq1}\sum_{n\geq1} N_f^m M^n f_{mn}(r),\nonumber\\
& & N^{\rm eff}_f(r) = N_f + \sum_{m\geq1}\sum_{n\geq0} N_f^m M^n g_{mn}(r),
\end{eqnarray}
wherein  the type IIB axion $C_0 =N_f^{\rm eff} \frac{\left(\psi - \phi_1-\phi_2\right)}{4\pi}$,
at the end of a Seiberg-like duality cascade, $N_{\rm eff}(r_0\in\rm IR)=0$ and writing the ten-dimensional warp factor $h \sim \frac{L^4}{r^4}$,  the length scale $L$ in the IR will be given by:
\begin{eqnarray}
\label{length-IR}
& & L\sim\sqrt[4]{M}N_f^{\frac{3}{4}}\sqrt{\left(\sum_{m\geq0}\sum_{n\geq0}N_f^mM^nf_{mn}(r_0)\right)}\left(\sum_{l\geq0}\sum_{p\geq0}N_f^lM^p g_{lp}(r_0)\right)^{\frac{1}{4}} L_{\rm KS},
\end{eqnarray}
$L_{KS}\sim \sqrt[4]{g_sM}\sqrt{\alpha^\prime}$. The relation (\ref{length-IR}) implies that  in the IR, relative to KS, there is a color-flavor enhancement of the length scale in the MQGP limit. Hence,  in the IR, even for $N_c^{\rm IR}=M=3$ and $N_f=2(u/d)+1(s)$, upon inclusion of of $n,m>1$  terms in
$M_{\rm eff}$ and $N_f^{\rm eff}$ in (\ref{NeffMeffNfeff}), $L\gg L_{\rm KS}(\sim L_{\rm Planck})$ in the MQGP limit (\ref{MQGP_limit}), implying that {\it the stringy corrections are suppressed and one can trust supergravity calculations}.

 \item
{\bf Obtaining} ${\bf N_c=3}$: As explained in \cite{Misra+Gale}, in the IR, at the end of a Seiberg-like duality cascade,  the number of colors $N_c$ gets identified with $M$, which in the `MQGP limit' (\ref{MQGP_limit})  can be tuned to equal 3. This is briefly summarized now. One can identify $N_c$ with the effective number $N_{\rm eff}$ of $D3$-branes and the effective number $M_{\rm eff}$ of $D5$-branes as:
$N_C = N_{\rm eff}(r) + M_{\rm eff}(r)$.  $N_{\rm eff}(r)$ is defined via
$\tilde{F}_5\equiv dC_4 + B_2\wedge F_3 = {\cal F}_5 + *{\cal F}_5,$
wherein ${\cal F}_5\equiv N_{\rm eff}\times{\rm Vol}({\rm Base\ of\ Resolved\ Warped\ Deformed\ Conifold})$. Similarly, $M_{\rm eff}$ is defined via  $M_{\rm eff} = \int_{S^3}\tilde{F}_3$ (the $S^3$ being dual to $\ e_\psi\wedge\left(\sin\theta_1 d\theta_1\wedge d\phi_1 - B_1\sin\theta_2\wedge d\phi_2\right)$,  $B_1$ being an `asymmetry factor' defined in \cite{metrics}; $e_\psi\equiv d\psi + {\rm cos}~\theta_1~d\phi_1 + {\rm cos}~\theta_2~d\phi_2$) and \cite{M(r)N_f(r)-Dasgupta_et_al}: $\tilde{F}_3 (\equiv F_3 - \tau H_3)\propto M(r)\equiv M\frac{1}{1 + e^{\alpha(r-{\cal R}_{D5/\overline{D5}})}}, \alpha\gg1.$
 As $N_{\rm eff}$  varies between $N\gg1$ in the UV and 0 in the deep IR, and  $M_{\rm eff}$  varies between 0 in the UV and $M$ in the deep IR, $N_c$ varies between $M$ in the deep IR and a large value [ in the MQGP limit of (\ref{MQGP_limit}) for a large value of $N$] in the UV.  Therefore, at very low energies, the number of colors $N_c$ can be approximated by $M$, which in the MQGP limit is  finite and can hence be taken to be equal to three. Additionally, one can set $N_f=2(u/d)+1(s)$. Hence, in the IR, this is somewhat like the Veneziano limit in which $\frac{N_f}{N_c}$ is fixed (but, unlike \cite{metrics,MQGP}, $N_c,N_f\rightarrow\infty$ in the Veneziano limit  in, e.g., s\cite{Nitti_et_al}) as (in the IR) $\frac{N_f}{N_c}\sim\frac{N_f}{M}$ in \cite{metrics}.

Thus, under the aforementioned Seiberg-like duality cascade, the $N\ D3$-branes are cascaded away and there is a finite $M$ left at the end corresponding to a strongly coupled IR-confining $SU(M)$ gauge theory; the finite temperature version of this $SU(M)$ gauge theory is what was considered in \cite{metrics}. So, at the end of the Seiberg-like duality cascade in the IR, the number of colors $N_c$, identified with $M$,  in the `MQGP limit' can be tuned to equal 3.

\item
{\bf The MQGP limit, Type IIA Strominger-Yau-Zaslow (SYZ) mirror  of \cite{metrics} and its  ${\cal M}$-theory  uplift at intemediate gauge coupling}:
\begin{itemize}
\item
For constructing a holographic dual of thermal QCD-like theories, one would be required to consider intemediate gauge coupling (as well as finite number of colors) $-$ dubbed as the `MQGP limit' in \cite{MQGP}. From the perspective of gauge-gravity duality, this would hence require looking at the strong-coupling/non-perturbative limit of string theory - M theory. The MQGP limit in \cite{MQGP, NPB} was defined as:
\begin{equation}
\label{MQGP_limit}
g_s\sim\frac{1}{{\cal O}(1)}, M, N_f \equiv {\cal O}(1),\ N \gg1,\ \frac{g_s M^2}{N}\ll1.
\end{equation}

\item
The  ${\cal M}$-theory  uplift of the type IIB holographic dual of \cite{metrics} was constructed in \cite{MQGP, NPB} by working out the Strominger-Yau-Zaslow (SYZ) type IIA mirror of \cite{metrics} effected via a triple T duality along a local special Lagrangian (sLag) $T^3$ $-$ which could be identified with the $T^2$-invariant sLag of \cite{M.Ionel and M.Min-OO (2008)} with a large base ${\cal B}(r,\theta_1,\theta_2)$ (of a $T^3(\phi_1,\phi_2,\psi)$-fibration over ${\cal B}(r,\theta_1,\theta_2)$) \cite{NPB,EPJC-2}\footnote{As explained in \cite{Misra+Gale} also as a footnote, consider $D5$-branes wrapping the resolved $S^2$ of a resolved conifold geometry \cite{Zayas-Tseytlin}, which one knows, globally, breaks  SUSY.  As in \cite{SYZ-free-delocalization}, to begin with, SYZ is implemented wherein the pair of $S^2$s are replaced by a pair of $T^2$s in the delocalized limit, and the correct T-duality coordinates are identified. Then,  when uplifting the mirror to M theory, it is found
 that a $G_2$-structure  can be chosen which is free of the delocalization. For the SYZ mirror of the resolved warped deformed conifold which figures in the gravitational dual of large-$N$ thermal QCD of \cite{metrics}, that gets uplifted to  ${\cal M}$-theory  with $G_2$ structure worked out in \cite{MQGP}, the idea is exactly the same.} Let us briefly discuss the basic idea. Consider the aforementioned $N$ D3-branes  with its world-volume directions $x^{0, 1, 2, 3}$ at the tip of conifold. Further, assuming the $M\ D5$-branes to be parallel to these $D3$-branes as well as wrapping the vanishing $S^2(\theta_1,\phi_1)$, a single T-dual along $\psi$  yields $N\ D4$-branes wrapping the $\psi$ circle and $M\ D4$-branes straddling a pair of orthogonal $NS5$-branes. The wold volumes of these pair of $NS5$-branes correspond to the vanishing $S^2(\theta_1,\phi_1)$ and the blown-up $S^2(\theta_2,\phi_2)$ with a non-zero resolution parameter $a$ (the radius of the blown-up $S^2(\theta_2,\phi_2)$). Two more T-dualities along $\phi_i$ and $\phi_2$, then convert the aforementioned pair of orthogonal $NS5$-branes into a pair of orthogonal Taub-NUT spaces, the $N\ D4$-branes into $N$ color $D6$-branes and the $M$ straddling $D4$-branes also to $D6$-branes. Also, in the presence of the aforementioned $N_f$ flavor $D7$-branes (embedded holomorphically via the Ouyang embedding), oriented parallel to the $D3$-branes and ``wrapping" a non-compact four-cycle $\Sigma^{(4)}(r, \psi, \theta_1, \phi_1$), upon T-dualization yield $N_f$ $D6$-branes ``wrapping" a non-compact three-cycle  $\Sigma^{(3)}(r, \theta_1, \phi_2$). An uplift to  ${\cal M}$-theory  of the SYZ type IIA mirror so obtained, converts  the $D6$-branes to KK monopoles that are variants of Taub-NUT spaces.  Therefore, all the branes are converted to geometry and fluxes and one  ends up with  ${\cal M}$-theory  on a $G_2$-structure manifold. Similarly, one may perform identical three T-dualities on the gravity dual on the type IIB side, which is a resolved warped-deformed conifold with fluxes,  to obtain another $G_2$ structure manifold, giving us the MQGP holographic dual of  \cite{MQGP,NPB}.
\end{itemize}

Hence, the type IIB model of \cite{metrics} make it an ideal holographic dual of thermal QCD because: (i) it is UV conformal (Landau poles are absent), (ii) it is IR confining, (iii) the quarks transform in the fundamental representation of flavor and color groups, and (iv) it  is defined for the full range of temperature - both low and high.

\item {\bf Conceptual Physics issues miscellanea}
\begin{itemize}

\item {\it Regime of validity of the top-down holographic model and $\Lambda_{\rm QCD}$}: A natural question that one would like to answer is the range of scales
in which the top-down holographic model elaborated upon, is expected to match QCD. 
The basic idea, using the notations introduced earlier on in this section, that answers this question is that the range of variation of the radial coordinate in the supergravity dual corresponding to the energy scale in QCD-like theories is determined by: $\left\{\left.r\right| N_{\rm eff}(r) = \int_{M_5(\theta_{1,2},\phi_{1,2},\psi)}\left(F_5 + B_2\wedge C_3\right)\approx0\right\}\cap\left\{\left.r\right|M_{\rm eff}\equiv{\cal O}(1)\right\} $ - $M_5(\theta_{1,2},\phi_{1,2},\psi)$ being the base of the non-K\"{a}hler resolved warped deformed conifold -  because this will ensure that the (effective) number of colors can be set to be ${\cal O}(1)$, and in fact 3. One can show that in the IR (wherein $|\log r|\gg1$) in the MQGP limit, $N_{\rm eff}(r)\sim N\left[1 + g_s\frac{g_s M^2(g_s N_f)}{N}\log^3r\right]$ \cite{Bulk-Viscosity}) and from (\ref{NeffMeffNfeff}), estimating: $M\left[1 + \frac{3g_s N_f M \log r}{2\pi}\right]\equiv {\cal O}(1)$, or $\log r =  - \frac{2\pi\left(-{\cal O}(1) + M\right)}{3g_s M N_f}$ and substituting into $N_{\rm eff}$ yields, e.g., for $N\sim10^2, N_f=3$, 
$M\sim{\cal O}(1)$ as in the MQGP limit. 

The computations in this paper are in the low temperature limit, i.e., for temperatures below the deconfinement temperature whereat one has bound states of quarks such as mesons which is what we are interested in looking at. To therefore understand the upper limit in energy on the QCD side or $r$ in the gravitational dual side, we therefore need to understand the non-perturbative QCD scale, $\Lambda_{\rm QCD}$, in terms of the geometrical data of our top-down holographic model. Let us remind ourselves that (before the Seiberg-like dualities) the $SU(M\times N)$ and $SU(N)$ gauge couplings $g_{SU(M+N)}$ and $g_{SU(N)}$ satisfy \cite{metrics}:\\
$4\pi^2\left(\frac{1}{g_{SU(M+N)}^2} + \frac{1}{g_{SU(N)}^2}\right) e^{\phi^{\rm IIB}(r,\theta_{10},\theta_{20})}= \pi ;\ 4\pi^2\left(\frac{1}{g_{SU(M+N)}^2} - \frac{1}{g_{SU(N)}^2}\right) e^{\phi^{\rm IIB}(r,\theta_{10},\theta_{20})}\\
\sim\frac{1}{2\pi\alpha^\prime}\int_{S^2(\theta_1,\phi_1)}B^{\rm IIB}\sim \left.g_s M_{\rm eff}(r)N_f^{\rm eff}(r)\log r\right|_{r\in\rm IR-UV\ interpolating\ region}$. Note, everywhere $r, a$ are in fact $\frac{r}{{\cal R}_{D5/\overline{D5}}}, \frac{a}{{\cal R}_{D5/\overline{D5}}}$ where ${\cal R}_{D5/\overline{D5}}$ is taken to be $\sqrt{3}a$ ($a$ being the resolution parameter of the blown-up $S^2$) \cite{Bulk-Viscosity}. Hence, $\Lambda_{\rm QCD} = {\cal R}_{D5/\overline{D5}}\sim\sqrt{3}a$ because near $r={\cal R}_{D5/\overline{D5}}$, the aforementioned gauge couplings become very large indicating the onset of non-perturbative QCD. Recall, in the UV, i.e., $\forall r > {\cal R}_{D5/\overline{D5}}$, we obtain the conformal $SU(N+M)\times SU(N+M)$ Klebanov-Witten-like (asymptotically supersymmetric as the complexified three-form type IIB flux $G_3$ becomes ISD for $r\gg {\cal R}_{D5/\overline{D5}}$) gauge theory, and in the IR, i.e., $\forall r < {\cal R}_{D5/\overline{D5}}$, we obtain the non-conformal  non-supersymmetric $SU(M+N)\times SU(N)$ gauge theory, which after a Seiberg-like duality cascade, is expected to yield IR confining QCD-like gauge theory.

\item
{\it Hierarchy of scales in the gravitational dual}: There is an IR cut-off denoted by $r_0$ indicative of the boundary of the deep IR.  The $D5-\overline{D5}$ separation ${\cal R}_{D5/\overline{D5}}$ provides another scale so that at energies larger than ${\cal R}_{D5/\overline{D5}}$, the $D5-\overline{D5}$ strings become massive and the gauge group, as already mentioned above, becomes the Klebanov-Witten-like UV-conformal $SU(M+N)\times SU(M+N)$, and at energies less than ${\cal R}_{D5/\overline{D5}}$ the gauge group is Higgsed down to the Klebanov-Strassler-like $SU(M+N)\times SU(N)$\footnote{In principle, there is another scale that exists in the type IIB holographic dual, which is provided by the embedding of the $D7$-branes via the Ouyang's embedding (\ref{Ouyang-definition}), where the modulus of the Ouyang embedding parameter $|\mu|^{\frac{2}{3}}$ gives a measure of the radial separation of the ``deepest" embedded flavor $D7$-branes in the IR from the color $D3$-branes corresponding hence to the mass of the lightest quark. Hence, one could have a refinement of the hierarchy and assume the IR to be given by $r\in[r_0, |\mu|^{\frac{2}{3}}]$, the IR-UV interpolating region given by $r\in[|\mu|^{\frac{2}{3}}, {\cal R}_{D5/\overline{D5}}]$ and the UV 
corresponding to $r\in[{\cal R}_{D5/\overline{D5}}, r_{\rm UV}]$. However, for simplicity of calculations, we have merged the first two in this paper.}; there is also a $U(1)^M$ for massless strings starting and ending on the same $\overline{D5}$-brane. For all practical purposes, $r>{\cal R}_{D5/\overline{D5}}$ is in the UV. As one is always working in the ``near-horizon" limit in which the ten-dimensional warp factor
$h(r,\theta_{1,2})\sim \frac{L^4}{r^4}\left(1 + f\left(r; N, M, N_f\right)\right)$, even in the UV (i.e., one drops the 1 that figures, e.g., in the warp factor in the Klebanov-Witten supergravity dual: $h = 1 + \frac{L^4}{r^4}$), where $L\equiv\left(4\pi g_s N\alpha^\prime\right)^{\frac{1}{4}}$, hence, the UV cut-off
$r_{\rm UV}\stackrel{<}{\sim}L$. So, in our holographic dual, the gravitational analog of $\Lambda_{\rm QCD}$ and $r_{\rm UV}$ are separated.
\end{itemize}

\item {\bf Holographic QCD Phenomenology}
\begin{itemize}
\item
 In \cite{Sil+Yadav+Misra-glueball}, two of the authors (VY and AM) along with K.Sil,  studied the glueball spectra and evaluated the masses of $0^{++}, 0^{-+},0^{--},1^{++},2^{++}$ glueballs in type IIB/type IIA/ M theory supergravity backgrounds using WKB quantization and by imposing Neumann/Dirichlet boundary conditions at the IR cut-off.  It was found that WKB quantization produces masses of $0^{++}, 0^{-+},0^{--},1^{++},2^{++}$ glueballs very close to the lattice results.
\item In \cite{Yadav+Misra+Sil-Mesons}, two of the authors (VY and AM) along with K.Sil  calculated (pseudo-)vector and (pseudo-)scalar meson spectra at finite coupling (part of the ‘MQGP’ limit), and compared their result with PDG data. It was found that masses of the (pseudo-)vector $(\rho[770], a1[1260], \rho[1450], a1[1640])$ and (pseudo-)scalar $(f0[980]/a0[980], f0[1370], f0[1450)$ mesons were closer to the PDG data than previously obtained in the literature.
\item In \cite{VA-Glueball-decay}, two of the authors (VY and AM) studied (exotic) scalar glueball $0^{++}_E$ which correspond to metric fluctuations of the M theory uplift  at finite coupling. Using the same (exotic) scalar glueball $0^{++}_E$-meson interaction Lagrangian linear in (exotic) scalar glueball and quartic in meson fields was derived. Decay widths of the processess  $G_E \rightarrow 2\pi, G_E \rightarrow 2\rho, \rho \rightarrow 2\pi, G_E \rightarrow 4\pi, G_E \rightarrow \rho + 2\pi$ as well as indirect four $\pi$ decay associated with $G_E \rightarrow \rho + 2\pi \rightarrow 4\pi$ and $G_E \rightarrow 2\rho \rightarrow 4\pi$, were also obtained. By appropriate choice of combination of constants of integration appearing in the solutions to the EOMs of the profile functions of the $\pi$ and $\rho$ mesons and six metric perturbation,s these decay widths were shown to match exactly with PDG data.
\end{itemize}

\item 
{\bf ${\cal O}(R^4)$-Corrected ${\cal M}$-theory uplift and $G$-Structure Torsion Classes}
\begin{itemize}
\item
 In \cite{OR4-Yadav+Misra}, two of the authors (V.Yadav and A.Misra), worked out the $O(l_p^6)$ corrections to the aforementioned  ${\cal M}$-theory  metric arising from terms quartic in the eleven dimensional supergravity action.  

\item
The $SU(3)/G_2/SU(4)/Spin(7)$-structure torsion classes of the relevant six-, seven- and eight-folds associated with the aforementioned M theory uplift (near $\psi = 0/2\pi/4\pi$ coordinare patches) were worked out which can be summarized in Table 2.
\begin{table}[h]
\begin{center}
\begin{tabular}{|c|c|c|c|}\hline
S. No. & Manifold & $G$-Structure & Non-Trivial Torsion Classes \\ \hline
1. & $M_6(r,\theta_1,\theta_2,\phi_1,\phi_2,\psi)$ & $SU(3)$ & $T^{\rm IIA}_{SU(3)} = W_1 \oplus W_2 \oplus W_3 \oplus W_4 \oplus W_5: W_4 \sim W_5$ \\ \hline
2. & $M_7(r,\theta_1,\theta_2,\phi_1,\phi_2,\psi,x^{10})$ & $G_2$ & $T^{\cal M}_{G_2} = W_{14} \oplus W_{27}$ \\ \hline
3. & $M_8(x^0,r,\theta_1,\theta_2,\phi_1,\phi_2,\psi,x^{10})$ & $SU(4)$ & $T^{\cal M}_{SU(4)} = W_2 \oplus W_3 \oplus W_5$ \\ \hline
4. & $M_8(x^0,r,\theta_1,\theta_2,\phi_1,\phi_2,\psi,x^{10})$ & $Spin(7)$ & $T^{\cal M}_{Spin(7)} = W_1 \oplus W_2$ \\ \hline
\end{tabular}
\end{center}
\caption{Summary of IR $G$-Structure Torsion Classes of Six-/Seven-/Eight-Folds in the type IIA/${\cal M}$-Theory Duals of Thermal QCD}
\end{table}
\end{itemize}
\end{itemize}

\section{Obtaining ${\cal L}_{\chi PT}^{(4)}[\pi,\rho]$ from DBI on Flavor $D6$ Branes}

In this section, similar to the discussion in \cite{HARADA} using the ``HLS formalism",  starting from the DBI action on flavor $D6$-branes (obtained from the SYZ type IIA mirror of the type IIB holographic dual of large-$N$ thermal QCD as constructed in \cite{metrics}) we review obtaining the $\chi$PT Lagrangian for $\pi, \rho$ mesons and their flavor partners, up to ${\cal O}(p^4)$ wherein the coupling constants are obtained as appropriate radial integrals.

The type IIB dual corresponding to high temperatures, i.e., $T>T_c$, will involve a black hole with the metric given by \cite{metrics}:
\begin{eqnarray}
\label{BH-T>Tc}
& & \hskip -0.5in ds^2 = \frac{1}{\sqrt{h(r,\theta_{1,2})}}\left(-g(r)dt^2 + \left(dx^1\right)^2 +  \left(dx^2\right)^2 + \left(dx^3\right)^2 \right)
+ \sqrt{h(r,\theta_{1,2})}\left(\frac{dr^2}{g(r)} + r^2 ds^2(\theta_{1,2},\phi_{1,2},\psi)\right),\nonumber\\
& &
\end{eqnarray}
where $g(r) = 1 - \frac{r_h^4}{r^4}$, and for low temperatures, i.e., $T<T_c$, is given by the thermal gravitational dual:
\begin{eqnarray}
\label{Thermal-T<Tc}
& & \hskip -0.5in ds^2 = \frac{1}{\sqrt{h(r,\theta_{1,2})}}\left(-dt^2 + \left(dx^1\right)^2 +  \left(dx^2\right)^2 + \tilde{g}(r)\left(dx^3\right)^2 \right)
+ \sqrt{h(r,\theta_{1,2})}\left(\frac{dr^2}{\tilde{g}(r)} + r^2 ds^2(\theta_{1,2},\phi_{1,2},\psi)\right)\nonumber\\
& &
\end{eqnarray}
where $\tilde{g}(r) = 1 - \frac{r_0^4}{r^4}$. One notes that $t\rightarrow x^3,\ x^3\rightarrow t$ in (\ref{BH-T>Tc}) following by a Double Wick rotation in the new $x^3, t$ coordinates obtains (\ref{Thermal-T<Tc}); $h(r,\theta_{1,2})$ is the ten-dimensional warp factor \cite{metrics, MQGP}. This amounts to:  $-g_{tt}^{\rm BH}(r_h\rightarrow r_0) = g_{x^3x^3}\ ^{\rm Thermal}(r_0),$ $ g_{x^3x^3}^{\rm BH}(r_h\rightarrow r_0) = -g_{tt}\ ^{\rm Themal}(r_0)$ in the results of
\cite{VA-Glueball-decay, OR4-Yadav+Misra} (See \cite{Kruczenski et al-2003} in the context of Euclidean/black $D4$-branes in type IIA).  

In (\ref{Thermal-T<Tc}), we will assume the spatial part of the solitonic $D3$ brane world volume to be given by $\mathbb{R}^2(x^{1,2})\times S^1(x^3)$ where the period of $S^1(x^3)$ is given by a very large: $\frac{2\pi}{M_{\rm KK}}$, where the very small $M_{\rm KK} = \frac{2r_0}{ L^2}\left[1 + {\cal O}\left(\frac{g_sM^2}{N}\right)\right]$, $r_0$ being the very small IR cut-off in the thermal background (See also \cite{Armoni et al-2020}) and $L = \left( 4\pi g_s N\right)^{\frac{1}{4}}$. So, $\lim_{M_{\rm KK}\rightarrow0}\mathbb{R}^2(x^{1,2})\times S^1(x^3) = \mathbb{R}^3(x^{1,2,3})$, thereby recovering 4D Physics.

As explained in \cite{Knauf-thesis}, the $T^3$-valued $(x, y, z)$ (used for effecting SYZ mirror via a triple T-dual in \cite{MQGP, NPB}) are defined via:
\begin{eqnarray}
\label{xyz-definitions}
& & \phi_1 = \phi_{10} + \frac{x}{\sqrt{h_2}\left[h(r_0,\theta_{10,20})\right]^{\frac{1}{4}} \sin\theta_{10}\ r_0},\nonumber\\
& & \phi_2 = \phi_{20} + \frac{y}{ \sqrt{h_4}
\left[h( r_0,\theta_{10,20})\right]^{\frac{1}{4}}\sin\theta_{20}\ r_0}\nonumber\\
& & \psi = \psi_0 + \frac{z}{\sqrt{h_1} \left[h( r_0,\theta_{10,20})
\right]^{\frac{1}{4}}\ r_0},
\end{eqnarray}
$h_{1,2,4}$ defined in \cite{metrics}, and one works up to linear order in $(x, y, z)$. Up to linear order in $r$, i.e., in the IR, it can be shown \cite{theta0-theta} that $\theta_{10,20}$ can be promoted to global coordinates $\theta_{1,2}$ in all the results in the paper.
 The meson sector in the type IIA dual background of top-down holographic type IIB setup is given by the flavor $D6$-branes action.
 For
\begin{equation}
\label{theta12-deloc}
\theta_{1}=\frac{\alpha_{\theta_{1}}}{N^{\frac{1}{5}}}, \ \theta_{2}=\frac{\alpha_{\theta_{2}}}{N^{\frac{3}{10}}},
\end{equation}
i.e., restricting to the Ouyang embedding (\ref{Ouyang-definition}) for a vanishingly small $|\mu_{\rm Ouyang}|$,
 one will assume that the embedding of the $D6$-brane will be given by $\iota :\Sigma ^{1,6}\Bigg( R^{1,3},r,\theta_{2}\sim\frac{\alpha_{\theta_{2}}}{N^{\frac{3}{10}}},y\Bigg)\hookrightarrow M^{1,9}$, effected by: $z=z(r)$.  As obtained in \cite{Yadav+Misra+Sil-Mesons} one sees that $z$=constant is a solution and by choosing $z=\pm {\cal C}\frac{\pi}{2}$, one can choose the $D6/\overline{D6}$-branes to be at ``antipodal" points along the z coordinate. As in \cite{Yadav+Misra+Sil-Mesons}, we will be working with redefined $(r,z)$ in terms of new variables $(Y,Z)$:
\[r=r_{0}e^{\sqrt{Y^{2}+Z^{2}}}\]
\[z={\cal C}\arctan\frac{Z}{Y}.\]
 Vector mesons are obtained by considering gauge fluctuations of a background gauge field along the world volume of the embedded flavor $D6$-branes (with world volume ${\Sigma}_7(x^{0,1,2,3},Z,\theta_2,\tilde{y}) = {\Sigma}_2(\theta_2,\tilde{y})\times{\Sigma}_5(x^{0,1,2,3},Z)$). Turning on a gauge field fluctuation $\tilde{F}$ about a small background gauge field $F_0$ and the backround $i^*(g+B) [i:\Sigma_7\hookrightarrow M_{10}$, $M_{10}$ being the ten-dimensional ambient space-time]. This implies:
{\footnotesize
\begin{eqnarray}
\label{DBI expansion}
& & {\rm Str}\left.\sqrt{{\rm det}_{{\Sigma}_7(x^{0,1,2,3},Z,\theta_2,\tilde{y})}\left(i^*(G+B) + (F_0 + \tilde{F})\right)}\right|_{Y=0}\delta\left(\theta_2 - \frac{\alpha_{\theta_2}}{N^{\frac{3}{10}}}\right)\nonumber\\
& & = \sqrt{{\rm det}_{{\Sigma}_2(\theta_2,\tilde{y})}\left(i^*(g+B)\right)}\left. {\rm Str}\sqrt{{\rm det}_{{\Sigma}_5(x^{0,1,2,3},Z)}\left(i^*(g+B) + (F_0 + \tilde{F})\right)}\right|_{Y=0}\delta\left(\theta_2 - \frac{\alpha_{\theta_2}}{N^{\frac{3}{10}}}\right)
\nonumber\\
& & = \left.\sqrt{{\rm det}_{{\Sigma}_2(\theta_2,\tilde{y})}\left(i^*(g+B)\right)}\sqrt{{\rm det}_{{{\Sigma}_5(x^{0,1,2,3},Z)}}(i^*g)} {\rm Str}\left({\bf 1}_3 - \frac{1}{2}\left[(i^*g)^{-1}\left((F_0 + \tilde{F})\right)\right]^2 + ....\right)\right|_{Y=0}\delta\left(\theta_2 - \frac{\alpha_{\theta_2}}{N^{\frac{3}{10}}}\right),
\nonumber\\
\end{eqnarray}
}
where $Y=0$ is the SYZ mirror of the Ouyang embedding \cite{Yadav+Misra+Sil-Mesons}.

Picking up terms quadratic in $\tilde{F}$:
{\footnotesize
\begin{equation}
\label{DBI action}
{\rm S}^{IIA}_{D6}=\frac{T_{D_6}(2\pi\alpha^\prime)^{2}}{4} \left(\frac{\pi L^2}{r_0}\right){\rm Str}\int \prod_{i=0}^3dx^i dZd\theta_{2}dy \delta\Bigg(\theta_{2}-\frac{\alpha_{\theta_{2}}}{N^{\frac{3}{10}}}\Bigg) e^{-\Phi} \sqrt{-{\rm det}_{\theta_{2}y}(\iota^*(g+B))}\sqrt{{\rm det}_{{\mathbb  R}^{1,3},Z}(\iota^*g)}g^{\tilde{\mu}\tilde{\nu}}\tilde{F}_{\tilde{\nu}\tilde{\rho}}g^{\tilde{\rho}\tilde{\sigma}}\tilde{F}_{\tilde{\sigma}\tilde{\mu}},
\end{equation}
}
where $\tilde{\mu}=i(=0,1,2,3),Z$

To begin with, for simplicity let us assume the absence of any external (axial-)vector fields. Expanding $A_\mu(x^\nu,Z) = \sum_{n=1}^\infty \rho^{(n)}_\mu(x^\nu)\psi_n(Z), A_Z(x^\nu,Z) = \sum_{n=0}^\infty\pi^{(n)}(x^\nu)\phi_n(Z) $, one obtains (as also earlier discussed in \cite{SS1, Dasgupta_et_al_Mesons, VA-Glueball-decay}):
\begin{eqnarray}
\label{full-expansion}
& &  -\frac{{\cal V}_{\Sigma_2}}{4}\int d^3x dZ \sum_{n,m}{\rm tr}\Biggl({\cal V}_2(Z)\tilde{F}^{(n)}_{\mu\nu}\tilde{F}^{(m)\mu\nu}\psi_m(Z)\psi_n(Z) + {\cal V}_1(Z)\rho^{(m)}_\mu \rho^{(n)\mu }\dot{\psi}_m\dot{\psi}_n\nonumber\\
&& - {\cal V}_1(Z)\partial_{\mu}\pi^{(n)}{\rho^{(m)}}^{\mu}\phi_{n}\dot{\psi}_{m}- {\cal V}_1(Z)\partial_{\mu}\pi^{(m)}{\rho^{(n)}}^{\mu}\phi_{m}\dot{\psi}_{n}\Biggr).
\end{eqnarray}
The terms quadratic in $\psi/\dot{\psi}(\ ^.\equiv\frac{d}{dZ})$ in (\ref{full-expansion}) are given as:
\begin{eqnarray}
\label{Ftildesq-i}
& & \hskip -0.3in -\frac{{\cal V}_{\Sigma_2}}{4}\int d^3x dZ \sum_{n,m}{\rm tr}\left({\cal V}_2(Z)\tilde{F}^{(n)}_{\mu\nu}\tilde{F}^{(m)\mu\nu}\psi_m(Z)\psi_n(Z) + {\cal V}_1(Z)\rho^{(m)}_\mu \rho^{(n)\mu }\dot{\psi}_m\dot{\psi}_n\right),
\end{eqnarray}
where:
\noindent \[ F_{\mu\nu}(x^\rho,Z) = \sum_n \tilde{F}_{\mu\nu}^{(n)}(x^\rho)\psi_n(Z), \]
\noindent \[ {\cal V}_{\Sigma_2} = -\frac{T_{D_6}(2\pi\alpha^\prime)^{2}}{4}\int dyd\theta_{2}\delta\Bigg(\theta_{2}-\frac{\alpha_{\theta_{2}}}{N^{\frac{3}{10}}}\Bigg),\]
and,
\begin{eqnarray}
\label{V1V2-defs}
& & \mathcal{V}_1(z)= 2\sqrt{h}g^{zz} e^{-\phi} \sqrt{-{\rm det}_{\theta_{2}y}(\iota^*(g+B))}\sqrt{{\rm det}_{{\mathbb  R}^{1,3},Z}(\iota^*g)}, \nonumber\\
& & \mathcal{V}_2(z) = he^{-\phi} \sqrt{-{\rm det}_{\theta_{2}y}(\iota^*(g+B))}\sqrt{{\rm det}_{{\mathbb  R}^{1,3},Z}(\iota^*g)}.
\end{eqnarray}

The EOM satisfied by $\rho_\mu(x^\nu)^{(n)}$ is: $\partial_\mu \tilde{F}^{\mu\nu}_{(n)} + \partial_\mu\log\sqrt{g_{t,\mathbb{R}^{1,2},|Z|}}\tilde{F}^{\mu\nu}_{(n)} = \partial_\mu\tilde{F}^{\mu\nu}_{(n)} = {\cal M}_{(n)}^2\rho^\nu_{(n)}$. After integrating by parts once, and utilizing the EOM for $\rho^{(n)}_\mu$, one writes :
\begin{eqnarray}
\label{Ftildesq-ii}
& & \int d^3x dZ\  {\rm tr}\left(-2 {\cal V}_2(Z) {\cal M}_{(m)}^2\psi_n^{\rho_\mu}\psi_m^{\rho_\mu} + {\cal V}_1(Z)\dot{\psi}_n^{\rho_\mu}\dot{\psi}_m^{\rho_\mu}\right)\rho^{\mu (n)}\rho_{\mu}^{(m)},
\end{eqnarray}
which yields the following equation of motion for $\psi^\mu_{(m)}$:
\begin{eqnarray}
\label{eoms_psi_n_rhomu}
& & \psi^{\mu}_{(m)}: \frac{d}{dZ\ }\left({\cal V}_1(Z) \dot{\psi}_{(m)}^{\mu}\right) + 2 {\cal V}_2(Z){\cal M}_{(m)}^2\psi^{\mu}_m = 0.
\end{eqnarray}

The normalization condition of $\psi_{(n)}$ are given as
\begin{eqnarray}
\label{norm_psi}
&&{\cal V}_{\Sigma_2}\int dZ\ {\cal V}_{2}(Z)\ \psi_{n}\psi_{(m)}=\delta_{nm}\nonumber\\
&&\frac{{\cal V}_{\Sigma_2}}{2}\int dZ\ {\cal V}_{1}(Z)\ \partial_{Z}\psi_{n} \partial_{Z}\psi_{(m)}={\cal M}_{(n)}^2\delta_{nm}.
\end{eqnarray}
Thus the action for vector meson part for all $n\ge 1$can be wriiten as
\begin{eqnarray}
& & -\int d^3x  \sum_{n}{\rm tr}\left(\frac{1}{4}\tilde{F}^{(n)}_{\mu\nu}\tilde{F}^{(n)\mu\nu}+\frac{{\cal M}_{(n)}^2}{2}\rho^{(n)}_\mu \rho^{(n)\mu }\right).
\end{eqnarray}

To normalize the kinetic term for $\pi ^{(n)}$, we impose the normalization condition for all $n$ corresponding to $\pi ^{(n)}$ which ranges from 0 to $\infty $
\begin{eqnarray}
\label{norm_scalar}
&&\frac{{\cal V}_{\Sigma_2}}{2}\int dZ\ {\cal V}_{1}(Z)\ \phi_{n}\phi_{m}=\delta_{nm}.
\end{eqnarray}
From (\ref{norm_psi}), it is seen that we can choose $\phi_{n}={\cal M}_{(n)}^{-1}\dot{\psi}_{n}$ for all $n\ge 1$. For $n=0$ corresponding to $\phi_0$ we choose its form such as it is orthogonal to $\dot{\psi}_{n}$ for all $n\ge 1$. By writing $\phi_{0}=\frac{C}{{\cal V}_{1}(Z)}$, we have
$$(\phi_{0},\phi_{n})\propto\int\ dZ\ C\partial_{Z}\psi\ =0.$$
Thus the cross component in (\ref{full-expansion}) vanishes for $n=0$, and the remaining cross components can be absorbed in the $\rho_{\mu}^{(n)}$ by following a specific gauge transformation given as,
$$\rho_{\mu}^{(n)}\rightarrow\rho_{\mu}^{(n)}+{\cal M}_{(n)}^{-1}\partial_{\mu}\pi^{(n)}.$$
Then the action becomes:
\begin{eqnarray}
& & -\int d^3x\ \ {\rm tr} \left[\frac{1}{2}\partial_{\mu}\pi^{(0)}\partial^{\mu}\pi^{(0)} + \sum_{n\ge 1}\left(\frac{1}{4}\tilde{F}^{(n)}_{\mu\nu}\tilde{F}^{(n)\mu\nu}+\frac{m_{n}^2}{2}\rho^{(n)}_\mu \rho^{(n)\mu }\right)\right].
\end{eqnarray}

Working in the $A_Z(x^\mu,Z)=0$-gauge, integrating out all higher order vector and axial vector meson fields except keeping only the lowest vector meson field \cite{HARADA} \\ $V_\mu^{(1)}(x^\mu) = g \rho_\mu(x^\mu) = \left(\begin{array}{ccc}
\frac{1}{\sqrt{2}}\left(\rho_\mu^0 + \omega_\mu\right) & \rho_\mu^+ & K_\mu^{*+} \\
\rho_\mu^- & -\frac{1}{\sqrt{2}}\left(\rho_\mu^0 - \omega_\mu\right) & K_\mu^{*0} \\
K_\mu^{*-} & {\bar K}_\mu^{*0}  & \phi_\mu
\end{array} \right)$ and lightest pseudo-scalar meson field i.e. $\pi = \frac{1}{\sqrt{2}}\left(\begin{array}{ccc}
\frac{1}{\sqrt{2}}\pi^0 + \frac{1}{\sqrt{6}}\eta_8 + \frac{1}{\sqrt{3}}\eta_0 & \pi^+ & K^+ \\
\pi^- & -\frac{1}{\sqrt{2}}\pi^0 + \frac{1}{\sqrt{6}}\eta_8 + \frac{1}{\sqrt{3}}\eta_0 & K^0 \\
K^- & {\bar K}^0 & -\frac{2}{\sqrt{6}}\eta_8 + \frac{1}{\sqrt{3}}\eta_0
\end{array}\right)$ meson, the gauge field $A_\mu(x^\nu,Z)$ up to ${\cal O}(\pi)$ is given by:
\begin{eqnarray}
\label{Amu-exp}
& & A_\mu(x^\nu,Z) = \frac{\partial_\mu\pi}{F_\pi}\psi_0(Z) - V_\mu(x^\nu)\psi_1(Z),
\end{eqnarray}
where $\psi_0(z) = \int^Z_0 dZ^\prime \phi_0(Z^\prime), V_\mu^{(1)}(x^\nu) = \rho^{(1)}_\mu - \frac{1}{{\cal M}_{(1)}}\partial_\mu\pi^{(1)}$.

To introduce external vector ${\cal V}_\mu$ and axial vector fields ${\cal A}_\mu$, one could use the Hidden Local Symmetry (HLS) formalism of \cite{HARADA} and references therein, wherein  $\frac{1}{F_\pi} \partial_\mu \pi\rightarrow\hat{\alpha}_{\mu \perp}= \frac{1}{F_\pi} \partial_\mu \pi + {\cal A}_\mu - \frac{i}{F_\pi}[{\cal V}_\mu,\pi] + \cdots$ (refer to (\ref{al perp exp})), and one also works with $\hat{\alpha}_{||} \equiv -V_\mu + {\cal V}_\mu - \frac{i}{2F_\pi^2}[\partial_\mu\pi,\pi] + \cdots$ (refer to (\ref{al para exp})).
 To obtain the low energy effective theory of QCD, again truncating the KK spectrum at certain level because mode expansion of the gauge field contains infinite number of vector meson fields $V_\mu^{(n)}(x^\mu)$ and axial vector meson fields $A_\mu^{(n)}(x^\mu)$ \cite{HARADA},
\begin{equation}
A_\mu(x^\mu,z) = \hat{\alpha}_{\mu \perp}(x^\mu) \psi_0(z)
  + (\hat{\alpha}_{\mu ||}(x^\mu) + V_\mu^{(1)}(x^\mu)  )
  + \hat{\alpha}_{\mu ||}(x^\mu)  \psi_1(z),
  \end{equation}
implying therefore
\begin{eqnarray}
\label{F mu nu}
& &   F_{\mu\nu} = -V_{\mu\nu} \psi_1 +v_{\mu\nu}(1+\psi_1) + a_{\mu\nu} \psi_0 - i[\hat{\alpha}_{\mu ||},\hat{\alpha}_{\nu ||}]\psi_1(1+\psi_1)  + i[\hat{\alpha}_{\mu \perp},\hat{\alpha}_{\nu \perp}](1+\psi_1-\psi_0^{2}) \nonumber\\
  & &
  -i([\hat{\alpha}_{\mu \perp},\hat{\alpha}_{\nu ||}]+[\hat{\alpha}_{\mu ||},\hat{\alpha}_{\nu \perp}])\psi_1 \psi_0.
 \end{eqnarray}
From appendix {\bf C} (based on \cite{HLS-Physics-Reports}), as regards a chiral power counting, one notes that $M_\rho\equiv{\cal O}(p)$ implying $\hat{\alpha}_{\nu ||}\equiv \frac{{\cal O}(p^3)}{M_\rho^2}\equiv{\cal O}(p),
\hat{\alpha}_{\nu\perp}\equiv {\cal O}(p)$. Further, $V_{\mu\nu}, a_{\mu\nu}$ and $v_{\mu\nu}$ are of ${\cal O}(p^2)$. Hence, using (\ref{F mu nu}), $\left(F_{\mu\nu}F^{\mu\nu}\right)^m$ is of ${\cal O}(p^{4m}), m\in\mathbb{Z}^+$. Therefore, one considers the kinetic term ($m=1$) at ${\cal O}(p^4)$, which yields the following expansion:
 \begin{eqnarray}
\label{F_munu F^munu}
& &  F_{\mu\nu}F^{\mu\nu} = \psi_1^2 V_{\mu\nu}V^{\mu\nu} - \psi_1(1+\psi_1) V_{\mu\nu}v^{\mu\nu} - \psi_0\psi_1 V_{\mu\nu}a^{\mu\nu} +i \psi_1^2(1+\psi_1)V_{\mu\nu}[\hat{\alpha}^\mu_{||},\hat{\alpha}^\nu_{ ||}] \nonumber\\
 &&
 - i \psi_1(1+\psi_1 - \psi_0^2)V_{\mu\nu}[\hat{\alpha}^\mu_{\perp},\hat{\alpha}^\nu_{\perp}] + i\psi_0\psi_1^2V_{\mu\nu}([\hat{\alpha}^{\mu}_{\perp},\hat{\alpha}^{\nu}_{||}]+[\hat{\alpha}^{\mu}_{||},\hat{\alpha}^{\nu}_{\perp}]) -  \psi_1(1+\psi_1) v_{\mu\nu}V^{\mu\nu} \nonumber\\
 && + (1+\psi_1)^2 v_{\mu\nu}v^{\mu\nu}+\psi_0(1+\psi_1) v_{\mu\nu}a^{\mu\nu} - i \psi_1(1+\psi_1)^2 v_{\mu\nu}[\hat{\alpha}^\mu_{||},\hat{\alpha}^\nu_{ ||}] \nonumber\\
 && +i(1+\psi_1) (1+\psi_1 - \psi_0^2)v_{\mu\nu}[\hat{\alpha}^\mu_{\perp},\hat{\alpha}^\nu_{\perp}] - i\psi_0\psi_1(1+\psi_1)v_{\mu\nu}([\hat{\alpha}^{\mu}_{\perp},\hat{\alpha}^{\nu}_{||}]+[\hat{\alpha}^{\mu}_{||},\hat{\alpha}^{\nu}_{\perp}]) \nonumber\\
 &&  - \psi_0\psi_1 a_{\mu\nu}V^{\mu\nu}+\psi_0(1+\psi_1) a_{\mu\nu}v^{\mu\nu} +\psi_0^2 a_{\mu\nu}a^{\mu\nu} - i\psi_0\psi_1(1+\psi_1) a_{\mu\nu}[\hat{\alpha}^\mu_{||},\hat{\alpha}^\nu_{ ||}] \nonumber\\
 &&  +i\psi_0(1+\psi_1-\psi_0^2)a_{\mu\nu}[\hat{\alpha}^\mu_{\perp},\hat{\alpha}^\nu_{\perp}] - i\psi_0^2\psi_1 a_{\mu\nu}([\hat{\alpha}^{\mu}_{\perp},\hat{\alpha}^{\nu}_{||}]+[\hat{\alpha}^{\mu}_{||},\hat{\alpha}^{\nu}_{\perp}]) \nonumber\\
 &&
  +i \psi_1^2(1+\psi_1)[\hat{\alpha}_{\mu ||},\hat{\alpha}_{\nu ||}]V^{\mu\nu} - i \psi_1(1+\psi_1)^2[\hat{\alpha}_{\mu ||},\hat{\alpha}_{\nu ||}]v^{\mu\nu} \nonumber\\
 &&  -i\psi_0\psi_1(1+\psi_1)[\hat{\alpha}_{\mu ||},\hat{\alpha}_{\nu ||}]a^{\mu\nu} - \psi_1^2(1+\psi_1)^2[\hat{\alpha}_{\mu ||},\hat{\alpha}_{\nu ||}][\hat{\alpha}^\mu_{||},\hat{\alpha}^\nu_{||}]\nonumber\\
 &&  +\psi_1(1+\psi_1)(1+\psi_1-\psi_0^2)[\hat{\alpha}_{\mu ||},\hat{\alpha}_{\nu ||}][\hat{\alpha}^\mu_{\perp},\hat{\alpha}^\nu_{\perp}] - \psi_0\psi_1^2(1+\psi_1)[\hat{\alpha}_{\mu ||},\hat{\alpha}_{\nu ||}]([\hat{\alpha}^{\mu}_{\perp},\hat{\alpha}^{\nu}_{||}]+[\hat{\alpha}^{\mu}_{||},\hat{\alpha}^{\nu}_{\perp}]) \nonumber\\
 &&  - i\psi_1(1+\psi_1-\psi_0^2)[\hat{\alpha}_{\mu \perp},\hat{\alpha}_{\nu \perp}]V^{\mu\nu} +i(1+\psi_1)(1+\psi_1-\psi_0^2)[\hat{\alpha}_{\mu \perp},\hat{\alpha}_{\nu \perp}]v^{\mu\nu} \nonumber\\
 &&  + \psi_1(1+\psi_1)(1+\psi_1-\psi_0^2)[\hat{\alpha}_{\mu \perp},\hat{\alpha}_{\nu \perp}][\hat{\alpha}^\mu_{||},\hat{\alpha}^\nu_{||}]
  -(1+\psi_1-\psi_0^2)^2[\hat{\alpha}_{\mu \perp},\hat{\alpha}_{\nu \perp}][\hat{\alpha}^\mu_{\perp},\hat{\alpha}^\nu_{\perp}] \nonumber\\
 &&  + \psi_0\psi_1(1+\psi_1-\psi_0^2)[\hat{\alpha}_{\mu \perp},\hat{\alpha}_{\nu \perp}]  ([\hat{\alpha}^{\mu}_{\perp},\hat{\alpha}^{\nu}_{||}]+[\hat{\alpha}^{\mu}_{||},\hat{\alpha}^{\nu}_{\perp}]) \nonumber\\
 &&  +i \psi_0\psi_1^2([\hat{\alpha}_{\mu \perp},\hat{\alpha}_{\nu ||}]+[\hat{\alpha}_{\mu ||},\hat{\alpha}_{\nu \perp}])V^{\mu\nu} - i\psi_0\psi_1(1+\psi_1)([\hat{\alpha}_{\mu \perp},\hat{\alpha}_{\nu ||}]+[\hat{\alpha}_{\mu ||},\hat{\alpha}_{\nu \perp}])v^{\mu\nu} \nonumber\\
 &&  -i\psi_0^2\psi_1([\hat{\alpha}_{\mu \perp},\hat{\alpha}_{\nu ||}]+[\hat{\alpha}_{\mu ||},\hat{\alpha}_{\nu \perp}])a^{\mu\nu} -\psi_0\psi_1^2(1+\psi_1)([\hat{\alpha}_{\mu \perp},\hat{\alpha}_{\nu ||}]+[\hat{\alpha}_{\mu ||},\hat{\alpha}_{\nu \perp}])[\hat{\alpha}^\mu_{||},\hat{\alpha}^\nu_{ ||}] \nonumber\\
 && +\psi_0\psi_1(1+\psi_1-\phi_0^2)([\hat{\alpha}_{\mu \perp},\hat{\alpha}_{\nu ||}]+[\hat{\alpha}_{\mu ||},\hat{\alpha}_{\nu \perp}])[\hat{\alpha}^\mu_{\perp},\hat{\alpha}^\nu_{\perp}]  \nonumber\\
 && - \psi_0^2\psi_1^2([\hat{\alpha}_{\mu \perp},\hat{\alpha}_{\nu ||}]+[\hat{\alpha}_{\mu ||},\hat{\alpha}_{\nu \perp}])([\hat{\alpha}^{\mu}_{\perp},\hat{\alpha}^{\nu}_{||}]+[\hat{\alpha}^{\mu}_{||},\hat{\alpha}^{\nu}_{\perp}]).
  \end{eqnarray}
Defining  parity as $x^i\rightarrow-x^i$, $i$ indexing the conformally Minkowskian spatial directions and $Z\rightarrow-Z$, given that $A_\mu(x, Z)$ will be odd, $\alpha_\perp$ will be even, $\alpha_{||}$ will be odd and $V_\mu$ will be odd implies $\psi_0(Z)$  will be odd and $\psi_1(Z)$ will be even. As coupling constants are assumed to be scalars and they are given by $Z$-integrals, the $Z$-dependent terms in the action must be separately of even-$Z$ parity. As  $\psi_0$ has odd $Z$-parity and $\psi_1$ has even $Z$-parity, therefore at ${\cal O}(p^4)$, terms with $(3\hat{\alpha} _{||}s\ ,\ 1\hat{\alpha}_\perp\ {\rm or}\ 3\hat{\alpha}_\perp s\ ,\ 1\hat{\alpha} _{||} )$, are dropped as they involve coefficients of the type $\psi_0^{2m+1}\psi_1^{2n}(Z)$ for appropriate postive integral values of $n, m$. Similarly, at ${\cal O}(p^2)$, $tr(\hat{\alpha}_{\mu \perp}\hat{\alpha}_{||}^{\mu})$  accompanied by $\dot{\psi_0}\dot{\psi_1}(\ ^. \equiv \frac{d}{dZ})$ of odd-$Z$ parity, is dropped.
At ${\cal O}(p^4)$, one hence obtains \cite{HARADA}:
\begin{eqnarray}
\label{Lagrangian-Op4}
 & & {\mathcal{L}}_{(4)}
\ni
y_1 \,
{\rm tr}[{\hat{\alpha}}_{\mu\perp}{\hat{\alpha}}^{\mu}_{\perp}
{\hat{\alpha}}_{\nu\perp}{\hat{\alpha}}^{\nu}_{\perp}]
+
y_2 \,
{\rm tr}[{\hat{\alpha}}_{\mu\perp}{\hat{\alpha}}_{\nu\perp}
{\hat{\alpha}}^{\mu}_{\perp}{\hat{\alpha}}^{\nu}_{\perp}]
+y_3 \,
{\rm tr}[{\hat{\alpha}}_{\mu||}{\hat{\alpha}}^{\mu}_{||}
{\hat{\alpha}}_{\nu||}{\hat{\alpha}}^{\nu}_{||}]
+y_4 \,
{\rm tr}[{\hat{\alpha}}_{\mu||}{\hat{\alpha}}_{\nu||}
{\hat{\alpha}}^{\mu}_{||}{\hat{\alpha}}^{\nu}_{||}] \nonumber\\
& &
+y_5 \,
{\rm tr}[{\hat{\alpha}}_{\mu\perp}{\hat{\alpha}}^{\mu}_{\perp}
{\hat{\alpha}}_{\nu||}{\hat{\alpha}}^{\nu}_{||}]
+y_6 \,
{\rm tr}[{\hat{\alpha}}_{\mu\perp}{\hat{\alpha}}_{\nu\perp}
{\hat{\alpha}}^{\mu}_{||}{\hat{\alpha}}^{\nu}_{||}]
+y_7 \,
{\rm tr}[{\hat{\alpha}}_{\mu\perp}{\hat{\alpha}}_{\nu\perp}
{\hat{\alpha}}^{\nu}_{||}{\hat{\alpha}}^{\mu}_{||}] \nonumber\\
& &
+y_8 \,
\left\{ {\rm tr}[{\hat{\alpha}}_{\mu\perp}{\hat{\alpha}}^{\nu}_{||}
{\hat{\alpha}}_{\nu\perp}{\hat{\alpha}}^{\mu}_{||}]
+
{\rm tr}[{\hat{\alpha}}_{\mu\perp}{\hat{\alpha}}^{\mu}_{||}
{\hat{\alpha}}_{\nu\perp}{\hat{\alpha}}^{\nu}_{||}\right\}
+y_9 \,
{\rm tr}[{\hat{\alpha}}_{\mu\perp}{\hat{\alpha}}_{\nu ||}
{\hat{\alpha}}^{\mu}_{\perp}{\hat{\alpha}}^{\nu}_{||}]\nonumber\\
& &
+z_1 \,
{\rm tr}[v_{\mu\nu}v^{\mu\nu}]
+z_2 \,
{\rm tr}[a_{\mu\nu}a^{\mu\nu}]
+z_3 \,
{\rm tr}[v_{\mu\nu}V^{\mu\nu}]
+iz_4 \,
{\rm tr}[V_{\mu\nu}{\hat{\alpha}}^{\mu}_{\perp}
{\hat{\alpha}}^{\nu}_{\perp}]\nonumber\\
& &
+iz_5 \,
{\rm tr}[V_{\mu\nu}{\hat{\alpha}}^{\mu}_{||}
{\hat{\alpha}}^{\nu}_{||}]
+iz_6 \,
{\rm tr}[v_{\mu\nu}
{\hat{\alpha}}^{\mu}_{\perp}{\hat{\alpha}}^{\nu}_{\perp}]
+iz_7 \,
{\rm tr}[v_{\mu\nu}
{\hat{\alpha}}^{\mu}_{||}{\hat{\alpha}}^{\nu}_{||}]
-iz_8 \,
{\rm tr}\left[a_{\mu\nu}
\left({\hat{\alpha}}^{\mu}_{\perp}{\hat{\alpha}}^{\nu}_{||}
+{\hat{\alpha}}^{\mu}_{||}{\hat{\alpha}}^{\nu}_{\perp}
\right)\right]
\end{eqnarray}
where:
\begin{equation}
 v_{\mu\nu}
= \frac{1}{2} \left(
\xi_R {\cal R}_{\mu\nu} \xi^\dag_R + \xi_L {\cal L}_{\mu\nu} \xi_L^\dag \right)  \hspace{0.5cm}
{\rm and} \hspace{0.5cm}
a_{\mu\nu}
=  \frac{1}{2} \left(
\xi_R {\cal R}_{\mu\nu} \xi^\dag_R - \xi_L {\cal L}_{\mu\nu} \xi_L^\dag
\right),
\end{equation}
${\cal L}_{\mu\nu} = \partial_{[\mu}{\cal L}_{\nu]} - i[{\cal L}_\mu,{\cal L}_\nu]$ and ${\cal R}_{\mu\nu} = \partial_{[\mu}{\cal R}_{\nu]} - i[{\cal R}_\mu,{\cal R}_\nu]$ and ${\cal L}_\mu = {\cal V}_\mu - {\cal A}_\mu$ where ${\cal R}_\mu =  {\cal V}_\mu + {\cal A}_\mu $, and $\xi_L^\dagger(x^\mu) = \xi_R(x^\mu) = e^{\frac{i \pi(x^\mu)}{F_\pi}}$; also, $V_{\mu\nu} = \partial_{[\mu}V_{\nu]}-i[V_\mu,V_\nu]$.

The various  couplings, using (\ref{F_munu F^munu}), are hence given by the following expressions \cite{HARADA}:
 \begin{eqnarray}
\label{y_i+z_i}
& &  F_\pi^{2} = -\frac{{\cal V}_{\Sigma_2}}{4}<<\dot\psi_0^{2}>>
\nonumber\\
& &
 aF_\pi^{2} = -\frac{{\cal V}_{\Sigma_2}}{4}<<\dot\psi_1^{2}>>
 \nonumber\\
& &
 \frac{1}{g^2} = \frac{{\cal V}_{\Sigma_2}}{2}<\psi_1^{2}>
 \nonumber\\
& &
  y_1 = -y_2 = -\frac{{\cal V}_{\Sigma_2}}{2}<(1+\psi_1-\psi_0)^2>
  \nonumber\\
& &
  y_3 = -y_4 = -\frac{{\cal V}_{\Sigma_2}}{2}<\psi_1^{2}(1+\psi_1)^2>
  \nonumber\\
& &
  y_5 = -{\cal V}_{\Sigma_2}<\psi_0^{2}\psi_1^{2}>
  \nonumber\\
& &
  y_6 = -y_7=-{\cal V}_{\Sigma_2}<\psi_1(1+\psi_1)(1+\psi_1-\psi_0^{2})>
 \nonumber\\
& &
  y_8 = -y_9= -{\cal V}_{\Sigma_2}<\psi_0^{2}\psi_1^{2}>
  \nonumber\\
& &
  z_1 = -\frac{{\cal V}_{\Sigma_2}}{4}<(1+\psi_1)^2>
  \nonumber\\
& &
  z_2 = -\frac{{\cal V}_{\Sigma_2}}{4}<(\psi_0)^2>
  \nonumber\\
& &
  z_3 = \frac{{\cal V}_{\Sigma_2}}{2}<\psi_1(1+\psi_1)>
 \nonumber\\
& &
  z_4 = {\cal V}_{\Sigma_2}<\psi_1(1+\psi_1-\psi_0^{2})
  \nonumber\\
& &
  z_5 = -{\cal V}_{\Sigma_2}<\psi_1^{2}(1+\psi_1)>
  \nonumber\\
& &
  z_6 = -{\cal V}_{\Sigma_2}<(1+\psi_1)(1+\psi_1-\psi_0^{2})>
  \nonumber\\
& &
  z_7 = {\cal V}_{\Sigma_2}<\psi_1(1+\psi_1)^{2}>
  \nonumber\\
& &
  z_8 = {\cal V}_{\Sigma_2}<\psi_0^{2}\psi_1>
  \end{eqnarray}
  where:
  \begin{equation}
\label{<<A>>}
  << A >> =  \int_{0}^{+\infty}{\mathcal{V}_1(z)} A dZ
\end{equation}
and
\begin{equation}
\label{<A>}
 < A > =  \int_{0}^{+\infty}{\mathcal{V}_2(z)} A dZ.
\end{equation}

\section{The Coupling Constants in ${\cal L}_{\chi PT}^{[4]}(\pi,\rho)$ from $S_{\rm DBI}^{D6}$ Incorporating ${\cal O}(R^4)$-Corrections from M Theory}

This section has the core of the main results of this paper. Inclusive of the ${\cal O}(R^4)$ corrections to the ${\cal M}$-theory uplift of large-$N$ thermal QCD as worked out in \cite{OR4-Yadav+Misra}, we show how to obtain lattice-compatible values of the coupling constants up to ${\cal O}(p^4)$ of the $\chi$PT Lagrangian of \cite{GL} in the chiral limit. The ${\cal O}(R^4)$-corrections - indicated by a $\tilde{}$ (e.g., the  ${\cal M}$-theory  metric: $\tilde{G}_{MN}^{\cal M}=G^{\rm MQGP}_{MN}\left(1+f_{MN}\right)$ \cite{OR4-Yadav+Misra}) in (\ref{V_12}) - to the $D6$-brane DBI action is described in Appendix D.

The ${\cal O}(\beta)$-corrected ${\cal M}$-theory metric of \cite{MQGP} in the MQGP limit adapted to the thermal background (\ref{Thermal-T<Tc}) near the $\psi=2n\pi, n=0, 1, 2$-branches up to ${\cal O}((r-r_0)^2)$ [and up to ${\cal O}((r-r_0)^3)$ for some of the off-diagonal components along the delocalized $T^3(x,y,z)$] - the components which do not receive an ${\cal O}(\beta)$ corrections, are not listed in (\ref{ M-theory-metric-psi=2npi-patch}) - is given below \cite{OR4-Yadav+Misra}:
{\scriptsize
\begin{eqnarray}
\label{ M-theory-metric-psi=2npi-patch}
 \hskip -0.5in \tilde{G}^{\cal M}_{x^3x^3} & = & G^{\rm MQGP}_{x^3x^3}\Biggl[1 + \frac{1}{4}  \frac{4 b^8 \left(9 b^2+1\right)^3 \left(4374 b^6+1035 b^4+9 b^2-4\right) \beta  M \left(\frac{1}{N}\right)^{9/4} \Sigma_1
   \left(6 a^2+  {r_0}^2\right) \log (  {r_0})}{27 \pi  \left(18 b^4-3 b^2-1\right)^5  \log N ^2   {N_f}   {r_0}^2
   \alpha _{\theta _2}^3 \left(9 a^2+  {r_0}^2\right)} (r-  {r_0})^2\Biggr]
\nonumber\\
\tilde{G}^{\cal M}_{x^{0,1,2}x^{0,1,2}} & = &  G^{\rm MQGP}_{x^{0,1,2}x^{0,1,2}}
\Biggl[1 - \frac{1}{4} \frac{4 b^8 \left(9 b^2+1\right)^4 \left(39 b^2-4\right) M \left(\frac{1}{N}\right)^{9/4} \beta  \left(6 a^2+{r_0}^2\right) \log
   ({r_0})\Sigma_1}{9 \pi  \left(3 b^2-1\right)^5 \left(6 b^2+1\right)^4 \log N ^2 {N_f} {r_0}^2 \left(9 a^2+{r_0}^2\right) \alpha
   _{\theta _2}^3} (r - {r_0})^2\Biggr]\nonumber\\
\tilde{G}^{\cal M}_{rr} & = & G^{\rm MQGP}_{rr}\Biggl[1 + \Biggl(- \frac{2 \left(9 b^2+1\right)^4 b^{10} M   \left(6 a^2+{r_0}^2\right) \left((r-{r_0})^2+{r_0}^2\right)\Sigma_1}{3 \pi
   \left(-18 b^4+3 b^2+1\right)^4 \log N  N^{8/15} {N_f} \left(-27 a^4+6 a^2 {r_0}^2+{r_0}^4\right) \alpha _{\theta
   _2}^3}\nonumber\\
& & +{C_{zz}}^{(1)}-2 {C_{\theta_1z}}^{(1)}+2 {C_{\theta_1x}}^{(1)}\Biggr)\beta\Biggr]\nonumber\\
 \tilde{G}^{\cal M}_{\theta_1x} & = & G^{\rm MQGP}_{\theta_1x}\Biggl[1 + \Biggl(
- \frac{\left(9 b^2+1\right)^4 b^{10} M  \left(6 a^2+{r_0}^2\right) \left((r-{r_0})^2+{r_0}^2\right)
   \Sigma_1}{3 \pi  \left(-18 b^4+3 b^2+1\right)^4 \log N  N^{8/15} {N_f} \left(-27 a^4+6 a^2
   {r_0}^2+{r_0}^4\right) \alpha _{\theta _2}^3}+{C_{\theta_1x}}^{(1)}
\Biggr)\beta\Biggr]\nonumber\\
\tilde{G}^{\cal M}_{\theta_1z} & = & G^{\rm MQGP}_{\theta_1z}\Biggl[1 + \Biggl(\frac{16 \left(9 b^2+1\right)^4 b^{12}  \beta  \left(\frac{(r-{r_0})^3}{{r_0}^3}+1\right) \left(19683
   \sqrt{3} \alpha _{\theta _1}^6+3321 \sqrt{2} \alpha _{\theta _2}^2 \alpha _{\theta _1}^3-40 \sqrt{3} \alpha _{\theta _2}^4\right)}{243
   \pi ^3 \left(1-3 b^2\right)^{10} \left(6 b^2+1\right)^8 {g_s}^{9/4} \log N ^4 N^{7/6} {N_f}^3 \left(-27 a^4 {r_0}+6 a^2
   {r_0}^3+{r_0}^5\right) \alpha _{\theta _1}^7 \alpha _{\theta _2}^6}+C_{\theta_1z}^{(1)}\Biggr)\Biggr]\nonumber\\
   \tilde{G}^{\cal M}_{\theta_2x} & = & G^{\rm MQGP}_{\theta_2x}\Biggl[1 + \Biggl(
   \frac{16 \left(9 b^2+1\right)^4 b^{12} \left(\frac{(r-{r_0})^3}{{r_0}^3}+1\right) \left(19683 \sqrt{3}
   \alpha _{\theta _1}^6+3321 \sqrt{2} \alpha _{\theta _2}^2 \alpha _{\theta _1}^3-40 \sqrt{3} \alpha _{\theta _2}^4\right)}{243 \pi ^3 \left(1-3
   b^2\right)^{10} \left(6 b^2+1\right)^8 {g_s}^{9/4} \log N ^4 N^{7/6} {N_f}^3 \left(-27 a^4 {r_0}+6 a^2
   {r_0}^3+{r_0}^5\right) \alpha _{\theta _1}^7 \alpha _{\theta _2}^6}+C_{\theta_2x}^{((1)}\Biggr)\beta\Biggr]\nonumber\\
\tilde{G}_{\theta_2y} & = & G^{\rm MQGP}_{\theta_2y}\Biggl[1 +  \frac{3 b^{10} \left(9 b^2+1\right)^4 M \beta \left(6 a^2+{r_0}^2\right) \left(1-\frac{(r-{r_0})^2}{{r_0}^2}\right) \log
   ({r_0}) \Sigma_1}{\pi  \left(3 b^2-1\right)^5 \left(6 b^2+1\right)^4 \log N ^2 N^{7/5} {N_f} \left(9 a^2+{r_0}^2\right) \alpha
   _{\theta _2}^3}\Biggr]\nonumber\\
\tilde{G}^{\cal M}_{\theta_2z} & = & G^{\rm MQGP}_{\theta_2z}\Biggl[1 + \Biggl(\frac{3 \left(9 b^2+1\right)^4 b^{10} M  \left(6 a^2+{r_0}^2\right) \left(1-\frac{(r-{r_0})^2}{{r_0}^2}\right) \log
   ({r_0}) \left(19683 \sqrt{6} \alpha _{\theta _1}^6+6642 \alpha _{\theta _2}^2 \alpha _{\theta _1}^3-40 \sqrt{6} \alpha _{\theta
   _2}^4\right)}{\pi  \left(3 b^2-1\right)^5 \left(6 b^2+1\right)^4 {\log N}^2 N^{7/6} {N_f} \left(9 a^2+{r_0}^2\right) \alpha
   _{\theta _2}^3}\nonumber\\
& & +{C_{\theta_2z}}^{(1)}\Biggr)\beta\Biggr]\nonumber\\
\tilde{G}^{\cal M}_{xy} & = & G^{\rm MQGP}_{xy}\Biggl[1 + \Biggl(\frac{3 \left(9 b^2+1\right)^4 b^{10} M  \left(6 a^2+{r_0}^2\right) \left(\frac{(r-{r_0})^2}{{r_0}^2}+1\right) \log
   ({r_0}) \alpha _{\theta _2}^3\Sigma_1}{\pi  \left(3 b^2-1\right)^5 \left(6 b^2+1\right)^4 \log N ^2 N^{21/20} {N_f} \left(9
   a^2+{r_0}^2\right) \alpha _{\theta _{2 l}}^6}+C_{xy}^{(1)}\Biggr)\beta\Biggr]\nonumber\\
\tilde{G}^{\cal M}_{xz}  & = & G^{\rm MQGP}_{xz}\Biggl[1 + \frac{18 b^{10} \left(9 b^2+1\right)^4 M \beta  \left(6 a^2+{r_0}^2\right)
   \left(\frac{(r-{r_0})^2}{{r_0}^2}+1\right) \log ^3({r_0}) \Sigma_1}{\pi  \left(3b^2-1\right)^5 \left(6 b^2+1\right)^4 \log N ^4 N^{5/4} {N_f} \left(9 a^2+{r_0}^2\right) \alpha
   _{\theta _2}^3}\Biggr]\nonumber\\
\tilde{G}^{\cal M}_{yy} & = & G^{\rm MQGP}_{yy}\Biggl[1  - \frac{3 b^{10} \left(9 b^2+1\right)^4 M \left(\frac{1}{N}\right)^{7/4} \beta  \left(6 a^2+{r_0}^2\right) \log ({r_0})\Sigma_1
   \left(\frac{(r-{r_0})^2}{r_0^2}+1\right)}{\pi  \left(3 b^2-1\right)^5 \left(6 b^2+1\right)^4 \log N ^2 {N_f} {r_0}^2 \left(9
   a^2+{r_0}^2\right) \alpha _{\theta _2}^3}\Biggr]\nonumber\\
 \tilde{G}^{\cal M}_{yz} & = & G^{\rm MQGP}_{yz}\Biggl[1 + \Biggl(\frac{64 \left(9 b^2+1\right)^8 b^{22} M \left(\frac{1}{N}\right)^{29/12}  \left(6 a^2+{r_0}^2\right)
   \left(\frac{(r-{r_0})^3}{{r_0}^3}+1\right) \log ({r_0}) }{27 \pi ^4 \left(3 b^2-1\right)^{15} \left(6 b^2+1\right)^{12}
   {g_s}^{9/4} \log N ^6  {N_f}^4 {r_0}^3 \left({r_0}^2-3 a^2\right) \left(9 a^2+{r_0}^2\right)^2 \alpha
   _{\theta _1}^7 \alpha _{\theta _2}^9}\nonumber\\
& & \hskip -0.3in \times \left(387420489 \sqrt{2} \alpha _{\theta _1}^{12}+87156324 \sqrt{3}
   \alpha _{\theta _2}^2 \alpha _{\theta _1}^9+5778054 \sqrt{2} \alpha _{\theta _2}^4 \alpha _{\theta _1}^6-177120 \sqrt{3} \alpha _{\theta
   _2}^6 \alpha _{\theta _1}^3+1600 \sqrt{2} \alpha _{\theta _2}^8\right)+C_{yz}^{(1)}\Biggr)\beta\Biggr]\nonumber\\
\tilde{G}^{\cal M}_{zz} & = & G^{\rm MQGP}_{zz}\Biggl[1 + \Biggl(C_{zz}^{(1)}-\frac{b^{10} \left(9 b^2+1\right)^4 M \left({r_0}^2-\frac{(r-{r_0})^3}{{r_0}}\right) \log ({r_0})
   \Sigma_1}{27 \pi ^{3/2} \left(3 b^2-1\right)^5 \left(6 b^2+1\right)^4 \sqrt{{g_s}} \log N ^2 N^{23/20} {N_f} \alpha
   _{\theta _2}^5}\Biggr)\beta\Biggr]\nonumber\\
\tilde{G}^{\cal M}_{x^{10}x^{10}} & = & G^{\rm MQGP}_{x^{10}x^{10}}\Biggl[1 -\frac{27 b^{10} \left(9 b^2+1\right)^4 M \left(\frac{1}{N}\right)^{5/4} \beta  \left(6 a^2+{r_0}^2\right)
   \left(1-\frac{(r-{r_0})^2}{{r_0}^2}\right) \log ^3({r_0}) \Sigma_1}{\pi  \left(3 b^2-1\right)^5 \left(6 b^2+1\right)^4 \log N ^4
   {N_f} {r_0}^2 \left(9 a^2+{r_0}^2\right) \alpha _{\theta _2}^3}\Biggr],
\end{eqnarray}
   }
   where:
\begin{eqnarray}
\label{Sigma_1-def}
& & \hskip -0.8in\Sigma_1 \equiv 19683
   \sqrt{6} \alpha _{\theta _1}^6+6642 \alpha _{\theta _2}^2 \alpha _{\theta _1}^3-40 \sqrt{6} \alpha _{\theta _2}^4\nonumber\\
   & & \hskip -0.8in\stackrel{\rm Global}{\longrightarrow} N^{\frac{6}{5}}\left(19683
   \sqrt{6} \sin^6\theta_1+6642 \sin^2{\theta _2} \sin^3{\theta _1}-40 \sqrt{6} \sin^4{\theta _2}\right),
\end{eqnarray}
${\cal C}^{(1)}_{MN}$ are constants of integration that figure in (\ref{ M-theory-metric-psi=2npi-patch}) after solving the EOMs for the ${\cal O}(\beta)$ metric perturbations $f_{MN}$, and $G^{\rm MQGP}_{MN}$ are the ${\cal M}$ theory metric components in the MQGP limit at ${\cal O}(\beta^0)$ 
\cite{VA-Glueball-decay}. The explicit dependence on $\theta_{10,20}$ of the ${\cal M}$-theory metric components up to ${\cal O}(\beta)$, using (\ref{theta12-deloc}), is effected by the replacemements: 
$\alpha_{\theta_1}\rightarrow N^{\frac{1}{5}}\sin\theta_{10},\ \alpha_{\theta_2}\rightarrow N^{\frac{3}{10}}\sin\theta_{20}$ in (\ref{ M-theory-metric-psi=2npi-patch}). Also, see (\ref{xyz-definitions}).  The main Physics-related take-away  is the following. From (\ref{ M-theory-metric-psi=2npi-patch}), one notes that in the IR: $r = \chi r_0, \chi\equiv {\cal O}(1)$, up to ${\cal O}(\beta)$:
\begin{equation}
\label{IR-beta-N-suppressed-logrh-rh-neg-exp-enhanced}
f_{MN} \sim \beta\frac{\left(\log r_0\right)^{m}}{r_0^n N^{\beta_N}},\ m\in\left\{0,1,3\right\},\ n\in\left\{0,2,5,7\right\},\
\beta_N>0.
\end{equation}
As estimated in \cite{Bulk-Viscosity}, $|\log \left(\frac{r_0}{{\cal R}_{D5/\overline{D5}}}\right)|\sim N^{\frac{1}{3}}$, implying there is a competition between Planckian and large-$N$ suppression and infra-red enhancement arising from $m,n\neq0$ in (\ref{IR-beta-N-suppressed-logrh-rh-neg-exp-enhanced}).

Now, using the standard Witten's prescription of reading off the type IIA metric (inclusive of the ${\cal O}(R^4)$ corrections):
\begin{eqnarray}
\label{TypeIIA-from-M-theory-Witten-prescription}
\hskip -0.1in ds_{11}^2 & = & e^{-\frac{2\phi^{\rm IIA}}{3}}\Biggl[\frac{1}{\sqrt{h(r,\theta_{1,2})}}\left(-dt^2 + \left(dx^1\right)^2 +  \left(dx^2\right)^2 + \tilde{g}(r)\left(dx^3\right)^2 \right)
\nonumber\\
& & \hskip -0.1in+ \sqrt{h(r,\theta_{1,2})}\left(\frac{dr^2}{\tilde{g}(r)} + ds^2_{\rm IIA}(r,\theta_{1,2},\phi_{1,2},\psi)\right)
\Biggr] + e^{\frac{4\phi^{\rm IIA}}{3}}\left(dx^{11} + A_{\rm IIA}^{F_1^{\rm IIB} + F_3^{\rm IIB} + F_5^{\rm IIB}}\right)^2,
\end{eqnarray}
where $A_{\rm IIA}^{F^{\rm IIB}_{i=1,3,5}}$ are the type IIA RR 1-forms obtained from the triple T/SYZ-dual of the type IIB $F_{1,3,5}^{\rm IIB}$ fluxes in the type IIB holographic dual of \cite{metrics}.

Turning now to obtaining the EOM for the profile function of the vector mesons, we will use (\ref{eoms_psi_n_rhomu}). Using (\ref{V1V2-defs}) and (\ref{O4-corrections}), (\ref{CMN}) and (\ref{E1010F1010}), one first obtains:
\begin{eqnarray}
\label{expressions-O(R^4)-corrections}
& & \hskip -0.5in {\cal V}_1 = {\cal V}_1^{\rm LO} + {\cal V}_1^{{\cal O}(R^4)}\ {\rm where}:\nonumber\\
& & \hskip -0.5in {\cal V}_1^{{\cal O}(R^4)}=\frac{\sqrt{h} e^{-2 Z} (2 e^{-\phi} G^{\cal M}_{x^{10}x^{10}} G^{\cal M}_{rr}  {Tr} C.F+e^{-\phi^{\rm IIA}}\sqrt{g^{\rm IIA}}
   (- {{\cal F}_{x^{10}x^{10}} } G^{\cal M}_{rr}-2  {{\cal F}_{rr} } G^{\cal M}_{x^{10}x^{10}}+2 G^{\cal M}_{x^{10}x^{10}} G^{\cal M}_{rr}))}{G^{\cal M}_{x^{10}x^{10}}\ ^{3/2} G^{\cal M}_{rr}\ ^2 r_0 ^2}\nonumber\\
& & \hskip -0.5in = -\frac{3  {g_s} M \sqrt[5]{\frac{1}{N}}  {N_f}^2 e^{-4 Z} \left(e^{4 Z}-1\right) (  {\cal C}_{zz}^{(1)}-2
     {\cal C}_{\theta_1z}^{(1)}+2   {\cal C}_{\theta_1x}^{(1)}) \log \left(r_0  e^Z\right) \left(72 a^2 r_0  e^Z \log \left(r_0  e^Z\right)+3 a^2+2
   r_0 ^2 e^{2 Z}\right)}{8 \pi ^2 \alpha _{\theta _1} \alpha _{\theta _2}^2};\nonumber\\
& & \hskip -0.5in {\cal V}_2 = {\cal V}_2^{\rm LO} + {\cal V}_2^{{\cal O}(R^4)}\ {\rm where}:\nonumber\\
& & \hskip -0.5in {\cal V}_2^{{\cal O}(R^4)} = e^{-\phi^{\rm IIA}}h Tr({\cal C}.{\cal F}) + {\cal E}_{x^{10}x^{10}} h \sqrt{g^{\rm IIA}}
\nonumber\\
& & \hskip -0.5in = \frac{3  {g_s}^2 M N^{4/5}  {N_f}^2 e^{-2 Z} (  {\cal C}_{zz}^{(1)}-2   {\cal C}_{\theta_1z}^{(1)}+2   {\cal C}_{\theta_1x}^{(1)}) \log \left(r_0  e^Z\right) \left(72 a^2 r_0  e^Z \log
   \left(r_0  e^Z\right)-3 a^2+2 r_0 ^2 e^{2 Z}\right)}{4 \pi  r_0 ^2 \alpha _{\theta _1} \alpha _{\theta _2}^2},
\end{eqnarray}
where ${\cal V}_{1,2}^{\rm LO}$ are the LO terms as obtained in \cite{VA-Glueball-decay}. The equation of motion (\ref{eoms_psi_n_rhomu}) satisfied by the profile function of the vector meson $\psi_1(z)$, using ideas similar to \cite{VA-Glueball-decay}, can be rewritten as a Sch\"{o}dinger-like equation with a potential ${\cal V} = {\cal V}^{\rm LO} + {\cal V}^{{\cal O}(R^4)}$ where (${\cal M}_{(1)} = m_0\frac{r_0}{\sqrt{4\pi g_sN}}$) and $ {\cal V}^{\rm LO}$ is the LO potential as given in  \cite{VA-Glueball-decay}. Further,
{
\begin{eqnarray}
\label{Schroedinger-like-psi}
& & \hskip -0.6in  {\cal V}^{{\cal O}(R^4)} = -2{\cal M}_{(1)}^2\frac{{\cal V}_1^{{\cal O}(R^4)}{\cal V}_2}{{\cal V}_1^2} + 2{\cal M}_{(1)}^2\frac{{\cal V}_2^{{\cal O}(R^4)}}{{\cal V}_1} - \frac{{\cal V}_1^{{\cal O}(R^4)}\left({\cal V}_1^\prime\right)\ ^2}{{\cal V}_1^3}
+ \frac{{\cal V}_1^\prime\left({\cal V}_1^{{\cal O}(R^4)}\right)^\prime}{2{\cal V}_1^2} + \frac{{\cal V}_1^{{\cal O}(R^4)}{\cal V}_1^{\prime\prime}}{2{\cal V}_1^2} - \frac{\left({\cal V}_1^{{\cal O}(R^4)}\right)^{\prime\prime}}{2{\cal V}_1}  \nonumber\\
& & \hskip -0.6in  =\beta  N \Biggl(\frac{2 \pi  {g_s}^{7/2} \log N  {\cal M}_{(1)}^2 {N_f} e^{-4 Z} ({\cal C}_{zz}^{(1)}-2   {\cal C}_{\theta_1z}^{(1)}+2   {\cal C}_{\theta_1x}^{(1)})
   \left(72 a^2 {r_0} e^Z \log \left({r_0} e^Z\right)-3 a^2+2 {r_0}^2 e^{2 Z}\right)}{{r_0}^8 \left(e^{4 Z}-1\right)
   \left(\frac{{g_s} e^{-4 Z}}{{r_0}^4}\right)^{3/2} \tilde{\Omega}(Z)}\nonumber\\
& & \hskip -0.6in +\frac{2 \pi  {g_s}^2 \log N  {\cal M}_{(1)}^2 {N_f} e^{4 Z}
   ({\cal C}_{zz}^{(1)}-2   {\cal C}_{\theta_1z}^{(1)}+2   {\cal C}_{\theta_1x}^{(1)}) \left(72 a^2 {r_0} e^Z \log \left({r_0} e^Z\right)+3 a^2+2 {r_0}^2 e^{2
   Z}\right)}{81 {r_0}^2 \left(e^{4 Z}-1\right) \alpha _{\theta _1}^2
   \tilde{\Omega}(Z)\ ^2}\Biggr)\nonumber\\
& & \hskip -0.6in \times  \Biggl[243 a^2 e^{-2 Z} \alpha _{\theta _1}^2 \Biggl(3 \log \left({r_0} e^Z\right) \biggl({g_s} {N_f} \left(8
   \log N  {r_0} e^Z+1\right)+32 \pi  {r_0} e^Z\biggr)-{g_s} (\log N +3) {N_f}\nonumber\\
& & \hskip -0.6in-72 {g_s} {N_f} {r_0}
   e^Z \log ^2\left({r_0} e^Z\right)-4 \pi \Biggr)+162 {r_0}^2 \alpha _{\theta _1}^2 \left({g_s} \log N  {N_f}-3
   {g_s} {N_f} \log \left({r_0} e^Z\right)+4 \pi \right)\Biggr]\nonumber\\
& & \hskip -0.6in  = -\beta\frac{\left(3 b^2-2\right) \log N  {m_0}^2 ({\cal C}_{zz}^{(1)}-2   {\cal C}_{\theta_1z}^{(1)}+2   {\cal C}_{\theta_1x}^{(1)})}{4 \left(3
   b^2+2\right) (\log N -3 \log ({r_0}))Z} + {\cal O}(Z^0),
\end{eqnarray}
}
where $a = \left(b + \gamma \frac{g_sM^2}{N}\left(1+\log r_0\right)\right)r_0$, and,
{
\begin{eqnarray}
\label{tildeOmegaZ}
& &\tilde{\Omega}(Z) \equiv 3 \log \left({r_0} e^Z\right) \left(3 a^2 \left({g_s}
   {N_f} \left(8 \log N  {r_0} e^Z-1\right)+32 \pi  {r_0} e^Z\right)-2 {g_s} {N_f} {r_0}^2 e^{2 Z}\right)\nonumber\\
& & +3
   a^2 ({g_s} (\log N -3) {N_f}+4 \pi )-216 a^2 {g_s} {N_f} {r_0} e^Z \log ^2\left({r_0} e^Z\right)+2
   {r_0}^2 e^{2 Z} ({g_s} \log N  {N_f}+4 \pi ).\nonumber\\
& &
\end{eqnarray}
}

As in \cite{VA-Glueball-decay}, one defines $g(Z) \equiv \sqrt{{\cal V}_1(Z)}\psi_1(Z)$ where $g(Z)$ satisfies the following Schr\"{o}dinger-like equation that, as mentioned above, is obtained from (\ref{eoms_psi_n_rhomu}):
\begin{equation}
\label{g(Z)-EOM}
g^{\prime\prime}(Z) +  \left(\frac{\omega_1 + \beta  {\cal C}^{zz}_{\ \ \theta_1z\ \theta_1x}}{Z}+\omega_2+\frac{1}{4 Z^2}\right)g(Z) = 0,
\end{equation}
wherein:
{\footnotesize
\begin{eqnarray}
\label{omega_1-and-2_defs}
& & \omega_1\equiv \frac{1}{4} \left({m_0}^2-3 b^2 \left({m_0}^2-2\right)\right)+18 b^2 {r_h} \log
   ({r_h})-\frac{3 b \gamma  {g_s} M^2 \left({m_0}^2-2\right) \log ({r_h})}{2 N}+\frac{36 b
   \gamma  {g_s} M^2 {r_h} \log ^2({r_h})}{N},\nonumber\\
& & \omega_2\equiv -\frac{4}{3}+\frac{3}{2} b^2 \left({m_0}^2+72 {r_h}-4\right)-36 b^2 {r_h} \log ({r_h})+\frac{3 b \gamma
   {g_s} M^2 \left({m_0}^2-4\right) \log ({r_h})}{N}-\frac{72 b \gamma  {g_s} M^2 {r_h}
   \log ^2({r_h})}{N},\nonumber\\
& & 
 {\cal C}^{zz}_{\ \ \theta_1z\ \theta_1x} = -\frac{\left(3 b^2-2\right) \log N  {m_0}^2 ({\cal C}_{zz}^{(1)}-2 {\cal C}_{\theta_1z}^{(1)}+2 {\cal C}_{\theta_1x}^{(1)})}{4 \left(3
   b^2+2\right) (\log N -3 \log ({r_0}))},
\end{eqnarray}}
and whose solution (using arguments similar to the ones in  \cite{VA-Glueball-decay}) is given in terms of Whittaker functions:
\begin{equation}
\label{Whittaker-solution}
g(Z) = c_{\psi_1}^{(1)}  M_{-\frac{i (\omega_1 +  \beta {\cal C}^{zz}_{\ \ \theta_1z\ \theta_1x} )}{2 \sqrt{\omega_2}},0}\left(2 i \sqrt{\omega_2} Z\right)+c_{\psi_1}^{(2)}
   W_{-\frac{i (\omega_1 +  \beta {\cal C}^{zz}_{\ \ \theta_1z\ \theta_1x}  )}{2 \sqrt{\omega_2}},0}\left(2 i \sqrt{\omega_2} Z\right).
\end{equation}
We pause here and note that the effect of the inclusion of the ${\cal O}(R^4)$ corrections into the EOM for the radial profile function $\psi_1(Z)$  for the $\rho$ meson is a shift in the residue of the simple pole in the potential of the Schr\"{o}dinger-like equation satisfied by the redefined $\rho$ meson profile function $g(Z)$.

Using arguments similar to the ones in \cite{VA-Glueball-decay}, implementing Neumann boundary condition ($\psi_1^\prime(Z=0)=0$) one sees that in the IR (i.e., near $Z=0$):
\begin{equation}
\label{psi1(Z)}
\psi_1(Z) = \sqrt{2} {\cal C}_{\psi_1}^{(1)\ {\rm IR}} \sqrt{i \sqrt{\omega_2}} \left[1-Z \left(\beta  {\cal C}^{zz}_{\ \ \theta_1z\ \theta_1x}+\omega_1\right)\right]; {\cal C}_{\psi_1}^{(1)\ {\rm IR}} \equiv {\cal C}_{\psi_1}^{{\rm IR}}=N^{-\Omega_{\psi_1}},\ \Omega_{\psi_1}>0.
\end{equation}

Now, as explained in Section {\bf 3},
\begin{eqnarray}
\label{phi0}
& &  \phi_0(Z) = \frac{{\cal C}_{\phi_0}^{\rm IR}}{{\cal V}_1(Z)} = \frac{\phi_0^{(-1)}}{Z} + \phi_0^{({\rm constant})} + \phi_0^{(1)}Z + \phi_0^{(2)}Z^2 + {\cal O}(Z^3).
\end{eqnarray}
By requiring:
\begin{eqnarray}
\label{alphatheta2-alphatheta1}
& & \alpha _{\theta _2} = \frac{9 \sqrt{\log N -3 \log r_0 }}{\sqrt{2} \sqrt{\log N +3 \log r_0 }}N^{\frac{1}{10}} \alpha _{\theta _1},\ \log N > |\log r_0|,\nonumber\\
& & b = \frac{1}{\sqrt{3}} + \epsilon,
\end{eqnarray}
for a very tiny $\epsilon$ to be ascertained later, one can set: $\phi^{(-1)}_0 = \phi_0^{(1)} = 0$ and one obtains:
\begin{eqnarray}
\label{phi0(Z)}
& & \phi_0(Z) =\frac{\pi ^2 {\cal C}_{\phi_0}^{\rm IR}\  N^{2/5} \alpha _{\theta _1}^3 (\log N -3 \log r_0 ) \left(\frac{27}{8 b^2 {g_s}
   \log N  (\log N +3 \log r_0 )}-\frac{81 b^2 \beta  ({\cal C}_{zz}^{(1)}-2   {\cal C}_{\theta_1z}^{(1)}+2   {\cal C}_{\theta_1x}^{(1)})}{8 \log
   ({r_0})}\right)}{{g_s} M {N_f}^2 {r_0}^3 (\log N +3 \log r_0 )}\nonumber\\
& & -\frac{\pi ^2 {\cal C}_{\phi_0}^{IR} N^{2/5}
   \alpha _{\theta _1}^3 (\log N -3 \log r_0 ) \left(\frac{9 \left(3 b^2+1\right) \beta  ({\cal C}_{zz}^{(1)}-2   {\cal C}_{\theta_1z}^{(1)}+2   {\cal C}_{\theta_1x}^{(1)})}{4 \log r_0 }+\frac{1944 b^4}{\left(3 b^2+2\right)^4}\right)}{{g_s} \log r_0  M {N_f}^2 {r_0}^2
   (\log N +3 \log r_0 )} Z^2 + {\cal O}(Z^3).\nonumber\\
\end{eqnarray}
One can similarly show that one obtains the following profile functions in the UV:
\begin{eqnarray}
\label{profile-functions-UV}
& & \psi_1^{\rm UV}(Z) = {\cal C}_{\psi_1}^{\rm UV}\frac{e^{-2Z}}{Z^{\frac{3}{2}}},\nonumber\\
& & \phi_0^{\rm UV}(Z) = {\cal C}_{\phi_0}^{\rm UV}\frac{e^{-2Z}}{Z^2}.
\end{eqnarray}

Let us discuss the normalization conditions on $\psi_1(Z)$ and $\phi_0(Z)$ and the consequent constraints on ${\cal C}_{\psi_1}^{\rm UV}$ and ${\cal C}_{\phi_0}^{\rm UV}(Z)$.

\begin{itemize}
\item
The normalization condition on $\psi_1(Z)$ (\ref{norm_psi}): $\left.{\cal V}_{\Sigma_2}\int_{0}^\infty dZ {\cal V}_2\left(\psi_1(Z)\right)^2\right|_{b = \frac{1}{\sqrt{3}} + \epsilon} = 1$ obtains:
{\footnotesize
\begin{eqnarray}
\label{Cpsi1UV}
& & \hskip -0.8in {\cal C}_{\psi_1}^{\rm UV}=\frac{\sqrt{\frac{\sqrt{7} ({f_{r_0}}-1) {f_{r_0}} {g_s}^2 M {N_f}^2 {\cal V}_{\Sigma_2}
  {\cal C}_{\psi_1}^{\rm IR}\ ^2 \log N  \left(7 \beta  {{({\cal C}_{zz}^{(1)}-2 {\cal C}_{\theta_1z}^{(1)}+2 {\cal C}_{\theta_1x}^{(1)})}} {f_{r_0}}^2 \gamma ^2 {g_s}^2 M^4 \log ^2(N)+3456
   \epsilon ^2 ({f_{r_0}}+1) N^2\right)}{2\ 3^{3/4} \epsilon ^{3/2} ({f_{r_0}}+1) N^{7/5} \alpha _{\theta _1}^3}-93312 \pi
   }}{24 \sqrt{-\frac{({f_{r_0}}-1) {g_s}^2 M N^{3/5} {N_f}^2 {\cal V}_{\Sigma_2}}{\epsilon ^2 ({f_{r_0}}+1) \alpha
   _{\theta _1}^3 \log N }}},
\end{eqnarray}
}
wherein, similar to \cite{Bulk-Viscosity}, the IR cut-off $r_0$ is assumed to be given as $r_0 = N^{-\frac{f_{r_0}}{3}}$. Let us impose ${\cal C}_{\psi_1}^{\rm UV} = 0$ which is equivalent to:
{\footnotesize
\begin{eqnarray}
\label{Vol2}
& & \hskip -0.8in {\cal V}_{\Sigma_2}=\frac{186624\ 3^{3/4} \pi  \epsilon ^{3/2} ({f_{r_0}}+1) N^{7/5} \alpha _{\theta _1}^3}{\sqrt{7} ({f_{r_0}}-1)
   {f_{r_0}} {g_s}^2 {\log N}   M {N_f}^2 {\cal C}_{\psi_1}^{\rm IR}\ ^2 \left(7 \beta  {({\cal C}_{zz}^{(1)}-2 {\cal C}_{\theta_1z}^{(1)}+2 {\cal C}_{\theta_1x}^{(1)})} {f_{r_0}}^2 \gamma ^2
   {g_s}^2 {\log N}  ^2 M^4+3456 \epsilon ^2 ({f_{r_0}}+1) N^2\right)}.
\end{eqnarray}
}

\item
The normalization condition on $\phi_0(Z)$ (\ref{norm_scalar}): $\frac{{\cal V}_{\Sigma_2}}{2}\int_{0}^\infty dZ {\cal V}_1\left(\phi_0(Z)\right)^2=1$ obtains:
\begin{eqnarray}
\label{Cphi0UV}
& & {\cal C}_{\phi_0}^{\rm UV}=\frac{243 \sqrt[4]{3} \pi ^2 {\cal C}_{\phi_0}^{\rm IR} \sqrt{\epsilon }
   N^{{f_{r_0}}+\frac{2}{5}} \sqrt{({f_{r_0}}+1) (-\beta {({\cal C}_{zz}^{(1)}-2 {\cal C}_{\theta_1z}^{(1)}+2 {\cal C}_{\theta_1x}^{(1)})}+2 {f_{r_0}}+2)}}{32 ({f_{r_0}}-1)^2
   {g_s}^2 \left(\log N\right) ^2 M {N_f}^2} + {\cal O}(\epsilon^{\frac{3}{2}}).\nonumber\\
& &
\end{eqnarray}
Being proportional to $\sqrt{\epsilon}$ and assuming $\epsilon\ll1$ (for black-hole gravity dual, $\epsilon< r_h^2\left(\log r_h\right)^{\frac{9}{2}}N^{-\frac{9}{10}}$ \cite{OR4-Yadav+Misra}), we will henceforth be approximating ${\cal C}_{\phi_0}^{\rm UV}\approx0$, i.e.,  $\psi_1^{\rm UV}(Z)\approx0, \phi_0^{\rm UV}(Z)\approx0$.
\end{itemize}

To evaluate $y_{1,3,5,7}$ and $z_{1,...,8}$ using (\ref{y_i+z_i}) along with (\ref{<<A>>}) and (\ref{<A>}), one will be splitting the radial integral into the IR and the UV, e.g., $\langle A\rangle[\tilde{G}_{MN}^{\rm IR}] + \langle A\rangle[G_{MN}^{\rm UV}]$, where using the results of Appendix {\bf A}, $f_{MN}^{\rm UV}$'s are vanishingly small (impling $\tilde{G}_{MN}^{\rm UV} = G_{MN}^{\rm UV}$). Using (\ref{psi1(Z)}) and (\ref{phi0(Z)}), one arrives at the  expressions for the
coupling constants $y_{1,3,5,7}$ and $z_{1,...,8}$ as explained in Appendix {\bf B}:
\begin{eqnarray}
\label{y_i-z_j}
& & \left.y_{1,...,7},\ z_{1,...,8}\right|_{\sim\rm IIB\ Ouyang}\nonumber\\
& & = {\cal V}_{\Sigma_2} \left({\cal C}_{\psi_1}^{\rm IR}\right)^{n_{y_i/z_j}}
 \left({\cal C}_{\phi_0}^{\rm IR}\right)^{m_{y_i/z_j}}\nonumber\\
& & \times\Biggl({\cal F}_{y_i/z_j}(r_0; M, N, N_f ) + \beta \left({\cal C}_{zz}^{(1)}-2 {\cal C}_{\theta_1z}^{(1)}+2 {\cal C}_{\theta_1x}^{(1)}\right){\cal H}_{y_i/z_j}(r_0; M, N, N_f) \Biggr).
\end{eqnarray}
Further,
{\footnotesize
\begin{eqnarray}
\label{Fpisq}
& & \hskip -0.4in  F_\pi^2 = {\cal V}_{\Sigma_2}\Biggl(\frac{243 \pi ^2 \beta  \left({\cal C}_{zz}^{(1)} - 2 {\cal C}_{\theta_1z}^{(1)} + 2 {\cal C}_{\theta_1x}^{(1)}\right) {\cal C}_{\phi_0}^{\rm IR}\ ^2 {f_{r_0}} ({f_{r_0}}+1) \log ^2(3) \alpha _{\theta _1}^3 N^{\frac{4
   {f_{r_0}}}{3}+\frac{2}{5}}}{8192 ({f_{r_0}}-1)^3 {g_s}^3 \left(\log N\right) ^3 M N_f ^2}\nonumber\\
& & \hskip -0.4in +\frac{81 \sqrt{3} \pi ^2 {\cal C}_{\phi_0}^{\rm IR}\ ^2
   \epsilon  {f_{r_0}} ({f_{r_0}}+1)^2 \log (3) (\log (243)-6) \alpha _{\theta _1}^3 N^{\frac{4 {f_{r_0}}}{3}+\frac{2}{5}}}{2048
   ({f_{r_0}}-1)^3 {g_s}^3 \left(\log N\right) ^3 M N_f ^2} -\frac{243 \pi ^2 {\cal C}_{\phi_0}^{\rm IR}\ ^2 {f_{r_0}} ({f_{r_0}}+1)^2 \log ^2(3) \alpha
   _{\theta _1}^3 N^{\frac{4 {f_{r_0}}}{3}+\frac{2}{5}}}{4096 ({f_{r_0}}-1)^3 {g_s}^3 \left(\log N\right) ^3 M N_f ^2}\Biggr)\nonumber\\
& &
\end{eqnarray}
}
and
\begin{eqnarray}
\label{gsq}
& & g_{\rm YM}^2 =\frac{{\log N}   N \left(7 {({\cal C}_{zz}^{(1)}-2 {\cal C}_{\theta_1z}^{(1)}+2 {\cal C}_{\theta_1x}^{(1)})} {f_{r_0}}^2 \gamma ^2 {g_s}^2 M^4 \log ^2(N)+3456 ({f_{r_0}}+1)
   \lambda_{\epsilon}^2\right)}{288 \lambda_{\epsilon}^2 \alpha _{\theta _1}^2 \log N  \left(\sqrt{3} \beta ^{3/2} {({\cal C}_{zz}^{(1)}-2 {\cal C}_{\theta_1z}^{(1)}+2 {\cal C}_{\theta_1x}^{(1)})}
    \lambda_{\epsilon} m_0^2-12 ({f_{r_0}}+1)  N\right)}.
\end{eqnarray}

In the chiral limit, the ${\cal O}(p^4)$ $SU(3)\ \chi$PT Lagrangian is given by \cite{GL}:
\begin{eqnarray}
\label{ChPT-Op4}
& & L_1 \left({\rm Tr}(\nabla_\mu U^\dagger \nabla^\mu U)\right)^2 + L_2\left({\rm Tr}(\nabla_\mu U^\dagger \nabla_\nu U)\right)^2 + L_3{\rm Tr} \left(\nabla_\mu U^\dagger \nabla^\mu U\right)^2\nonumber\\
& & - i L_9 Tr\left({\cal L}_{\mu\nu}\nabla^\mu U \nabla^\nu U^\dagger + {\cal R}_{\mu\nu}\nabla^\mu U \nabla^\nu U^\dagger\right) + L_{10} Tr\left(U^\dagger {\cal L}_{\mu\nu}U{\cal R}^{\mu\nu}\right) + H_1 Tr\left({\cal L}_{\mu\nu}^2 + {\cal R}_{\mu\nu}^2\right),\nonumber\\
& &
\end{eqnarray}
where $\nabla_\mu U\equiv \partial_\mu U - i {\cal L}_\mu U + i U {\cal R}_\mu,\ U=e^{\frac{2i\pi}{F_\pi}}$. For completeness, we have reviewed the arguments of \cite{HLS-Physics-Reports} to obtain (\ref{ChPT-Op4})-like terms from the HLS Lagrangian by integrating out the $\rho$ mesons in Appendix {\bf C}. Using hence (\ref{GL-LEC-integrate-rho-out}) along with results of \cite{Relationship-Li-yi-zi}, one obtains relationships between the LECs $y_i, z_i$ of (\ref{Lagrangian-Op4}) and the $L_i$s of (\ref{ChPT-Op4}) - (\ref{L1}) for $L_1$, (\ref{L9}) for $L_9$ and (\ref{L10}) for $L_{10}$.

The parameters $L_i$ and $H_i$ are renormalized at one-loop level with all vertices in one-loop diagrams arising from the ${\cal O}(p^2)$ terms. Using dimensional regularization and performing renormalizations of the parameters via \cite{GL}:
\begin{equation}
L_i = L_i^r(\mu) + \Gamma_i \lambda(\mu) \ , \qquad
H_i = H_i^r(\mu) + \Delta_i \lambda(\mu) \ ,
\end{equation}
where $\mu$ is the renormalization point,
and $\Gamma_i$ and $\Delta_i$ are certain numbers given later;
$\lambda(\mu)$ is the divergent part given by
\begin{equation}
\lambda(\mu) = - \frac{1}{2\left(4\pi\right)^2}
\left[   \frac{1}{\bar{\epsilon}} - \ln \mu^2 + 1 \right]
\ ,
\end{equation}
where
\begin{equation}
\frac{1}{\bar{\epsilon}} = \frac{2}{4-d}
- \gamma_E + \ln 4\pi 
\ ,
\end{equation}
$d$ being the non-radial non-compact space-time dimensionality to be set to four.
The constants $\Gamma_i$ and $\Delta_i$ for $SU(3)$ $\chi$PT theory were worked out in \cite{GL}:
\begin{equation}
\begin{array}{ccccc}
\Gamma_1 = \frac{3}{32} \ ,
& \Gamma_2 = \frac{3}{16} \ ,
& \Gamma_3 = 0 \ , 
& \Gamma_4 = \frac{1}{8} \ ,
& \Gamma_5 = \frac{3}{8} \ ,
\\
\Gamma_6 = \frac{11}{144} \ ,
& \Gamma_7 = 0 \ ,
& \Gamma_8 = \frac{5}{48} \ ,
& \Gamma_9 = \frac{1}{4} \ , 
& \Gamma_{10} = - \frac{1}{4} \ ;
\\
\Delta_1 = - \frac{1}{8} \ ,
& \Delta_2 = \frac{5}{24} \ .
& & &
\end{array}
\end{equation}
The analog of the 1-loop renormalization in $\chi$PT can be understood on the gravity dual side by noting that the latter requires holographic renormalization. This can be seen as follows. It can be shown \cite{Tc-HD} that  the bulk on-shell $D=11$ supergravity action inclusive of ${\cal O}(R^4)$-corrections is given by:
\begin{equation}
\label{on-shell-D=11-action-up-to-beta}
\hskip -0.3in S_{D=11}^{\rm on-shell} = \frac{1}{2}\Biggl[-\frac{7}{2}S_{\rm EH}^{(0)} + 2 S_{\rm GHY}^{(0)}+ \beta\left(\frac{20}{11}S_{\rm EH} - \frac{117}{22}\int_{M_{11}}\sqrt{-g^{(1)}}R^{(0)}
+ 2 S_{\rm GHY} - \frac{2}{11}\int_{M_{11}}\sqrt{-g^{(0)}}g_{(0)}^{MN}\frac{\delta J_0}{\delta g_{(0)}^{MN}}\right)\Biggr].
\end{equation}
The UV divergences of the various terms in (\ref{on-shell-D=11-action-up-to-beta}) are summarized below:
\begin{eqnarray}
\label{UV_divergences}
& &\left. \int_{M_{11}}\sqrt{-g}R\right|_{\rm UV-divergent},\ \left.\int_{\partial M_{11}}\sqrt{-h}K\right|_{\rm UV-divergent} \sim r_{\rm UV}^4 \log r_{\rm UV},\nonumber\\
& & \left.\int_{M_{11}} \sqrt{-g}g^{MN}\frac{\delta J_0}{\delta g^{MN}}\right|_{\rm UV-divergent} \sim
\frac{r_{\rm UV}^4}{\log r_{\rm UV}}. 
\end{eqnarray}
It can be shown \cite{Tc-HD} that an appropriate linear combination of the boundary  terms: $\left.\int_{\partial M_{11}}\sqrt{-h}K\right|_{r=r_{\rm UV}}$ and $\left.\int_{\partial M_{11}}\sqrt{-h}h^{mn}\frac{\partial J_0}{\partial h^{mn}}\right|_{r=r_{\rm UV}}$ serves as the appropriate counter terms to cancel the UV divergences (\ref{UV_divergences}) \footnote{For consistency, one needs to impose the following relationship between the UV-valued effective number of flavor $D7$-branes of the parent type IIB dual, $N_f^{\rm UV}$ and $\log r_{\rm UV}$: $N_f^{\rm UV} = \frac{\left(\log r_{\rm UV}\right)^{\frac{15}{2}}}{\log N}$.}.

We will now discuss how to match our results with the experimental values of the 1-loop renormalized coupling constants $L_{1,2,3,9,10}^r$ in (\ref{ChPT-Op4}) and $F_\pi^2$ and $g_{\rm YM}^2$; the experimental value of $H_1$ apparently is unavailable. Table 3 contains
the values of the 1-loop renormalized values of the coupling constants in (\ref{ChPT-Op4}) \cite{Ecker-2015}.
\newpage
\begin{table}
\centering
\begin{tabular}{c|c|c|c|c}
\hline
\textbf{LECs} & \textbf{GL 1985} \cite{GL} & \textbf{NLO 2014} & \textbf{NNLO free fit} & \textbf{NNLO BE14} \cite{BE14} \\
\hline
$10^{3}$ $L_1^{r}$ &0.7(3) & 1.0(1) & 0.64)06 & 0.53(06) \\
\hline
$10^{3}$ $L_2^{r}$ & 1.3(7) & 1.6(2) & 0.59(04) & 0.81(04) \\
\hline
$10^{3}$ $L_3^{r}$ & -4.4(2.5) & -3.8(3) & -2.80(20) & -3.07(20) \\
\hline
$10^{3}$ $L_4^{r}$ & -0.3(5) & 0.0(3) & 0.76(18) & 0.3 \\
\hline
$10^{3}$ $L_5^{r}$ & 1.4(5) & 1.2(1) & 0.50(07) & 1.01(06) \\
\hline
$10^{3}$ $L_6^{r}$ & -0.2(3) & 0.0(4) & 0.49(25) & 0.14(05) \\
\hline
$10^{3}$ $L_7^{r}$ & -0.4(2) & -0.3(2) & -0.19(08) & -0.34(09) \\
\hline
$10^{3}$ $L_8^{r}$ & 0.9(3) & 0.5(2) & 0.17(11) & 0.47(10) \\
\hline
\end{tabular}
\caption{Various fits for NLO LECs $L_i^{r}(i=1,2,...,8)$}.
\end{table}

Table 4 elaborates upon the column titled {\bf GL 1985} \cite{GL}.
\begin{table}
\centering
\begin{tabular}{c|c|c}
 \hline
i & $L_i^{r}(M_{\rho})$ $10^{3}$ & Source \\
\hline
 1 & 0.4 $\pm$ 0.3 & $K_{e4}$, $\pi\pi \rightarrow \pi\pi$\\
  \hline
 2 & 1.4 $\pm$ 0.3 & $K_{e4}$, $\pi\pi \rightarrow \pi\pi$\\
  \hline
 3 & -3.5 $\pm$ 1.1 & $K_{e4}$, $\pi\pi \rightarrow \pi\pi$\\
  \hline
 4 & -0.3 $\pm$ 0.5 & Zweig rule \\
  \hline
 5 & 1.4 $\pm$ 0.5 & $F_K : F_\pi$ \\
  \hline
 6 & -0.2 $\pm$ 0.3 & Zweig rule \\
  \hline
 7 & -0.4 $\pm$ 0.2 & Gell-Mann-Okubo, $L_5$, $L_8$ \\
  \hline
 8 & 0.9 $\pm$ 0.3 & $M_{K0} - M_{K_+}$, $L_5$, $(m_s - \hat{m}):(m_d - m_u)$\\
  \hline
 9 & 6.9 $\pm$0.7 & $<r^2>_V^\pi$\\
 \hline
 10 & -5.5 $\pm$ 0.7 & $\pi \rightarrow e\nu\gamma$ \\
  \hline
\end{tabular}
\caption{Phenomenological Values of the 1-loop renormalised couplings $L_i^{r}(M_{\rho})$ of (\ref{ChPT-Op4}) \cite{Pich}. Last column shows the source to extract this information }
\end{table}

We will now show how, in five steps, it is possible to match the phenomenological values of the ${\cal O}(p^4)$  $SU(3)$ $\chi$PT Lagrangian \cite{GL} one-loop renormalized LECs $L_{1,9,10}^r$ as well as $F_\pi^2, g_{\rm YM}(\Lambda_{\rm QCD}=0.4 {\rm GeV}, \Lambda=1.1 {\rm Gev}, \mu=M_\rho)$ where $\Lambda$ is the ``HLS-QCD" matching scale \cite{HLS-Physics-Reports} and $\mu$ is the renormalization scale, as well as the order of magnitude and signs of $L_{2,3}^r$. 

\newpage

\begin{itemize}
\item
{\bf Step 1: Matching $L_{1,2,3}^r$}

Using (\ref{Vol2}), (\ref{gsq}), (\ref{y_i}), (\ref{z_i}) and $y_2=-y_1$ one obtains:
{
\begin{eqnarray}
\label{L1}
& & \hskip -0.4in  L_1^r = \frac{L_2^r}{2} = - \frac{L_3^r}{6} = \frac{1}{g_{\rm YM}^2} - z_4 + y_2 \\
& &  \hskip -0.4in = \frac{1}{143360 \sqrt{7} ({f_{r_0}}-1) {f_{r_0}}
   {g_s}^8 \log N  M^4 N_f ^8 \alpha _{\theta _1}^3 {\cal C}_{\psi_1}^{\rm IR}\ ^2 \Omega}\Biggl\{3 \pi  ({f_{r_0}}+1) N^{7/5}\nonumber\\
& & \hskip -0.4in \times \Biggl(-\frac{4822335 \sqrt{2} \sqrt[8]{3} \sqrt[4]{7} \pi ^3 \beta  \left({\cal C}_{zz}^{(1)} - 2 {\cal C}_{\theta_1z}^{(1)} + 2 {\cal C}_{\theta_1x}^{(1)}\right) {\cal C}_{\phi_0}^{\rm IR}\ ^2 \epsilon ^{9/4}
   {f_{r_0}} ({f_{r_0}}+1) {g_s}^4 M^2 N_f ^4 \alpha _{\theta _1}^9 {\cal C}_{\psi_1}^{\rm IR}\  N^{2 {f_{r_0}}+\frac{7}{5}}}{({f_{r_0}}-1)^3
   \left(\log N\right) ^3}\nonumber\\
& & \hskip -0.4in +\frac{430080 \sqrt{21} \beta  \left({\cal C}_{zz}^{(1)} - 2 {\cal C}_{\theta_1z}^{(1)} + 2 {\cal C}_{\theta_1x}^{(1)}\right) \epsilon ^3 ({f_{r_0}}-1) {f_{r_0}} {g_s}^8 \log N  M^4 {m_0}^2 N^{3/5}
   N_f ^8 \alpha _{\theta _1}^5 {\cal C}_{\psi_1}^{\rm IR}\ ^2}{\pi  ({f_{r_0}}+1)}\nonumber\\
& & \hskip -0.4in +\frac{64 \sqrt{\frac{7}{\pi }} ({f_{r_0}}-1)^2 {g_s}^9 \log r_0
   M^4 N_f ^8 {\cal C}_{\psi_1}^{\rm IR}\ ^2 N^{\frac{2 {f_{r_0}}}{3}-\frac{3}{5}}\Omega}{({f_{r_0}}+1)^2} +\frac{26880 \sqrt{7} ({f_{r_0}}-1) {g_s}^9 \log r_0  M^4
   N_f ^8 {\cal C}_{\psi_1}^{\rm IR}\ ^2 N^{\frac{2 {f_{r_0}}}{3}-\frac{2}{5}}\Omega}{({f_{r_0}}+1)^2}\nonumber\\
& & \hskip -0.4in -\frac{8960 \sqrt{7} ({f_{r_0}}-1) {f_{r_0}} {g_s}^9 \log N  M^4 N_f ^8
   \alpha _{\theta _1}^2 {\cal C}_{\psi_1}^{\rm IR}\ ^2 N^{\frac{2 {f_{r_0}}}{3}-\frac{2}{5}} \Omega}{{f_{r_0}}+1}\nonumber\\
& & \hskip -0.4in -\frac{132269760 \sqrt{2} \sqrt[8]{3} \sqrt[4]{7} \pi ^3 {\cal C}_{\phi_0}^{\rm IR}\ ^2
   \epsilon ^{17/4} {f_{r_0}} ({f_{r_0}}+1)^2 {g_s}^4 M^2 N_f ^4 \alpha _{\theta _1}^9 {\cal C}_{\psi_1}^{\rm IR}\  N^{2
   {f_{r_0}}+\frac{7}{5}}}{({f_{r_0}}-1)^3 \left(\log N\right) ^3}\nonumber\\
& & \hskip -0.4in -\frac{5160960 \sqrt{7} \epsilon ^2 ({f_{r_0}}-1) {f_{r_0}} {g_s}^8 \log N  M^4 N^{3/5}
   N_f ^8 \alpha _{\theta _1}^5 {\cal C}_{\psi_1}^{\rm IR}\ ^2}{\pi }\nonumber\\
& & \hskip -0.4in-\frac{195259926456 \sqrt[4]{3} \pi ^7 {\cal C}_{\phi_0}^{\rm IR}\ ^4 \epsilon ^{13/2} {f_{r_0}}
   ({f_{r_0}}+1)^4 \alpha _{\theta _1}^{15} N^{4 {f_{r_0}}+\frac{11}{5}}}{({f_{r_0}}-1)^7 \left(\log N\right) ^7}\Biggr)
\Biggr\} ,
\end{eqnarray}
}
where:
\begin{equation}
\label{Omega}
 \Omega \equiv \left(7 \beta  \left({\cal C}_{zz}^{(1)} - 2 {\cal C}_{\theta_1z}^{(1)} + 2 {\cal C}_{\theta_1x}^{(1)}\right) {f_{r_0}}^2 \gamma ^2 {g_s}^2
   \left(\log N\right) ^2 M^4+3456 \epsilon ^2 ({f_{r_0}}+1) N^2\right).
\end{equation}
Now, as we are working up to ${\cal O}(\beta)$ and further due to the smallness of $\epsilon$ assuming working up to ${\cal O}(\epsilon^2)$, one sees that (\ref{L1}) gets simplified to read:
\begin{eqnarray}
\label{L1_b}
& & \hskip -0.4in  L_1^r = \frac{1}{143360 \sqrt{7} ({f_{r_0}}-1) {f_{r_0}}
   {g_s}^8 \log N  M^4 N_f ^8 \alpha _{\theta _1}^3 }\nonumber\\
& & \hskip -0.4in \times\Biggl\{3 \pi  ({f_{r_0}}+1) N^{7/5} \Biggl(\frac{64 \sqrt{\frac{7}{\pi }} ({f_{r_0}}-1)^2 {g_s}^9 \log r_0
   M^4 N_f ^8 N^{\frac{2 {f_{r_0}}}{3}-\frac{3}{5}}}{({f_{r_0}}+1)^2} \nonumber\\
& & \hskip -0.4in +\frac{26880 \sqrt{7} ({f_{r_0}}-1) {g_s}^9 \log r_0  M^4
   N_f ^8  N^{\frac{2 {f_{r_0}}}{3}-\frac{2}{5}}}{({f_{r_0}}+1)^2} -\frac{8960 \sqrt{7} ({f_{r_0}}-1) {f_{r_0}} {g_s}^9 \log N  M^4 N_f ^8
   \alpha _{\theta _1}^2  N^{\frac{2 {f_{r_0}}}{3}-\frac{2}{5}} }{{f_{r_0}}+1}\nonumber\\
& & \hskip -0.4in -\frac{5160960 \sqrt{7}  ({f_{r_0}}-1) {f_{r_0}} {g_s}^8 \log N  M^4 N^{3/5}
   N_f ^8 \alpha _{\theta _1}^5}{\pi}\left(\frac{\epsilon ^2}{ \Omega}\right)\Biggr)
\Biggr\} 
\end{eqnarray}
As will be shown below ((\ref{f_{r_0}}) and (\ref{fr0-omega})), to match the experimental value of $L_9^r$, one needs to set $f_{r_0}=1-\kappa, 0<\kappa\ll1$. Hence:
\begin{eqnarray}
\label{L1_c}
& & \hskip -0.4in  L_1^r = \frac{1}{143360 \sqrt{7} {f_{r_0}}
   {g_s}^8 \log N  M^4 N_f ^8 \alpha _{\theta _1}^3 }\Biggl\{3 \pi  ({f_{r_0}}+1) N^{7/5} \Biggl(\frac{26880 \sqrt{7}  {g_s}^9 \log r_0  M^4
   N_f ^8  }{({f_{r_0}}+1)^2} \nonumber\\
& & \hskip -0.4in  -\frac{8960 \sqrt{7}  {f_{r_0}} {g_s}^9 \log N  M^4 N_f ^8
   \alpha _{\theta _1}^2   }{{f_{r_0}}+1} -\frac{5160960 \sqrt{7}  {f_{r_0}} {g_s}^8 \log N  M^4 N^{\frac{1}{3}}
   N_f ^8 \alpha _{\theta _1}^5}{\pi}\left(\frac{\epsilon ^2}{ \Omega}\right)\Biggr)
\Biggr\} 
\end{eqnarray}
Further, as the $\log r_0$ in (\ref{L1_c}) is in fact $\log\left(\frac{r_0}{{\cal R}_{D5/\overline{D5}}}\right)$ - ${\cal R}_{D5/\overline{D5}}>r_0$ being the $D5-\overline{D5}$ separation - one sees from (\ref{L1_c}) that in order to obtain a positive value (as required from phenomenological value of $L_1^r$), $\Omega<0$. Note, as shown below in (\ref{alphatheta1}), matching with the experimental value of the pion decay constant $F_\pi$ requires an $N$-suppression in $\alpha_{\theta_1}$, implying the $N$ enhancement in the last term in (\ref{L1_c}) is artificial.  So, to ensure one does not pick up an ${\cal O}\left(\frac{1}{\beta}\right)$ contribution in $L_1$ from $\frac{\epsilon^2}{\Omega}$ in (\ref{L1_c}) and also to ensure that the third term in (\ref{L1_c}) required to partly compensate the first two negative terms in the same (as explained above) to produce a positive term, is not vanishingly small, from (\ref{Omega}), one needs to set: 
\begin{equation}
\label{epsilon-sqrtbeta-over-N-lambdaeps}
\epsilon = \lambda_{\epsilon}\frac{\sqrt{\beta}}{N}.
\end{equation}
Finally, combining the above observations with the requirement to match the experimental value $L_1^{\rm exp}=0.64\times10^{-3}$,
one requires to implement the following constraint:
\begin{eqnarray}
\label{CCsO4}
& & \left({\cal C}_{zz}^{(1)} - 2 {\cal C}_{\theta_1z}^{(1)} + 2 {\cal C}_{\theta_1x}^{(1)}\right) = -\frac{493.7 (\delta +1) ({f_{r_0}}+1) \lambda_{\epsilon}^2}{{f_{r_0}}^2 \gamma ^2 {g_s}^2 \left(\log N\right) ^2 M^4},\nonumber\\
& & \delta = \frac{0.053 \alpha _{\theta _1}^3 N^{-\frac{2 {f_{r_0}}}{3}-1}}{{g_s}}.
\end{eqnarray}

{\it We hence see from (\ref{epsilon-sqrtbeta-over-N-lambdaeps}) that $\epsilon$ provides an expansion parameter connecting the $\frac{1}{N}$ and $\beta$ expansions. Also, together with (\ref{CCsO4}), this is the first connection between large-$N$ and higher derivative corrections in the context of ${\cal M}$-theory dual of large-$N$ thermal QCD-like theories.}

From (\ref{L1}) upon comparison with experimental values of $L_{2,3}^r$, we see that one can obtain a match with their order of magnitude and sign, but not the exact numerical value.

\item
{\bf Step 2: Matching $F_\pi^2$}

Now, using (\ref{CCsO4}) and (\ref{Vol2}), one can show that the difference of (\ref{Fpisq}) and the experimental value of $F_\pi^2 = \frac{0.0037 N^{-\frac{2 {f_{r_0}}}{3}-1}}{{g_s}}$\footnote{$F_\pi^2$ can be made to match the experimental value of $92.3 MeV$ wherein from $0^{++}$-glueball mass \cite{Misra+Gale}, one identifies:
$\frac{1700}{4} MeV \equiv \frac{r_0}{\sqrt{4 \pi g_s N}}$.} vanishes for:
\begin{eqnarray}
\label{alphatheta1}
& & \alpha_{\theta_1} = \frac{0.03 \sqrt[24]{\beta } {g_s}^{11/3} \sqrt[12]{\lambda_{\epsilon}} \left(\log N\right) ^{2/3} \sqrt[3]{M} N_f ^{2/3}
   \sqrt[3]{{\cal C}_{\psi_1}^{\rm IR}\ } {(1-{f_{r_0}})^{2/3} } N^{-\frac{{f_{r_0}}}{3}-\frac{13}{60}}}{\sqrt[3]{{\cal C}_{\phi_0}^{\rm IR}\ }
   \sqrt[3]{{f_{r_0}}+1}}.
\end{eqnarray}

\item
{\bf Step 3: Matching $L_9^r$}

Using (\ref{GL-LEC-integrate-rho-out}) and \cite{Relationship-Li-yi-zi}, one can show that:
\begin{equation}
\label{L9}
L_9^r = \frac{1}{8}\left(\frac{2}{g_{\rm YM}^2} - 2z_3 - z_4 - z_6\right).
\end{equation}
By using  (\ref{CCsO4}), (\ref{Vol2}) and (\ref{z_i}) one obtains:
\begin{equation}
\label{L9r-match}
L_9^r = -\frac{0.0031 ({f_{r_0}}-1) {g_s} N^{\frac{2 {f_{r_0}}}{3}+\frac{4}{5}}}{({f_{r_0}}+1) \alpha _{\theta _1}^3}.
\end{equation}
Now, assuming:
\begin{equation}
\label{f_{r_0}}
{f_{r_0}} = 1 - \omega \alpha_{\theta_1}^3,
\end{equation}
where
\begin{equation}
\label{fr0-omega}
\omega = \frac{4.6 N^{-\frac{2 {f_{r_0}}}{3}-\frac{4}{5}}}{{g_s}},
\end{equation}
one sees one gets a match with the phenomenological/experimental value $L_9^{\rm exp}=6.9\times10^{-3}$. Substituting (\ref{f_{r_0}}) - (\ref{fr0-omega}) into (\ref{CCsO4}), one obtains: 
\begin{equation}
\label{CCsO4-lambda-epsilon}
\left({\cal C}_{zz}^{(1)} - 2 {\cal C}_{\theta_1z}^{(1)} + 2 {\cal C}_{\theta_1x}^{(1)}\right) \approx -\frac{987.4 \lambda_{\epsilon}^2}{ \gamma ^2 {g_s}^2 \left(\log N\right) ^2 M^4}.
\end{equation}
Consistency of (\ref{f_{r_0}}), (\ref{fr0-omega}) and (\ref{alphatheta1}), setting
$\sqrt[24]{\beta}\approx\beta^{0.04}$ to unity (because of the very small exponent of $\beta$), requires:
\begin{eqnarray}
\label{alphatheta1-ii}
\alpha_{\theta_1} = \frac{11.33 \sqrt[3]{{\cal C}^{\rm IR}_{\phi_0}} N^{55/36}}{{g_s} \sqrt[12]\lambda_{\epsilon} \sqrt[3]{M} {N_f}^{2/3} \sqrt[3]{{\cal C}^{\rm IR}_{\psi_1}} \log
   ^{\frac{2}{3}}(N)}.
\end{eqnarray}

\item
{\bf Step 4: Matching $g_{\rm YM}^2(\Lambda_{\rm QCD}=0.4$GeV, $\Lambda=1.1$GeV, $\mu=M_\rho$)}

Similarly, $g_{\rm YM}^2$  can be chosen to match the experimental value $36$ (at $\Lambda_{\rm QCD}=0.3$GeV and the HLS-QCD matching scale ``$\Lambda"=1.1$GeV \cite{HLS-Physics-Reports}) and renormalization scale $\mu=M_\rho$ by imposing:
\begin{equation}
\frac{12 \delta  N}{\alpha _{\theta _1}^2 \left(\frac{855.11 \beta ^{3/2} (\delta +1) \lambda_{\epsilon}^3 {m_0}^2}{{f_{r_0}}^2 \gamma ^2
   {g_s}^2 \left(\log N\right) ^2 M^4}+12 N\right)}=36,
\end{equation}
which can be effected by:
\begin{equation}
\label{gYMsq-gamma}
\gamma = \frac{ 10^{6} \lambda_{\epsilon}^{3/2} {m_0} \beta ^{3/4} \alpha _{\theta _1}\sqrt{ 8.4(\delta + 1)}}{{f_{r_0}}
   {g_s} \left(\log N\right)  M\sqrt{ N \left(3.27\times 10^9 \delta -1.18\times 10^{11} \alpha _{\theta _1}^2\right)}}
\end{equation}
By requiring the argument of the square root in the denominator of (\ref{gYMsq-gamma}) to be positive, one hence obtains an upper bound on ${\cal C}_{\phi_0}^{\rm IR}$ from (\ref{gYMsq-gamma}):
\begin{eqnarray}
\label{constant_upper_bound}
& & {\cal C}_{\phi_0}^{\rm IR} < \frac{1.04\times10^{-13} \sqrt[8]{\beta } ({f_{r_0}}-1)^2 {g_s}^2 \sqrt[4]{\lambda_{\epsilon}} M N_f ^2
   {\cal C}_{\psi_1}^{\rm IR}\  N^{-3 {f_{r_0}}-\frac{73}{20}} \log ^2(N)}{{f_{r_0}}+1}.
\end{eqnarray}

\item
{\bf Step 5: Matching $L_{10}^r$}

Now, using (\ref{Vol2}), (\ref{CCsO4}), (\ref{z_i}) and \cite{Relationship-Li-yi-zi}, \cite{Wilson-matching} (which provides also the UV-finite part of the $\rho-\pi$ one-loop correction via an ``$a(M_\rho)$" factor taken to be equal to 2) one can show that (for $N_f=3$):
\begin{eqnarray}
\label{L10}
& &  L_{10}^r = \frac{1}{4}\left(-\frac{1}{g_{\rm YM}^2} + 2z_3 - 2z_2 + 2z_1 \right) + \frac{11 N_f a(M_\rho)}{96(4\pi)^2}\nonumber\\
& &  = \frac{0.47 {g_s} N^{\frac{2 {f_{r_0}}}{3}+1}}{\alpha _{\theta _1}}-
\frac{8640. {g_s} N^{5/3} \alpha _{\theta _1}^2}{{g_s} N^{5/3}-1831.63 \alpha _{\theta _1}^3} + 4.4\times10^{-3}.
\nonumber\\
\end{eqnarray}
Now, using $L_{10}^r = -5.5\times10^{-3}$, one obtains the following value of the $\theta_1$ delocalization parameter $\alpha_{\theta_1}$:
\begin{eqnarray}
\label{alphatheta1-L10}
& & \alpha_{\theta_1} = -2.8\times10^{-14} g_s N^{\frac{5}{3}} + \frac{\sqrt{4.2\times10^{-6} - 2.9\times10^{-11}g_s^2N^{\frac{10}{3}}}}{2}.
\end{eqnarray}
For $g_s=0.1$, this implies $N<140$.  For numerical computation we set $N = 10^2, g_s=0.1, M=N_f=3$, and from (\ref{alphatheta1-ii}) and (\ref{alphatheta1-L10}) we obtain the following non-linear relation between ${\cal C}_{\psi_1}^{\rm IR}, \lambda_{\epsilon}$ and ${\cal C}_{\phi_0}^{\rm IR}$ (assumed to be also satisfying (\ref{constant_upper_bound})):
\begin{equation}
\label{alphatheta1-L10-ii}
\frac{1.6\sqrt[3]{{\cal C}^{\rm IR}_{\phi_0}} }{\lambda_{\epsilon}^{\frac{1}{12}}  \sqrt[3]{{\cal C}^{\rm IR}_{\psi_1}}} = 10^{-7}.
\end{equation}

\end{itemize}

\newpage

The key results of this section are summarized below.
\begin{enumerate}
\item
Fixing (eight non-zero) parameters:
\begin{enumerate}
\item  a linear combination of constants of integration appearing in the solutions of the EOMs of the ${\cal O}(R^4)$ ${\cal M}$-theory uplift's metric components $G^{\cal M}_{zz, \theta_1 z, \theta_1x} $\footnote{In principle, there are other costants of integration appearing in other ${\cal O}(R^4)$ ${\cal M}$-theory metric components (\ref{ M-theory-metric-psi=2npi-patch}), but it turns out that there is a specific combination of only three that appears while matching $\chi$PT LECs up to ${\cal O}(p^4)$. Even though it is unclear why a specific combination, but it is intuitively evident that it involves $G^{\cal M}_{zz, \theta_1 z, \theta_1x} $ as these essentially correspond to the $S^3$, part of the non-compact four-cycle wrapped by the flavor $D7$-branes in the type IIB dual of \cite{metrics}; these $D7$-flavor branes are (triple) T dualized to the type IIA $D6$ flavor branes. This, in fact, is an extremely non-trivial signature of the four-cycle wrapped by the type IIB $D7$-branes, that manifests itself as ${\cal O}(R^4)$-corrections to the MQGP background \cite{MQGP}/${\cal M}$-theory uplift of thermal QCD-like theories.}; 
\item
constants of integration 
$ {\cal C}^{\rm IR}_{\psi_1,\phi_0}$\footnote{ ${\cal C}^{\rm UV}_{\psi_1,\phi_0}$ can be self-consistently set to zero - see 2. and 3.} appearing in the solutions  of the EOMs in the IR of respectively the $\rho$ and $\pi$ mesons;  

\item
$\epsilon$ (or equivalently $\lambda_\epsilon$ as in  $\epsilon=\lambda_\epsilon\frac{\sqrt{\beta}}{N}), \gamma$ as in $b=\frac{1}{\sqrt{3}} + \epsilon$ in the relationship between the resolution parameter  $a$ (i.e., the radius of the blown-up $S^2$) and the IR cut-off $r_0$: $a = \left(b + {\bf \gamma} {\cal O}\left(\frac{g_sM^2}{N}\right)\right)r_0$; 

\item
$\theta_1$ delocalization parameter  $\alpha_{\theta_1}$; 

\item $D6$-brane tension or equivalently $\alpha^\prime$; 

\item
$ f_{r_0}$ as in $r_0 = N^{-\frac{f_{r_0}}{3}}$
\end{enumerate}
 [all for given values of $N, M, N_f$ in the MQGP limit (\ref{MQGP_limit})] of the top-down holographic dual by matching with experimental values of one-loop renormalized $\chi$PT Lagrangian's LECs $L_{1, 2, 3, 9, 10}^r, F_\pi^2, g_{SU(3)}$: table 5 lists out the same. 

Using the values of the parameters of our ${\cal M}$-theory dual of thermal QCD-like theories, one can also obtain the values of the LEC $H_1$ of (\ref{ChPT-Op4}), and in principle, the LECs  of the $\chi$PT Lagrangian at ${\cal O}(p^6)$ \cite{Op6}. We defer the latter in particular for later exploration.

{\footnotesize
\begin{table}[h]
\begin{center}
\begin{tabular}{|c|c|c|c|}\hline
S. No. & Quantities whose    & Parameters of the holographic dual  & 
Equation numbers \\
& experimental values & used for fitting & \\
& are fitted to & & \\ \hline
1. & $L_{1,2,3}^r$ & Specific linear combination of & (\ref{CCsO4}) \\
& & constants of integration appearing in & [using (\ref{epsilon-sqrtbeta-over-N-lambdaeps})] \\
& & solutions to ${\cal O}(R^4)$ corrections & \\
& & to ${\cal M}$-theory metric components & \\ 
& &   $G^{\cal M}_{zz, \theta_1 z, \theta_1x}$; $f_{r_0}; \gamma; \lambda_\epsilon$; & \\ \cline{3-4}
& & ${\cal O}(R^4)-\frac{1}{N}$ connection: &  (\ref{epsilon-sqrtbeta-over-N-lambdaeps}) \\
& & $a-r_0$ relation must have & \\
& & an $\epsilon\sim\frac{l_p^3}{N}$ contribution & \\
& & at ${\cal O}\left(\frac{g_s M^2}{N}\right)^0$ & \\ \hline
2. & $F_\pi^2;\ L_9^r$ & $f_{r_0}; \alpha_{\theta_1}; {\cal C}^{\rm IR}_{\phi_0}; {\cal C}^{\rm IR}_{\psi_1}; \lambda_\epsilon$ & (\ref{alphatheta1}), (\ref{L9r-match})-(\ref{alphatheta1-ii})[consistency \\
& & & check] \\ \hline
3. & $g_{SU(3)}$ & $\gamma$; upper bound on ${\cal C}^{\rm IR}_{\phi_0}$ &  (\ref{gYMsq-gamma}), (\ref{constant_upper_bound}) \\ \hline
4. & $L_{10}^r$ & $\alpha_{\theta_1}; {\cal C}^{\rm IR}_{\phi_0}; {\cal C}^{\rm IR}_{\psi_1}$ & (\ref{alphatheta1-L10}) - (\ref{alphatheta1-L10-ii}) [even though  \\
& & & specific values of $N, M, N_f, g_s$ \\
& & & chosen, but can find analog of \\
& & & (\ref{alphatheta1-L10-ii}) $\forall N<140$ (\ref{alphatheta1-L10}) \\
& & & respecting (\ref{MQGP_limit})] \\ \hline
\end{tabular}
\end{center}
\caption{Summary of Matching of Parameters of the ${\cal M}$-theory uplift and Experimental values of 1-Loop renormalized $\chi$PT LECs, $F_\pi^2, g_{SU(3)}^2$ }
\end{table}
}

\item
Further, the normalization condition of the $\rho$-meson profile function $\psi_1(Z)$ ($n=1$ mode) is used to determine the constant of integration ${\cal C}_{\psi_1}^{\rm UV}$ appearing in the solution to $\psi_1(Z)$ in the UV, in terms of ${\cal C}^{\rm IR}_{\psi_1}, f_{r_0}, \gamma, \epsilon, \lambda_\epsilon$ and
$T_{D6}$ or equivalently $\alpha^\prime$ (via ``${\cal V}_{\Sigma_2}$"); it was shown that one could self-consistently set ${\cal C}^{\rm UV}_{\psi_1}=0$ \footnote{The $\rho$-meson mass parameter $m_0$ can be fixed by imposing Dirichlet boundary condition on $\psi_1(Z)$ at $Z=0$ \cite{VA-Glueball-decay}.}.

\item
Substituting the expression for $T_{D6}/\alpha^\prime/{\cal V}_{\Sigma_2}$ obtained above from the normalization condition of $\psi_1(Z)$ into the normalization condition for the profile function $\phi_0(Z)$, one could self-consistently set ${\cal C}^{\rm UV}_{\phi_0}=0$.

\end{enumerate}  

\section{Summary}

In this paper, we have shown that ${\cal O}\left(\frac{1}{N}\right)$-corrections of the  ${\cal M}$-theory  uplift of holographic thermal QCD of \cite{metrics}  as worked out in \cite{MQGP,NPB} in conjunction with the ${\cal O}(R^4)$-corrections to the same as worked out in \cite{OR4-Yadav+Misra}, can be used to match the experimental values of the   coupling constants up to ${\cal O}(p^4)$  appearing in the $SU(3)$ Chiral Perturbation theory Lagrangian for the pion and rho vector mesons as well as their flavor partners, in the chiral limit.

The following are the key lessons learnt in this paper:
\begin{itemize}
\item
There is a particular combination of the constants of integration appearing in the solutions to the ${\cal O}(R^4)$ corrections to the  ${\cal M}$-theory  dual of thermal QCD that will appear in all the coupling constants of $\chi$PT at least up to ${\cal O}(p^4)$. As an example, in working in the $\psi=0,2\pi,4\pi$ patches close to the type IIB Ouyang embedding effected by working near small $\theta_{1,2}$ the aforementioned combination is ${\cal C}_{zz}^{(1)} - 2 {\cal C}_{\theta_1z}^{(1)} + 2 {\cal C}_{\theta_1x}^{(1)}$, where ${\cal C}_{MN}^{(1)}$s are the constants of integration appearing in the solutions to the EOMs for the ${\cal O}(R^4)$-corrections $G_{MN}^{{\cal M}, (1)} = G_{MN}^{{\cal M},(0)}f_{MN},$ $G_{MN}^{{\cal M},(0)}$ being the MQGP metric of \cite{metrics,NPB}. This dependence is expected to change in the $\psi \neq 2n\pi, n=0, 1, 2$-patches.
\item
Matching the result obtained from our  ${\cal M}$-theory  ${\cal O}(R^4)$-corrected holographic computation with the experimental values of one-loop renormalized $L_{1,2,3}^r$, one sees one that one is required to do two things. One, the ${\cal O}\left(\frac{1}{N}\right)$ correction to the leading order (in $N$) result in expressing the resolution parameter in terms of the IR cut-off, also must involve a term proportional to $\frac{l_p^3}{N}$, i.e., $a = \left(b + \gamma \frac{g_sM^2}{N}\left(1+\log r_0\right)\right)r_0\rightarrow \left(\tilde{b} + \lambda \frac{l_p^3}{N} + \gamma\frac{g_sM^2}{N}(1+\log r_0)\right)r_0$. The second, the value of the aforementioned linear combination of integration constants figuring in the solutions to the ${\cal O}(R^4)$ corrections to the MQGP metric of \cite{MQGP,NPB}, gets fixed in terms of $\lambda, \gamma, M, g_s$ and $N$. This is the first evidence of the relationship between the ${\cal O}\left(\frac{1}{N}\right)$ and ${\cal O}(\beta)$ corrections.
\item
Matching the experimental values of $F_\pi^2$ and the one-loop renormalized $L_9^r$ and internal consistency, determine the angular delocalization in the polar angles $\theta_{1,2}$ (\ref{theta12-deloc}) consistent with the type IIB Ouyang embedding of the flavor $D7$-branes in the type IIB holographic dual and its SYZ type IIA mirror \cite{Yadav+Misra+Sil-Mesons}. Note, similar to as explained in \cite{SYZ-free-delocalization}, the SYZ type IIA mirror (and hence its M theory uplift) is independent of the angular delocalization. In the context of obtaining the values of the $\chi$PT Lagrangian's coupling constants this is encoded in the fact even though the aforementioned angular delocalization  parameters $\alpha_{\theta_{1,2}}$ would change depending on the values of $\theta_{10,20}$ of (\ref{xyz-definitions}), the corresponding values of $\alpha_{\theta_{1,2}}$ can always be found.
\item
Matching the experimental value of $g_{\rm YM}^2$ at $\Lambda_{\rm QCD}=0.4$GeV, the HLS-QCD matching scale $\Lambda=1.1$GeV and renormalization scale $\mu=M_\rho$ with the value obtained from our setup, provide the following pair of results. One, the ${\cal O}\left(\frac{g_sM^2}{N}\right)$ correction appearing in the resolution parameter - IR cut-off relation turns out to also have an $l_p^{\frac{9}{2}}$ dependence apart from dependence on $g_s, N, M, N_f$, as well as the constants of integration appearing in the solutions to the radial profile functions of the $\pi$ and $\rho$ mesons (and their flavor partners) in the IR.  Second, one obtains an upper bound on the constant of integration appearing in the expression for the profile function of the $\pi$ meson (and its flavor partners) in the IR.
\item
Upon matching with the experimental value of the one-loop renormalized $L_{10}^r$, one obtains a non-linear relation between the constants of integration appearing in the radial profile functions for the $\pi$ and $\rho$ mesons in the IR, as well as the coefficient of the $\frac{l_p^3}{N}$ term required to exist in the resolution-parameter-IR-cut-off relation (discussed in the second bullet above) upon matching with the experimental values of $L_{1,2,3}$. For numerical clarity, we explicitly wrote down the same for $N=10^2, g_s=0.1, M=N_f=3$.
\end{itemize}

\newpage

\section*{Acknowledgement}

VY is supported by a Senior Research Fellowship (SRF) from the University Grants Commission,
Govt. of India. GY is supported by a Junior Research Fellowship (JRF) from the Council of Scientific and Industrial Research, Govt. of India. AM was partially supported by a grant from the Council of Scientific and Industrial Research, Government of India, grant number CSR-1477-PHY.

\appendix

\section{Equations of Motion for $f_{MN} $s, their large-$N$ large-$r$ (UV) Limit and Proof of $f_{MN}$s to be Vanishingly Small in the UV}
\setcounter{equation}{0} \seceqaa

In this appendix, we will show that $f_{MN}$ is vanishingly small in the UV  inclusive of the ${\cal O}(l_p^6 R^4)$ corrections to the ${\cal M}$-Theory uplift of large-$N$ thermal QCD at low temperatures (i.e. below the deconfinement temperature) near the $\psi=2n\pi, n=0, 1, 2$-patches. In the following:
{\footnotesize
\begin{eqnarray}
\label{g_2+a_i+b_j-defs}
& & \hskip -0.6in g_2(r) \equiv 1 - \frac{r_0^4}{r^4},\nonumber\\
& & \hskip -0.6in{a_1}(r)=\frac{3 \left(-{N_f^{\rm UV}}\  \log \left(9 a^2 r^4+r^6\right)+\frac{8 \pi }{{g_s}}+2 \log N  {N_f^{\rm UV}}\ -4 {N_f^{\rm UV}}\
   \log \left(\alpha _{\theta _1}\right)-4 {N_f^{\rm UV}}\  \log \left(\alpha _{\theta _2}\right)+4 {N_f^{\rm UV}}\  \log (4)\right)}{8 \pi }\nonumber\\
& & \hskip -0.6in{a_2}(r)=\frac{12 a^2 {g_s} M_{\rm UV}^2 {N_f^{\rm UV}}\  ({c_1}+{c_2} \log ({r_0}))}{9 a^2+r^2}\nonumber\\
& &\hskip -0.6in {a_3}(r)=\frac{r^2 {a_2}(r)}{2 {N_f^{\rm UV}}\  \left(6 a^2+r^2\right)} \nonumber\\
& &\hskip -0.6in {a_4}(r)=\frac{6 a^2+r^2}{9 a^2+r^2} \nonumber\\
& &\hskip -0.6in B(r)=\frac{3 {g_s} M_{\rm UV}^2 \log (r) \left(-{g_s} \log N  {N_f^{\rm UV}}\ +2 {g_s} {N_f^{\rm UV}}\  \log \left(\alpha _{\theta
   _1}\right)+2 {g_s} {N_f^{\rm UV}}\  \log \left(\alpha _{\theta _2}\right)+12 {g_s} {N_f^{\rm UV}}\  \log (r)+6 {g_s} {N_f^{\rm UV}}\ -2
   {g_s} {N_f^{\rm UV}}\  \log (4)+8 \pi \right)}{32 \pi ^2}\nonumber\\
& &\hskip -0.6in {b_1}(r)=\frac{1}{6
   \sqrt{2} \pi ^{5/4} r^2 \alpha _{\theta _1} \alpha _{\theta _2}^2}\Biggl\{{g_s}^{3/4} M_{\rm UV}\nonumber\\
& & \hskip -0.6in \times\Biggl[-{g_s} {N_f^{\rm UV}}\  \left(r^2-3 a^2\right) \log N  (2 \log (r)+1)+\log (r) \left(4
   {g_s} {N_f^{\rm UV}}\  \left(r^2-3 a^2\right) \log \left(\frac{1}{4} \alpha _{\theta _1} \alpha _{\theta _2}\right)+8 \pi
   \left(r^2-3 a^2\right)-3 {g_s} {N_f^{\rm UV}}\  r^2\right)\nonumber\\
& & \hskip -0.6in+2 {g_s} {N_f^{\rm UV}}\  \left(r^2-3 a^2\right) \log \left(\frac{1}{4}
   \alpha _{\theta _1} \alpha _{\theta _2}\right)+18 {g_s} {N_f^{\rm UV}}\  \left(r^2-3 a^2 (6 r+1)\right) \log ^2(r)\Biggr]\Biggr\}\nonumber\\
& & \hskip -0.6in {b_2}(r)=\frac{{g_s}^{7/4} M_{\rm UV} {N_f^{\rm UV}}\  \log (r) \left(36 a^2 \log (r)+r\right)}{3 \sqrt{2} \pi ^{5/4} r \alpha _{\theta
   _2}^3} ,\nonumber\\
& &
\end{eqnarray}
}
\noindent wherein $M_{\rm UV}\equiv M_{\rm eff}(r>{\cal R}_{D5/\overline{D5}})$ and similarly $N_f^{\rm UV}\equiv N_f^{\rm eff}(r>{\cal R}_{D5/\overline{D5}})$. We should keep in  mind that near the $\psi=\psi_0\neq2n\pi, n=0, 1, 2$-patch, some $G^{\cal M}_{rM}, M\neq r$ and $G^{\cal M}_{x^{10}N}, N\neq x^{10}$ components are non-zero, making this exercise much more non-trivial. As shown in \cite{OR4-Yadav+Misra}, the contributions from $E_8$ is sub-dominant as compared to the contributions from $J_0$ terms.

As the EOMs are too long, they have not been explicitly typed but their forms have been written out. The explicit forms of ${\cal F}_{MN}$ have been given.

Using (\ref{g_2+a_i+b_j-defs}), one obtains the following EOMs in the UV.:
{\footnotesize
\begin{eqnarray}
\label{IR-psi=2nPi-EOMs}
& & {\rm EOM}_{MN}:\nonumber\\
& & \sum_{{\cal M},{\cal N}=0}^{x^{10}} \sum_{p=0}^2{\cal H}^{{\cal M}{\cal N}\ (p)}_{MN}\left(r; r_0, a, N, M^{\rm UV}, N_f^{\rm UV}, g_s, \alpha_{\theta_{1,2}}\right)f_{{\cal M}{\cal N}}^{(p)}(r) +
\beta {\cal F}_{MN}\left(r; r_0, a, N, M, N_f, g_s, \alpha_{\theta_{1,2}}\right) = 0,\nonumber\\
& &
\end{eqnarray}
}
where $M, N$ run over the $D=11$ coordinates,   $f^{(p)}_{MN}\equiv \frac{d^p f_{MN}}{dr^p}, p=0, 1, 2$.

In the UV, and in the large-$N$ limit, one can show that EOMs are algebraic and are given as under:
\begin{eqnarray*}
&  & \hskip -0.7in EOM_{x^3x^3} : -\frac{49 r^2 {g_2} (r) (2 {f_{zz} } (r)-3 {f_{x^{10}x^{10}}} (r)-4 {f_{\theta_1z} } (r)-5 {f_{\theta_2y} } (r))}{3456 \sqrt{\pi } \sqrt{{g_s}} \alpha
   _{\theta _1}^2 \alpha _{\theta _2}^4 {b_1}(r)^2} -
\frac{7 a^8 \beta  M_{\rm UV} \left(\frac{1}{N}\right)^{9/4}\Sigma_1 \log \left({r_0}\right)}{216 \pi ^2 {g_s} \log N  ^2 {N_f^{\rm UV}}\  r^6 \alpha
   _{\theta _2}^3} = 0\nonumber\\
& & \nonumber\\
& & \hskip -0.7in EOM_{\mathbb{R}^{1,2}(t,x^{1,2})} :  \frac{49 r^2 (2 {f_{zz} } (r)-3 {f_{x^{10}x^{10}}} (r)-4 {f_{\theta_1z} } (r)-5 {f_{\theta_2y} } (r))}{3456 \sqrt{\pi } \sqrt{{g_s}} \alpha
   _{\theta _1}^2 \alpha _{\theta _2}^4 {b_1}(r)^2}  -\frac{7 a^8 \beta  M_{\rm UV} \left(\frac{1}{N}\right)^{9/4} \Sigma_1 \log (r)}{216 \pi ^2 {g_s} \log N  ^2 {N_f^{\rm UV}}\  r^6
   \alpha _{\theta _2}^3} = 0\nonumber\\
& & \nonumber\\
& & \hskip -0.7in  EOM_{rr} :  \frac{49 \sqrt{\pi } \sqrt{{g_s}} N {a_4}(r) (2 {f_{zz} } (r)-3 {f_{x^{10}x^{10}}} (r)-4
   {f_{\theta_1z} } (r)-5 {f_{\theta_2y} } (r))}{864 r^2 \alpha _{\theta _1}^2 \alpha _{\theta _2}^4 {b_1}(r)^2
   {g_2} (r)} -\frac{2 a^{12} \beta  M_{\rm UV} \left(\frac{1}{N}\right)^{9/10} \Sigma_1}{3 \sqrt{3} \pi
   ^{5/4} \sqrt[4]{{g_s}} \log N  ^2 {N_f^{\rm UV}}\  r^{14} \alpha _{\theta _2}^5} = 0\nonumber\\
& & \hskip -0.7in EOM_{\theta_1\theta_1} : -\frac{3969 \sqrt[5]{N} \alpha _{\theta _1}^2 (2 {f_{x^{10}x^{10}}} (r)+{f_{\theta_2y} } (r))}{512 \alpha _{\theta _2}^2}  + \frac{a^{12} \beta  {g_s}^{5/4} M_{\rm UV}^2 \left(\frac{1}{N}\right)^{17/10} (-\Sigma_1) \log ^2\left({r_0}\right)}{26244 \sqrt{2} \pi ^{11/4} \log N   r^{12} \alpha _{\theta
   _1}^6 \alpha _{\theta _2}^4} = 0\nonumber\\
& & \nonumber\\
& & \hskip -0.7in   EOM_{\theta_1\theta_2} : -\frac{3969 N^{3/10} \alpha _{\theta _1}^2 {b_2}(r) (2 {f_{x^{10}x^{10}}} (r)+{f_{\theta_2y} } (r))}{512 \alpha _{\theta
   _2}^2 {b_1}(r)}=0 \nonumber\\
& & \nonumber\\
& & \hskip -0.7in   EOM_{\theta_1x} : \frac{49 N^{21/20} (2 {f_{zz} } (r)-3 {f_{x^{10}x^{10}}} (r)-4 {f_{\theta_1z} } (r)-5 {f_{\theta_2y} } (r))}{1728 \alpha _{\theta _1}^2 \alpha
   _{\theta _2}^4 {b_1}(r)} \nonumber\\
& & +\frac{a^{12} \beta  {g_s} M_{\rm UV}^2 \left(\frac{1}{N}\right)^{17/20} \left(19683 \sqrt{6} \alpha _{\theta _1}^6+6642 \alpha
   _{\theta _2}^2 \alpha _{\theta _1}^3-40 \sqrt{6} \alpha _{\theta _2}^4\right) \log \left({r_0}\right)}{18 \sqrt{6} \pi ^3
   \log N   r^{12} \alpha _{\theta _1} \alpha _{\theta _2}^7} = 0\nonumber\\
& & \nonumber\\
& & \hskip -0.7in EOM_{\theta_1y} : \frac{49 N^{7/20} (2 {f_{zz} } (r)-3 {f_{x^{10}x^{10}}} (r)-4 {f_{\theta_1z} } (r)-5 {f_{\theta_2y} } (r))}{128 \sqrt{6} \alpha _{\theta
   _2}^3 {b_1}(r)}  - \frac{11 \sqrt{\frac{2}{3}} a^{12} \beta  M_{\rm UV} \left(\frac{1}{N}\right)^{23/20} (-\Sigma_1) \log
   ^2\left({r_0}\right)}{3 \pi ^{3/2} \sqrt{{g_s}} \log N  ^2 {N_f^{\rm UV}}\  r^{12} \alpha _{\theta _1}^3 \alpha
   _{\theta _2}^3} = 0 \nonumber\\
& & \hskip -0.7in EOM_{\theta_1z} : -\frac{49 N^{13/20} (2 {f_{zz} } (r)-3 {f_{x^{10}x^{10}}} (r)-4 {f_{\theta_1z} } (r)-5 {f_{\theta_2y} } (r))}{1152 \alpha _{\theta _2}^4
   {b_1}(r)} + \frac{a^{12} \beta  {g_s} M_{\rm UV}^2 \left(\frac{1}{N}\right)^{5/4} \alpha _{\theta _1}\Sigma_1 \log
   \left({r_0}\right)}{12 \sqrt{6} \pi ^3 \log N   r^{12} \alpha _{\theta _2}^7} = 0
\nonumber\\
& & \nonumber\\
\end{eqnarray*}
\begin{eqnarray}
\label{EOMs}
& &  \hskip -0.7in EOM_{\theta_2\theta_2} : {f_{zz} } (r) = {f_{x^{10}x^{10}}} (r)\nonumber\\
& & \nonumber\\
& & \hskip -0.7in EOM_{\theta_2x} : \frac{49 N^{23/20} {b_2}(r) (2 {f_{zz} } (r)-3 {f_{x^{10}x^{10}}} (r)-4 {f_{\theta_1z} } (r)-5 {f_{\theta_2y} } (r))}{1728 \alpha _{\theta _1}^2 \alpha _{\theta _2}^4
   {b_1}(r)^2} +
\frac{8 a^{14} \beta  M_{\rm UV} \left(\frac{1}{N}\right)^{7/20}\Sigma_1 \log \left({r_0}\right)}{243 \pi ^{3/2} \sqrt{{g_s}} \log N  ^2 {N_f^{\rm UV}}\  r^{14} \alpha _{\theta
   _1}^4 \alpha _{\theta _2}^5} = 0\nonumber\\
&& \nonumber\\
& & \hskip -0.7in EOM_{\theta_2y}: -\frac{49 N^{17/20} \left(7 \sqrt{2} \sqrt[4]{\pi } \sqrt[4]{{g_s}} {f_{zz} } (r)-15 \sqrt{2} \sqrt[4]{\pi } \sqrt[4]{{g_s}}
   {f_{x^{10}x^{10}}} (r)-14 \sqrt{2} \sqrt[4]{\pi } \sqrt[4]{{g_s}} {f_{\theta_1z} } (r)+5 \sqrt{2} \sqrt[4]{\pi } \sqrt[4]{{g_s}}
   {f_{\theta_2y} } (r)\right)}{15552 \alpha _{\theta _1}^4 \alpha _{\theta _2}^3 {b_1}(r)^2} \nonumber\\
& & +
\frac{8 a^{12} \beta  \left(\frac{1}{N}\right)^{13/20} \left(19683 \sqrt{6} \alpha _{\theta _1}^6+6642 \alpha _{\theta _2}^2 \alpha _{\theta
   _1}^3-40 \sqrt{6} \alpha _{\theta _2}^4\right)}{243 {g_s}^2 \log N  ^2 {N_f^{\rm UV}}\ ^2 r^{12} \alpha _{\theta _1}^6 \alpha _{\theta
   _2}} = 0\nonumber\\
& & \hskip -0.7 in EOM_{\theta_2z} :  -\frac{49 N^{3/4} {b_2}(r) (2 {f_{zz} } (r)-3 {f_{x^{10}x^{10}}} (r)-4 {f_{\theta_1z} } (r)-5 {f_{\theta_2y} } (r))}{1152 \alpha _{\theta
   _2}^4 {b_1}(r)^2}   - \frac{4 a^{14} \beta  M_{\rm UV} \left(\frac{1}{N}\right)^{3/4}(-\Sigma_1) \log (r)}{81 \pi ^{3/2} \sqrt{{g_s}} \log N  ^2
   {N_f^{\rm UV}}\  r^{14} \alpha _{\theta _1}^2 \alpha _{\theta _2}^5} = 0\nonumber\\
& & \hskip -0.7in  EOM_{xx} : \frac{144 \sqrt{3} \sqrt[4]{\pi } a^{10} \beta  \sqrt[4]{\frac{1}{N}}}{\log N  ^3 {N_f^{\rm UV}}\ ^2
   r^{10}} +\frac{{g_s}^{7/4} N^{9/10} \alpha _{\theta _2}^2 (\log N  -18 \log r )^2 ({f_{zz} } (r)-2
   {f_{\theta_1z} } (r)+2 {f_{\theta_1x} } (r)-{f_r}(r))}{\log r ^2 (\log N  -9 \log r )^2}=0\nonumber\\
& & \hskip -0.7in EOM_{xy} : \frac{49 N^{6/5} (2 {f_{zz} } (r)-3 {f_{x^{10}x^{10}}} (r)-4 {f_{\theta_1z} } (r)-5 {f_{\theta_2y} } (r))}{3888 \sqrt{6} \alpha _{\theta _1}^4
   \alpha _{\theta _2}^5 {b_1}(r)^2}  + \frac{8 a^{12} \beta  \left(\frac{1}{N}\right)^{3/10} \left(-\Sigma_1\right)}{243 \sqrt{3} \sqrt[4]{\pi } {g_s}^{9/4} \log N  ^3
   {N_f^{\rm UV}}\ ^2 r^{12} \alpha _{\theta _1}^6 \alpha _{\theta _2}^3} = 0\nonumber\\
& & \nonumber\\
& & \hskip -0.7in EOM_{xz} : -\frac{49 N^{3/2} (2 {f_{zz} } (r)-3 {f_{x^{10}x^{10}}} (r)-4 {f_{\theta_1z} } (r)-5 {f_{\theta_2y} } (r))}{34992 \alpha _{\theta _1}^4
   \alpha _{\theta _2}^6 {b_1}(r)^2} \nonumber\\
   & & \hskip -0.3in + \frac{16 a^{10} \beta  \left(\frac{1}{N}\right)^{13/20} \left(-4 {g_s} {N_f^{\rm UV}}\  \log \left(\alpha _{\theta
   _1}\right)-4 {g_s} {N_f^{\rm UV}}\  \log \left(\alpha _{\theta _2}\right)+6 {g_s} {N_f^{\rm UV}}\  \log
   \left(\frac{1}{r_0}\right)+4 {g_s} {N_f^{\rm UV}}\  \log (4)+8 \pi \right)}{27 \sqrt{3} \sqrt[4]{\pi } {g_s}^{13/4}
   \log N  ^4 {N_f^{\rm UV}}\ ^3 r^{10} \alpha _{\theta _1}^2 \alpha _{\theta _2}^4}\nonumber\\
& &  = 0\nonumber\\
& & \hskip -0.7in EOM_{yy} : -\frac{49 \sqrt{N} (16 {f_{zz} } (r)+3 {f_{x^{10}x^{10}}} (r)+4 {f_{\theta_1z} } (r)+5 {f_{\theta_2y} } (r)-36 {f_{yz} }(r)+18
   {f_{yy}}(r))}{1728 \alpha _{\theta _1}^2 \alpha _{\theta _2}^4 {b_1}(r)^2}\nonumber\\
& &  +
\frac{2 \sqrt{2} a^{12} \beta (-\Sigma_1)}{27 \sqrt[4]{\pi } {g_s}^{9/4} \log N  ^3 N {N_f^{\rm UV}}\ ^2 r^{12} \alpha _{\theta
   _1}^4 \alpha _{\theta _2}^2}=0\nonumber\\
& & \nonumber\\
& & \hskip -0.7in EOM_{yz} :  -\frac{49 N^{4/5} (2 {f_{zz} } (r)-3 {f_{x^{10}x^{10}}} (r)-4 {f_{\theta_1z} } (r)-5 {f_{\theta_2y} } (r))}{2592 \sqrt{6} \alpha _{\theta _1}^2
   \alpha _{\theta _2}^5 {b_1}(r)^2}
- \frac{8 a^{12} \beta  \left(\frac{1}{N}\right)^{21/20} \Sigma_1}{729 {g_s}^2 \log N  ^3 {N_f^{\rm UV}}\ ^2 r^{12} \alpha
   _{\theta _1}^4 \alpha _{\theta _2}} = 0\nonumber\\
& & \nonumber\\
& & \hskip -0.7in EOM_{zz} : \frac{49 N^{11/10} (2 {f_{zz} } (r)-3 {f_{x^{10}x^{10}}} (r)-4 {f_{\theta_1z} } (r)-5 {f_{\theta_2y} } (r))}{23328 \alpha _{\theta _1}^2
   \alpha _{\theta _2}^6 {b_1}(r)^2}
-\frac{8 \sqrt{2} a^{12} \beta  \left(\frac{1}{N}\right)^{2/5} (-\Sigma_1)}{6561 \sqrt[4]{\pi } {g_s}^{9/4}
   \log N  ^3 {N_f^{\rm UV}}\ ^2 r^{12} \alpha _{\theta _1}^4 \alpha _{\theta _2}^4} = 0\nonumber\\
& & \hskip -0.7in EOM_{x^{10}x^{10}} :  -\frac{49 \sqrt{N} (16 {f_{zz} } (r)+3 {f_{x^{10}x^{10}}} (r)-32 {f_{\theta_1z} } (r)-13 {f_{\theta_2y} } (r))}{1728 \alpha _{\theta _1}^2
   \alpha _{\theta _2}^4 {a_1}(r)^2 {b_1}(r)^2} \nonumber\\
& &  +
\frac{8 \sqrt{\pi } a^{10} \beta  (27 \log N  +14) M_{\rm UV} \left(\frac{1}{N}\right)^{7/4}\Sigma_1 \log (r)}{81
   \sqrt{{g_s}} \log N  ^5 {N_f^{\rm UV}}\ ^3 r^{10} \alpha _{\theta _2}^3} = 0\nonumber\\
& & 
\end{eqnarray}
where:
\begin{equation}
\label{Sigma_1}
\Sigma_1 \equiv  \left(-19683 \sqrt{6} \alpha _{\theta
   _1}^6-6642 \alpha _{\theta _2}^2 \alpha _{\theta _1}^3+40 \sqrt{6} \alpha _{\theta _2}^4\right).
\end{equation}

From (\ref{EOMs}), one sees that the ${\cal O}(\beta)$-terms are further large-$N$-large-$r$-suppressed. Hence, in the UV one sees that one can consistently assume $f_{MN}$ to be vanishingly small.

\section{Coupling Constants $y_{1,3,5,7}$ and $z_{1,....,8}$}
\setcounter{equation}{0} \seceqbb

In this appendix, we work out the IR and UV contributions to the coupling constants $y_{1,3,5,7}$ and $z_{1,2,...,8}$.
In the following, $\psi_0(Z) = \int_0^Z \phi_0(Z)dZ$. In the following, we will be splitting $\int dZ$ into $\int_{\rm IR} + \int_{\rm UV}$. Now, one can argue
that $\int_{\rm UV}\sim \left({\cal C}_{\phi_0}^{\rm UV}\right)^m\left({\cal C}_{\psi_1}^{\rm UV}\right)^n, m, n \in \mathbb{Z}^+$, and as argued in Section {\bf 4}, we can self-consistently set ${\cal C}_{\phi_0}^{\rm UV} = {\cal C}_{\psi_1}^{\rm UV} = 0$, and one can hence argue that one can disregard $\int_{\rm UV}$.

In the following calculations, one will need the value of ${\cal V}_2$ in the IR, which can be shown to be given by:
\begin{eqnarray}
\label{V2-IR}
& & \hskip -0.6in {\cal V}_2(Z\in{\rm IR}) = -\frac{3 \left(3 b^2-2\right)
   {g_s}^2 M N^{4/5} {N_f}^2 \log ({r_0}) (\log N -3 \log ({r_0}))}{2 \pi  \log N  \alpha _{\theta _1} \alpha _{\theta
   _2}^2}\nonumber\\
& & \hskip -0.6in \frac{3 \left(2-3 b^2\right) \beta  {g_s}^2 \log r_0  M N^{4/5} {N_f}^2 Z ({\cal C}_{zz}^{(1)}-2   {\cal C}_{\theta_1z}^{(1)}+2   {\cal C}_{\theta_1x}^{(1)})
  }{4 \pi  \alpha _{\theta _1} \alpha _{\theta _2}^2}  \left(\frac{9 b^2 {g_s}^2 M N^{4/5} {N_f}^2 \log ({r_0}) (\log N -3 \log ({r_0}))}{\pi
   \log N  \alpha _{\theta _1} \alpha _{\theta _2}^2}\right)\nonumber\\
\end{eqnarray}

Therefore, using (\ref{y_i+z_i}) and using $b=\frac{1}{\sqrt{3}} + \epsilon$ \cite{OR4-Yadav+Misra} and setting $r_0 = N^{-\frac{f_{r_0}}{3}}$ \cite{Bulk-Viscosity}, one obtains the following results for $y_{1,3,5,7}$:
\begin{eqnarray}
\label{y_i}
& & \hskip -0.5in \bullet\  y_1 = \int {\cal V}_2(Z)  \left(1 + \psi_1(Z) - \psi_0^2(Z)\right)^2\nonumber\\
& & \hskip -0.5in  = \frac{177147 \pi ^6 \beta  {\cal C}_{\phi_0}^{\rm IR}\ ^4 \epsilon ^6 {f_{r_0}} ({f_{r_0}}+1)^3 \log r_0  \alpha _{\theta _1}^6
   N^{4 {f_{r_0}}+\frac{14}{5}} \left({\cal C}_{zz}^{(1)} - 2 {\cal C}_{\theta_1z}^{(1)} + 2 {\cal C}_{\theta_1x}^{(1)}\right)}{8192 ({f_{r_0}}-1)^6 {g_s}^4
   \left(\log N\right) ^7 M^2 N_f ^4}\nonumber\\
& & \hskip -0.5in -\frac{4782969 \sqrt{3} \pi ^7 {\cal C}_{\phi_0}^{\rm IR}\ ^4 \epsilon ^5 {f_{r_0}} ({f_{r_0}}+1)^4
   \alpha _{\theta _1}^9 N^{4 {f_{r_0}}+\frac{11}{5}}}{20480 ({f_{r_0}}-1)^7 {g_s}^6 \left(\log N\right) ^7 M^3 N_f ^6}\nonumber\\
& & \hskip -0.5in \nonumber\\
& & \hskip -0.5in \bullet\  y_3 = \int dZ {\cal V}_2\psi_1^2(Z)\left(1 + \psi_1(Z)\right)^2\nonumber\\
& & \hskip -0.5in  = -\frac{21 {g_s}^2 \log r_0  M N^{4/5} {N_f}^2 {\cal C}_{\psi_1}^{\rm IR}\ ^4 (\log N -3 \log r_0 )}{8 \pi
   \log N  \alpha _{\theta _1} \alpha _{\theta _2}^2}\nonumber\\
& &  \hskip -0.5in+ \frac{3^{9/8} 7^{3/4} \sqrt[4]{\epsilon } {g_s}^2 \log r_0  M N^{4/5} {N_f}^2 {\cal C}_{\psi_1}^{\rm IR}\ ^3
   (\log N -3 \log r_0 )}{2 \sqrt{2} \pi  \log N  \alpha _{\theta _1} \alpha _{\theta _2}^2}\nonumber\\
& & \hskip -0.5in + 945 \sqrt[3]{3} \sqrt[6]{\frac{\pi }{2}} \beta  \epsilon  {g_s}^{3/2} \left(\frac{1}{\log N }\right)^{2/3}
   \log r_0  M N^{3/10} {N_f}^{4/3} {r_0}^2 {\cal C}_{\psi_1}^{\rm IR}\ ^4 ({\cal C}_{zz}^{(1)}-2 {\cal C}_{\theta_1z}^{(1)}+2{\cal C}_{\theta_1x}^{(1)})\nonumber\\
& & \hskip -0.5in \nonumber\\
& &  \hskip -0.5in \bullet\  y_5 = \int dZ {\cal V}_2\psi_0^2(Z)\left(\psi_1(Z)\right)^2) =\frac{6561 \sqrt[4]{3} \sqrt{7} \pi ^3 {\cal C}_{\phi_0}^{\rm IR}\ ^2 \epsilon ^{5/2} {f_{r_0}} ({f_{r_0}}+1)^3 \alpha _{\theta _1}^5 {\cal C}_{\psi_1}^{\rm IR}\ ^2 N^{2
   {f_{r_0}}+\frac{8}{5}}}{256 ({f_{r_0}}-1)^4 {g_s}^2 \left(\log N\right) ^3 M N_f ^2 \alpha _{\theta _2}^2}
\nonumber\\
& & \hskip -0.5in -\frac{63 \sqrt[4]{3} \sqrt{7} \pi ^3 \beta  {\cal C}_{\phi_0}^{\rm IR}\ ^2 \epsilon ^{3/2} {f_{r_0}} ({f_{r_0}}+1) \alpha _{\theta _1}^3 {\cal C}_{\psi_1}^{\rm IR}\ ^2 N^{2
   {f_{r_0}}-\frac{3}{5}} \left({\cal C}_{zz}^{(1)} - 2 {\cal C}_{\theta_1z}^{(1)} + 2 {\cal C}_{\theta_1x}^{(1)}\right) \left(6 \epsilon  N+{f_{r_0}} \gamma  {g_s} \log N  M^2\right)^2}{32768
   ({f_{r_0}}-1)^3 {g_s}^2 \left(\log N\right) ^3 M N_f ^2}\nonumber\\
& &  \hskip -0.5in\nonumber\\
& & \hskip -0.5in \bullet\  y_7= \int dZ {\cal V}_2\psi_1(Z) (1 + \psi_1(Z)) (1 + \psi_1(Z) - \psi_0(Z)^2)^2\nonumber\\
& &  \hskip -0.5in = \frac{4782969 \sqrt[4]{3} \sqrt{7} \pi ^7 {\cal C}_{\phi_0}^{\rm IR}\ ^4 \epsilon ^{9/2} {f_{r_0}} ({f_{r_0}}+1)^4 \alpha _{\theta _1}^9
   {\cal C}_{\psi_1}^{\rm IR}\ ^2 N^{4 {f_{r_0}}+\frac{11}{5}}}{40960 ({f_{r_0}}-1)^7 {g_s}^6 \left(\log N\right) ^7 M^3 N_f ^6}\nonumber\\
& & \hskip -0.5in -\frac{3720087
   \sqrt[4]{3} \sqrt{7} \pi ^7 \beta  {\cal C}_{\phi_0}^{\rm IR}\ ^4 \epsilon ^{9/2} {f_{r_0}} ({f_{r_0}}+1)^3 \alpha _{\theta _1}^9
   {\cal C}_{\psi_1}^{\rm IR}\ ^2 N^{4 {f_{r_0}}+\frac{11}{5}} \left({\cal C}_{zz}^{(1)} - 2 {\cal C}_{\theta_1z}^{(1)} + 2 {\cal C}_{\theta_1x}^{(1)}\right)}{262144 ({f_{r_0}}-1)^7
   {g_s}^6 \left(\log N\right) ^7 M^3 N_f ^6}.
\end{eqnarray}

Similarly, one obtains the following expressions for $z_{1,...,8}$:
{\footnotesize
\begin{eqnarray}
\label{z_i}
& &\hskip -0.3in \bullet\  z_1 = \int dZ {\cal V}_2 \left(1 + \psi_1(Z)\right)^2= \frac{3 \sqrt[4]{3} \sqrt{7} \sqrt{\epsilon } {g_s}^2 M N^{4/5} {N_f}^2 {\cal C}_{\psi_1}^{\rm IR}\ ^2 \log ({r_0})
   (\log N -3 \log ({r_0}))}{16 \pi  \alpha _{\theta _1} \alpha _{\theta _2}^2 \log N } \nonumber\\
& & \hskip -0.3in + \frac{7 \sqrt{7} \beta  \sqrt{\epsilon } {g_s}^2 \log r_0  M N^{4/5} {N_f}^2 {\cal C}_{\psi_1}^{\rm IR}\ ^2
   ({\cal C}_{zz}^{(1)}-2 {\cal C}_{\theta_1z}^{(1)}+2 {\cal C}_{\theta_1x}^{(1)}) \left(12 \log N +\left(-36+6 \log ^2(3)+\log (9) \log (27)-\log (27)
   \log (81)\right) \log ({r_0})\right)}{2048\ 3^{3/4} \pi  \alpha _{\theta _1} \alpha _{\theta _2}^2 (\log N -3 \log
   ({r_0}))}\nonumber\\
& &\hskip -0.3in \nonumber\\
& &\hskip -0.3in \bullet\  z_2 = \int dZ {\cal V}_2\psi_0^2(Z)  =\frac{81 \sqrt{3} \pi ^3 \beta  {\cal C}_{\phi_0}^{\rm IR}\ ^2 \epsilon ^3 ({f_{r_0}}+1)^2 \alpha _{\theta _1}^3 N^{\frac{5 {f_{r_0}}}{3}+\frac{7}{5}}
   \left({\cal C}_{zz}^{(1)} - 2 {\cal C}_{\theta_1z}^{(1)} + 2 {\cal C}_{\theta_1x}^{(1)}\right)}{128 ({f_{r_0}}-1)^2 {g_s} \left(\log N\right) ^2 M N_f ^2}\nonumber\\
& & \hskip -0.3in -\frac{81 \sqrt{3} \pi ^3
   {\cal C}_{\phi_0}^{\rm IR}\ ^2 \epsilon ^3 {f_{r_0}} ({f_{r_0}}+1)^2 \alpha _{\theta _1}^3 N^{2 {f_{r_0}}+\frac{7}{5}}}{256 ({f_{r_0}}-1)^3
   {g_s}^2 \left(\log N\right) ^3 M N_f ^2}\nonumber\\
& &\hskip -0.3in \nonumber\\
& &\hskip -0.3in \bullet\  z_3 = \int dZ {\cal V}_2\psi_1(1 + \psi_1)(Z) = \frac{3 \sqrt[4]{3} \sqrt{7} \sqrt{\epsilon } {g_s}^2 M N^{4/5} {N_f}^2 {\cal C}_{\psi_1}^{\rm IR}\ ^2 \log ({r_0})
   (\log N -3 \log ({r_0}))}{8 \pi  \alpha _{\theta _1} \alpha _{\theta _2}^2 \log N }\nonumber\\
& &\hskip -0.3in + \frac{7 \sqrt[4]{3} \sqrt{7} \beta  \sqrt{\epsilon } {g_s}^2 M N^{4/5} {N_f}^2 {\cal C}_{\psi_1}^{\rm IR}\ ^2 \log
   ({r_0}) ({\cal C}_{zz}^{(1)}-2 {\cal C}_{\theta_1z}^{(1)}+2 {\cal C}_{\theta_1x}^{(1)})}{512 \pi  \alpha _{\theta _1} \alpha _{\theta _2}^2}\nonumber\\
& &\hskip -0.3in \nonumber\\
& &\hskip -0.3in \bullet\  z_4 = \int dZ {\cal V}_2\psi_1(1 + \psi_1 - \phi_0^2)(Z)  = \frac{189\ 3^{3/8} \sqrt[4]{7} \pi ^3 \beta  {\cal C}_{\phi_0}^{\rm IR}\ ^2 \epsilon ^{3/4} {f_{r_0}} ({f_{r_0}}+1) \alpha _{\theta _1}^3
   {\cal C}_{\psi_1}^{\rm IR}\  N^{2 {f_{r_0}}+\frac{7}{5}} \left({\cal C}_{zz}^{(1)} - 2 {\cal C}_{\theta_1z}^{(1)} + 2 {\cal C}_{\theta_1x}^{(1)}\right)}{16384 \sqrt{2} ({f_{r_0}}-1)^3
   {g_s}^2 \left(\log N\right) ^3 M N_f ^2}\nonumber\\
& &\hskip -0.3in +\frac{81\ 3^{3/8} \sqrt[4]{7} \pi ^3 {\cal C}_{\phi_0}^{\rm IR}\ ^2 \epsilon ^{11/4} {f_{r_0}} ({f_{r_0}}+1)^2
   \alpha _{\theta _1}^3 {\cal C}_{\psi_1}^{\rm IR}\  N^{2 {f_{r_0}}+\frac{7}{5}}}{256 \sqrt{2} ({f_{r_0}}-1)^3 {g_s}^2 \left(\log N\right) ^3 M
   N_f ^2}\nonumber\\
& &\hskip -0.3in \nonumber\\
& &\hskip -0.3in \bullet\  z_5 = \int dZ {\cal V}_2\psi_1^2(1 + \psi_1)(Z),
  = -\frac{3 \sqrt[8]{3} 7^{3/4} \sqrt[4]{\epsilon } {g_s}^2 M N^{4/5} {N_f}^2 {\cal C}_{\psi_1}^{\rm IR}\ ^3 \log ({r_0}) (\log N -3 \log
   ({r_0}))}{4 \sqrt{2} \pi  \alpha _{\theta _1} \alpha _{\theta _2}^2 \log N }
\nonumber\\
& & \hskip -0.3in -\frac{21 \sqrt[8]{3} 7^{3/4} \beta  \sqrt[4]{\epsilon } {g_s}^2 M N^{4/5} {N_f}^2 {\cal C}_{\psi_1}^{\rm IR}\ ^3 \log ({r_0})
   ({\cal C}_{zz}^{(1)}-2 {\cal C}_{\theta_1z}^{(1)}+2 {\cal C}_{\theta_1x}^{(1)})}{256 \sqrt{2} \pi  \alpha _{\theta _1} \alpha _{\theta _2}^2}\nonumber\\
& &  \nonumber\\
& &\hskip -0.3in \bullet\  z_6 = \int dZ {\cal V}_2(1 + \psi_1)(1 + \psi_1 - \psi_0^2)(Z)  = \frac{567\ 3^{3/8} \sqrt[4]{7} \pi ^3 {\cal C}_{\phi_0}^{\rm IR}\ ^2 \epsilon ^{11/4} {f_{r_0}} ({f_{r_0}}+1) \alpha _{\theta _1}^3 {\cal C}_{\psi_1}^{\rm IR}\
   N^{2 {f_{r_0}}+\frac{7}{5}} \left({\cal C}_{zz}^{(1)} - 2 {\cal C}_{\theta_1z}^{(1)} + 2 {\cal C}_{\theta_1x}^{(1)}\right)}{8192 \sqrt{2} ({f_{r_0}}-1)^3 {g_s}^2 \left(\log N\right) ^3 M
   N_f ^2}\nonumber\\
& &\hskip -0.3in +\frac{81\ 3^{3/8} \sqrt[4]{7} \pi ^3 {\cal C}_{\phi_0}^{\rm IR}\ ^2 \epsilon ^{11/4} {f_{r_0}} ({f_{r_0}}+1)^2 \alpha _{\theta _1}^3
   {\cal C}_{\psi_1}^{\rm IR}\  N^{2 {f_{r_0}}+\frac{7}{5}}}{64 \sqrt{2} ({f_{r_0}}-1)^3 {g_s}^2 \left(\log N\right) ^3 M N_f ^2}\nonumber\\
& & \hskip -0.3in= \frac{567\ 3^{3/8} \sqrt[4]{7} \pi ^3 {\cal C}_{\phi_0}^{\rm IR}\ ^2 \epsilon ^{11/4} {f_{r_0}} ({f_{r_0}}+1) \alpha _{\theta _1}^3 {\cal C}_{\psi_1}^{\rm IR}\
   N^{2 {f_{r_0}}+\frac{7}{5}} \left({\cal C}_{zz}^{(1)} - 2 {\cal C}_{\theta_1z}^{(1)} + 2 {\cal C}_{\theta_1x}^{(1)}\right)}{8192 \sqrt{2} ({f_{r_0}}-1)^3 {g_s}^2 \left(\log N\right) ^3 M
   N_f ^2}\nonumber\\
& &\hskip -0.3in +\frac{81\ 3^{3/8} \sqrt[4]{7} \pi ^3 {\cal C}_{\phi_0}^{\rm IR}\ ^2 \epsilon ^{11/4} {f_{r_0}} ({f_{r_0}}+1)^2 \alpha _{\theta _1}^3
   {\cal C}_{\psi_1}^{\rm IR}\  N^{2 {f_{r_0}}+\frac{7}{5}}}{64 \sqrt{2} ({f_{r_0}}-1)^3 {g_s}^2 \left(\log N\right) ^3 M N_f ^2}\nonumber\\
& & \nonumber\\
& & \hskip -0.3in \bullet\  z_7 = \int dZ {\cal V}_2\psi_1(1 + \psi_1)^2(Z)  =
\frac{3 \sqrt[8]{3} 7^{3/4} \sqrt[4]{\epsilon } g_s ^2 M N^{4/5} {N_f}^2 {\cal C}_{\psi_1}^{\rm IR}\ ^3 \log (r_0 )
   (\log N -3 \log (r_0 ))}{4 \sqrt{2} \pi  \alpha _{\theta _1} \alpha _{\theta _2}^2 \log N }\nonumber\\
& &\hskip -0.3in -\frac{21 \sqrt[8]{3} 7^{3/4} \beta  \sqrt[4]{\epsilon } g_s ^2 M N^{4/5} {N_f}^2 {\cal C}_{\psi_1}^{\rm IR}\ ^3 \log
   (r_0 ) ( {\cal C}_{zz}^{(1)} - 2 {\cal C}_{\theta_1z}^{(1)} + 2 {\cal C}_{\theta_1x})}{256 \sqrt{2} \pi  \alpha _{\theta _1} \alpha _{\theta
   _2}^2} +\frac{48\ 3^{3/4} \sqrt{\frac{1}{\epsilon }} g_s ^2 M N^{4/5} {N_f}^2 r_0  \log N  \log ^2(r_0 )
   {\cal C}_{\psi_1}^{\rm UV}\ }{\pi  \log N  \alpha _{\theta _1} \alpha _{\theta _2}^2}\nonumber\\
& & \nonumber\\
& & \hskip -0.3in \bullet\  z_8 = \int dZ {\cal V}_2\psi_0^2\psi_1(Z) =  \frac{567\ 3^{3/8} \sqrt[4]{7} \pi ^3 \beta  {\cal C}_{\phi_0}^{\rm IR}\ ^2 \epsilon ^{11/4} {f_{r_0}} ({f_{r_0}}+1) \alpha _{\theta
   _1}^3 {\cal C}_{\psi_1}^{\rm IR}\  N^{2 {f_{r_0}}+\frac{7}{5}} \left({\cal C}_{zz}^{(1)} - 2 {\cal C}_{\theta_1z}^{(1)} + 2 {\cal C}_{\theta_1x}^{(1)}\right)}{8192
   \sqrt{2} ({f_{r_0}}-1)^3 {g_s}^2 \left(\log N\right) ^3 M N_f ^2}\nonumber\\
& & \hskip -0.3in +\frac{81\ 3^{3/8} \sqrt[4]{7} \pi ^3
   {\cal C}_{\phi_0}^{\rm IR}\ ^2 \epsilon ^{11/4} {f_{r_0}} ({f_{r_0}}+1)^2 \alpha _{\theta _1}^3 {\cal C}_{\psi_1}^{\rm IR}\  N^{2
   {f_{r_0}}+\frac{7}{5}}}{64 \sqrt{2} ({f_{r_0}}-1)^3 {g_s}^2 \left(\log N\right) ^3 M N_f ^2}.
\end{eqnarray}
}

\section{$G_{\rm global} \times H_{\rm local}$ HLS Formalism and Obtaining GL's Mesonic $\chi$PT Lagrangian up to  ${\cal O}(p^4)$ After Integrating Out the Vector Mesons from the HLS Lagrangian \cite{HLS-Physics-Reports}}
\setcounter{equation}{0} \seceqcc

In this appendix, we summarize the HLS formalism and the arguments of how to obtain the $SU(3)$ $\chi$PT Lagrangian of \cite{GL} up to (NLO in momentum below the chiral symmetry breaking scale) ${\cal O}(p^4)$ by integrating out the vector mesons from the HLS Lagrangian (both as discussed in detail in  \cite{HLS-Physics-Reports})

\subsection{HLS Formalism}

The HLS formalism describes a model based on $G_{\rm global} \times H_{\rm local}$ symmetry, where
$G = \mbox{SU($N_f$)}_{\rm L} \times 
\mbox{SU($N_f$)}_{\rm R}$  is the 
global chiral symmetry and 
$H = \mbox{SU($N_f$)}_{\rm V}$ is the H(idden) L(ocal) S(ymmetry). The building blocks of 
this model are  SU($N_f$)-matrix valued variables $\xi_{\rm L}$ and 
$\xi_{\rm R}$ which are introduced by splitting $U$ in the ChPT as
\begin{equation*}
\label{split-U-xiLR}
U = \xi_{\rm L}^\dagger \xi_{\rm R} \ .
\end{equation*}
Now, under $G_{\rm global} \times H_{\rm local}$ $\xi_{\rm L,R}(x)$ transform as follows:
\begin{equation*}
\label{xiLR-trans}
\xi_{\rm L,R}(x) \rightarrow \xi_{\rm L,R}^{\prime}(x) =
h(x) \cdot \xi_{\rm L,R}(x) \cdot g^{\dag}_{\rm L,R}
\ ,
\end{equation*}
where $h(x) \in H_{\rm local},\ 
g_{\rm L,R} \in G_{\rm global}$.  These variables are parameterized as
\begin{eqnarray*}
\label{def:xiLR}
&& \xi_{\rm L,R} = e^{i\sigma/F_\sigma} e^{\mp i\pi/F_\pi} \ ,
\quad
\left[ \, \pi = \pi^a T_a \,,\, \sigma = \sigma^a T_a \right] \ ,
\end{eqnarray*}
where the matrix-valued $\pi$ denotes the Nambu-Goldstone (NG) bosons associated with 
the spontaneous breaking of $G$ chiral symmetry and $\sigma$ denotes the NG bosons absorbed into the gauge bosons. Further, $F_\pi$ and $F_\sigma$ denote the relevant decay constants, and
the parameter $a$ is defined via: $a \equiv \frac{F_\sigma^2}{F_\pi^2}$.

From $\xi_{\rm L}$ and $\xi_{\rm R}$ the following are constructed:
\begin{eqnarray*}
\alpha_{\perp\mu} &=&
\left( 
  \partial_\mu \xi_{\rm R} \cdot \xi_{\rm R}^\dag -
  \partial_\mu \xi_{\rm L} \cdot \xi_{\rm L}^\dag 
\right)
/ (2i) \ ,
\label{def-alpha perp}
\\
\alpha_{\parallel\mu} &=&
\left( 
  \partial_\mu \xi_{\rm R} \cdot \xi_{\rm R}^\dag +
  \partial_\mu \xi_{\rm L} \cdot \xi_{\rm L}^\dag
\right)
/ (2i) \ ,
\label{def-alpha parallel}
\end{eqnarray*}
which transform under $H_{\rm local}$ as
\begin{eqnarray*}
&& \alpha_{\perp\mu} \rightarrow 
  h(x) \cdot \alpha_{\perp\mu} \cdot h^\dag (x) \ ,
\\
&& \alpha_{\parallel\mu} \rightarrow 
  h(x) \cdot \alpha_{\perp\mu} \cdot h^\dag (x)
  -i  \partial_\mu h(x) \cdot h^\dag(x) \ .
\end{eqnarray*}

The covariant derivatives of $\xi_{\rm L}$ and $\xi_{\rm R}$ can be obtained from the transformation properties of $\xi_{\rm L,R}(x)$ as:
\begin{equation*}
D_\mu \xi_{\rm L,R} = \partial_\mu \xi_{\rm L,R} - i V_\mu 
\xi_{\rm L,R} \ ,
\end{equation*}
where $V_\mu = V_\mu^a T_a$ are the gauge fields corresponding to $H_{\rm local}$.  
Now, $V_\mu$ transforms under $H_{\rm local}$ as:
\begin{equation*}
V_\mu \rightarrow
 h(x) \cdot V_\mu \cdot h^\dag (x)
  - i \partial_\mu h(x) \cdot h^\dag(x) 
\ .
\end{equation*}

The ``covariantized 1-forms" defined as
\begin{eqnarray*}
\label{def:al hat perp 0}
& & \widehat{\alpha}_{\perp\mu} =
\frac{1}{2i}
\left( 
  D_\mu \xi_{\rm R} \cdot \xi_{\rm R}^\dag -
  D_\mu \xi_{\rm L} \cdot \xi_{\rm L}^\dag 
\right) = \alpha_{\perp\mu}\ ,
\nonumber\\
& & \widehat{\alpha}_{\parallel\mu} =
\frac{1}{2i}
\left( 
  D_\mu \xi_{\rm R} \cdot \xi_{\rm R}^\dag +
  D_\mu \xi_{\rm L} \cdot \xi_{\rm L}^\dag
\right)= \alpha_{\parallel\mu} 
- V_\mu \  ,
\end{eqnarray*}
transform homogeneously under $H_{\rm local}$:
\begin{equation*}
\widehat{\alpha}_{\perp,\parallel}^\mu \rightarrow 
  h(x) \cdot \widehat{\alpha}_{\perp,\parallel}^\mu \cdot h^\dag (x) \ .
\end{equation*}

Thus, one can construct the following two invariants:
\begin{eqnarray*}
&& {\cal L}_{\rm A} \equiv
F_\pi^2 \, \mbox{tr} 
\left[ \widehat{\alpha}_{\perp\mu} \widehat{\alpha}_{\perp}^\mu
\right]
\ ,
\label{def:LA 0}
\\
&& a {\cal L}_{\rm V} \equiv
F_\sigma^2 \, \mbox{tr}
\left[ 
  \widehat{\alpha}_{\parallel\mu} \widehat{\alpha}_{\parallel}^\mu
\right]
=
F_\sigma^2 \, \mbox{tr}
\left[
  \left(
    V_\mu - \alpha_{\parallel\mu}
  \right)^2
\right]
\ .
\label{def:LV 0}
\end{eqnarray*}
Therefore, the most general Lagrangian
made out of $\xi_{\rm L,R}$ and $D_\mu\xi_{\rm L,R}$ with the lowest number of
derivatives, i.e., at ${\cal O}(p^2)$ (see \cite{GLF, GL, HLS-Physics-Reports} for power counting arguments), is  given by:
\begin{equation*}
\label{L-Op2}
{\cal L} = {\cal L}_{\rm A} + a {\cal L}_{\rm V} \ .
\end{equation*}

Using the EOM for $V_\mu$ at ${\cal O}(p^2)$, one obtains:
\begin{equation*}
\label{Vmu-EOM-Op2}
V_\mu = \alpha_{\parallel\mu} \ .
\end{equation*}
Further, with the relation
\begin{equation*}
\widehat{\alpha}_{\perp\mu} 
= \frac{1}{2i}\,\xi_{\rm L} \cdot \partial_\mu U \cdot 
  \xi_{\rm R}^\dag 
= \frac{i}{2} \,\xi_{\rm R} \cdot \partial_\mu 
  U^\dag \cdot \xi_{\rm L}^\dag
\label{perp U0}
\end{equation*}
substituted into (\ref{L-Op2}), one obtains the following ${\cal O}(p^2)$ term in the $\chi$PT Lagrangian:
\begin{equation*}
{\cal L} = {\cal L}_{\rm A} =
\frac{F_\pi^2}{4} \mbox{tr}
\left[ \partial_\mu U^\dag \partial^\mu U \right]
\ .
\end{equation*}
In the unitary gauge, $\sigma=0$\footnote{This unitary gauge is not preserved though under the $G_{\rm global}$
transformation, which in general has the following form
\begin{eqnarray*}
\xi \rightarrow \xi^{\prime} 
&=& \xi \cdot g_{\rm R}^\dag = g_{\rm L} \cdot \xi 
\nonumber\\
&=&
\exp\left[i\sigma^\prime( \pi, g_{\rm R}, g_{\rm L} )/F_\sigma \right]
\exp\left[i\pi^\prime/F_\pi \right]
\nonumber\\
&=&
\exp\left[i\pi^\prime/F_\pi\right]
\exp\left[- i\sigma^\prime(\pi, g_{\rm R}, g_{\rm L})/F_\sigma\right]
\ .
\end{eqnarray*}
However, $\exp\left[i\sigma^\prime( \pi, g_{\rm R}, g_{\rm L} )/F_\sigma
\right]$
can be eliminated if we simultaneously perform the $H_{\rm local}$
gauge transformation:
\begin{eqnarray*}
h = 
\exp\left[i\sigma^\prime( \pi, g_{\rm R}, g_{\rm L} )/F_\sigma
\right]
\equiv
h \left( \pi, g_{\rm R}, g_{\rm L}\right) \ .
\end{eqnarray*}

Therefore there is a global symmetry $G =
\mbox{SU($N_f$)}_{\rm L} \times 
\mbox{SU($N_f$)}_{\rm R}$ under the following combined transformation:
\begin{equation*}
G \ : \ \xi \rightarrow h \left( \pi, g_{\rm R}, g_{\rm L}\right)
\cdot \xi \cdot g_{\rm R}^\dag
= g_{\rm L} \cdot \xi \cdot 
h^\dag \left( \pi, g_{\rm R}, g_{\rm L}\right)
\ .
\label{com trans}
\end{equation*}}, two 
$SU(N_f)$-matrix valued variables $\xi_{\rm L}$ and 
$\xi_{\rm R}$ are related via:
\begin{equation*}
\xi_{\rm L}^\dag = \xi_{\rm R} \equiv \xi = e^{i\pi/F_\pi}
\ .
\end{equation*}


\subsection{Obtaining the ${\cal O}(p^2)$ $SU(3)$ $\chi$PT Lagrangian by Integrating out the Vector Mesons from the HLS Lagrangian}

First, we introduce the external gauge fields ${\cal L}_\mu$
and ${\cal R}_\mu$ which include $W$ boson, $Z$-boson and photon
fields.
This is done by gauging the $G_{\rm global}$ symmetry.
The transformation properties of ${\cal L}_\mu$
and ${\cal R}_\mu$, ${\cal L}_\mu\rightarrow g_L {\cal L}_\mu g_L^\dagger - i\partial_\mu g_L . g_L^\dagger$ and ${\cal R}_\mu\rightarrow g_R {\cal R}_\mu g_R^\dagger - i\partial_\mu g_R . g_R^\dagger$ are used to define covariant derivatives of $\xi_{\rm L,R}$:
\begin{eqnarray}
D_\mu \xi_{\rm L} &=&
\partial_\mu \xi_{\rm L} - i V_\mu \xi_{\rm L}
+ i \xi_{\rm L} {\cal L}_\mu \ ,
\nonumber\\
D_\mu \xi_{\rm R} &=&
\partial_\mu \xi_{\rm R} - i V_\mu \xi_{\rm R}
+ i \xi_{\rm R} {\cal R}_\mu \ .
\label{covder}
\end{eqnarray}
It should be noticed that in the HLS these external gauge fields are
included without assuming the vector dominance.
It is outstanding feature of the HLS model that $\xi_{\rm L,R}$ have
two independent source charges and hence two independent gauge bosons
are automatically introduced in the HLS model.
Both the vector meson fields and external gauge fields are
simultaneously incorporated into the Lagrangian fully consistent with
the chiral symmetry.
By using the above covariant derivatives two 
Maurer-Cartan 1-forms are constructed as
\begin{eqnarray}
\widehat{\alpha}_{\perp\mu} &=&
\left( 
  D_\mu \xi_{\rm R} \cdot \xi_{\rm R}^\dag -
  D_\mu \xi_{\rm L} \cdot \xi_{\rm L}^\dag 
\right)
/ (2i) \ ,
\nonumber\\
\widehat{\alpha}_{\parallel\mu} &=&
\left( 
  D_\mu \xi_{\rm R} \cdot \xi_{\rm R}^\dag +
  D_\mu \xi_{\rm L} \cdot \xi_{\rm L}^\dag 
\right)
/ (2i) \ .
\label{def:al hat}
\end{eqnarray}
These 1-forms upon expansion in a power series in $\pi$ yield:
\begin{eqnarray}
\label{al perp exp}
\widehat{\alpha}_{\perp\mu} &=&
\frac{1}{F_\pi} \partial_\mu \pi 
+ {\cal A}_\mu 
- \frac{i}{F_\pi} \left[ {\cal V}_\mu \,,\, \pi \right]
- \frac{1}{6F_\pi^3} 
  \Bigl[ \bigl[ \partial_\mu \pi \,,\, \pi \bigr] \,,\, \pi \Bigr]
+ \cdots
\ ,
\\
\label{al para exp}
\widehat{\alpha}_{\parallel\mu} &=&
\frac{1}{F_\sigma} \partial_\mu \sigma - V_\mu
+ {\cal V}_\mu 
- \frac{i}{2F_\pi^2} \bigl[ \partial_\mu \pi \,,\, \pi \bigr]
- \frac{i}{F_\pi} \left[ {\cal A}_\mu \,,\, \pi \right]
+ \cdots
\ ,
\end{eqnarray}
where
${\cal V}_\mu = \left({\cal R}_\mu + {\cal L}_\mu\right)/2$ and
${\cal A}_\mu = \left({\cal R}_\mu - {\cal L}_\mu\right)/2$.

The covariantized 1-forms in Eqs.~(\ref{def:al hat})
transform homogeneously:
\begin{equation}
\widehat{\alpha}_{\parallel,\perp}^\mu \rightarrow
h(x) \cdot \widehat{\alpha}_{\parallel,\perp}^\mu \cdot h^\dag(x) \ .
\label{hat al:trans}
\end{equation}
Then we can construct two independent terms with lowest derivatives
which are invariant under the full 
$G_{\rm global}\times H_{\rm local}$ symmetry as
\begin{eqnarray}
&& {\cal L}_{\rm A} \equiv
F_\pi^2 \, \mbox{tr} 
\left[ \hat{\alpha}_{\perp\mu} \hat{\alpha}_{\perp}^\mu \right]
=
\mbox{tr} \left[ \partial_\mu \pi \partial^\mu \pi \right]
+ \cdot 
\ ,
\label{def:LA}
\\
&& a {\cal L}_{\rm V} \equiv
F_\sigma^2 \, \mbox{tr}
\left[ 
  \hat{\alpha}_{\parallel\mu} \hat{\alpha}_{\parallel}^\mu
\right]
=
\mbox{tr} \biggl[
\left( \partial_\mu \sigma - F_\sigma V_\mu \right)
\left( \partial^\mu \sigma - F_\sigma V^\mu \right)
\biggr]
+ \cdots
\ ,
\label{def:LV}
\end{eqnarray}
where the expansions of the covariantized 1-forms in 
(\ref{al perp exp}) and (\ref{al para exp}) were substituted to
obtain the second expressions.
These expansions imply that ${\cal L}_A$ generates the kinetic term
of pseudoscalar meson, while
${\cal L}_V$ generates the kinetic term of the would-be NG boson
$\sigma$ in addition to the mass term of the vector meson.

The HLS gauge boson field strength defined by $
V_{\mu\nu} \equiv \partial_\mu V_\nu - \partial_\nu
V_\mu - i [ V_\mu , V_\nu ] $,
which also transforms homogeneously:
\begin{equation}
V_{\mu\nu} \rightarrow h(x) \cdot V_{\mu\nu} \cdot h^\dag(x) \ ,
\end{equation}
is another building block of the $\chi$PT Lagrangian. Then the simplest term with $V_{\mu\nu}$ is the kinetic term of the
gauge boson:
\begin{equation}
{\cal L}_{\rm kin}(V_\mu) = - \frac{1}{2g^2} \, \mbox{tr} 
\left[ V_{\mu\nu} V^{\mu\nu} \right] \ ,
\label{HLS gauge kinetic}
\end{equation}
where
$g$ is the HLS gauge coupling constant.

Now the Lagrangian with lowest derivatives
is given by:
\begin{eqnarray}
{\cal L} &=& 
{\cal L}_{\rm A} + a {\cal L}_{\rm V} + 
{\cal L}_{\rm kin}(V_\mu)
\nonumber\\
&=&
F_\pi^2 \, \mbox{tr} 
\left[ \hat{\alpha}_{\perp\mu} \hat{\alpha}_{\perp}^\mu \right]
+ F_\sigma^2 \, \mbox{tr}
\left[ 
  \hat{\alpha}_{\parallel\mu} \hat{\alpha}_{\parallel}^\mu
\right]
- \frac{1}{2g^2} \, \mbox{tr} 
\left[ V_{\mu\nu} V^{\mu\nu} \right] 
\ .
\label{leading Lagrangian 0}
\end{eqnarray}

\subsection{Obtaining the ${\cal O}(p^4)$ $SU(3)$ $\chi$PT Lagrangian by Integrating out the Vector Mesons from the HLS Lagrangian}

Integrating out the vector mesons in the Lagrangian of the
HLS given in (\ref{leading Lagrangian 0})
we obtain the Lagrangian for pseudoscalar mesons.
The resultant Lagrangian includes ${\cal O}(p^4)$ terms of the ChPT
in addition to 
${\cal O}(p^2)$ terms.
To perform this it is convenient to introduce the following quantities:
\begin{eqnarray}
\alpha_{\perp\mu} &=&
\left( 
  {\cal D}_\mu \xi_{\rm R} \cdot \xi_{\rm R}^\dag -
  {\cal D}_\mu \xi_{\rm L} \cdot \xi_{\rm L}^\dag 
\right)
/ (2i) \ ,
\nonumber\\
\alpha_{\parallel\mu} &=&
\left( 
  {\cal D}_\mu \xi_{\rm R} \cdot \xi_{\rm R}^\dag +
  {\cal D}_\mu \xi_{\rm L} \cdot \xi_{\rm L}^\dag 
\right)
/ (2i) \ ,
\end{eqnarray}
where ${\cal D}_\mu\xi_{\rm L}$ and ${\cal D}_\mu\xi_{\rm L}$ are
defined by
\begin{eqnarray}
&& {\cal D}_\mu\xi_{\rm L} =
  \partial_\mu \xi_{\rm L} + i \xi_{\rm L} {\cal L}_\mu \ ,
\nonumber\\
&& {\cal D}_\mu\xi_{\rm R} =
  \partial_\mu \xi_{\rm R} + i \xi_{\rm R} {\cal R}_\mu 
\ .
\label{covder:2}
\end{eqnarray}
The relations of these $\alpha_{\perp\mu}$ and $\alpha_{\parallel\mu}$
with $\widehat{\alpha}_{\perp\mu}$ and
$\widehat{\alpha}_{\parallel\mu}$ in (\ref{def:al hat})
are given by
\begin{eqnarray}
&& \widehat{\alpha}_{\perp\mu} = \alpha_{\perp\mu} \ , \nonumber\\
&& \widehat{\alpha}_{\parallel\mu} = \alpha_{\parallel\mu} - V_\mu \ .
\label{rel: al perp para}
\end{eqnarray}

From the Lagrangian in (\ref{leading Lagrangian 0})
the equation of motion for the vector meson is given by
\begin{equation}
F_\sigma^2 \left( V_\mu - \alpha_{\parallel\mu} \right)
- \frac{1}{g^2} \left( \partial^\nu V_{\mu\nu}
- i \left[ V^\nu \,,\, V_{\mu\nu} \right] \right) = 0 \ .
\label{para0}
\end{equation}
In the leading order of the derivative expansion the solution of
(\ref{para0}) is given by
\begin{equation}
\label{EOM V}
V_\mu = \alpha_{\parallel\mu} 
+ \frac{1}{M_\rho^2} {\cal O}(p^3) \ ,
\end{equation}
consistent with (\ref{Vmu-EOM-Op2}). Substituting this into the field strength of the HLS gauge boson and
performing the derivative expansion one can show that one obtains \cite{HLS-Physics-Reports}:
\begin{eqnarray}
V_{\mu\nu} &=& \xi_{\rm R} \left( {\cal R}_{\mu\nu} + U^\dag {\cal L}_{\mu\nu} U
+ \frac{i}{4} \nabla_\mu U^\dag \cdot \nabla_\nu U
- \frac{i}{4} \nabla_\nu U^\dag \cdot \nabla_\mu U
\right) \xi_{\rm R}^\dag
+ \frac{1}{M_\rho^2} {\cal O}(p^4) 
\ ,
\label{V0}
\end{eqnarray}
where,
\begin{equation}
\label{perpU}
\hat{\alpha}_{\perp\mu} 
= \frac{i}{2} \xi_{\rm L} \cdot \nabla_\mu U \cdot \xi_{\rm R}^\dag
= \frac{1}{2i} \xi_{\rm R} \cdot \nabla_\mu 
  U^\dag \cdot \xi_{\rm L}^\dag
\ ,
\end{equation}
has been used. By substituting (\ref{perpU}) 
into the HLS Lagrangian,
the first term in the HLS Lagrangian 
(\ref{leading Lagrangian 0}) becomes
the first term in
the leading order ChPT Lagrangian:
\begin{equation}
\left. {\cal L}_{(2)}^{\rm ChPT} \right\vert_{\chi=0}
=
\frac{F_\pi^2}{4} \mbox{tr}
\left[ \nabla_\mu U^\dag \nabla^\mu U \right]
\ .
\label{leading ChPT:2}
\end{equation}
In addition, 
the second term in (\ref{leading Lagrangian 0})
with (\ref{para0}) substituted
becomes of ${\cal O}(p^6)$ in the ChPT
and the third term (the kinetic term of the HLS gauge boson) 
with (\ref{V0}) becomes
of ${\cal O}(p^4)$ in the ChPT:
\begin{eqnarray}
{\cal L}_4^V &=&
\frac{1}{32g^2} 
\left( 
\mbox{tr}\left[ \nabla_\mu U \nabla^\mu U^\dag \right] \right)^2
+ \frac{1}{16g^2} 
\,\mbox{tr}\left[ \nabla_\mu U \nabla_\nu U^\dag \right]
\mbox{tr} \left[ \nabla^\mu U \nabla^\nu U^\dag \right]
\nonumber\\
&&
{} -  \frac{3}{16g^2} \, \mbox{tr} \left[
  \nabla_\mu U \nabla^\mu U^\dag \nabla_\nu U \nabla^\nu U^\dag
\right]
-  \frac{i}{4g^2} \, \mbox{tr}\left[
  {\cal L}_{\mu\nu} \nabla^\mu U \nabla^\nu U^\dag
  + {\cal R}_{\mu\nu} \nabla^\mu U^\dag \nabla^\nu U
\right]
\nonumber\\
&&
- \frac{1}{4g^2} \,\mbox{tr}
\left[ {\cal L}_{\mu\nu} U {\cal R}^{\mu\nu} U^\dag \right] - \frac{1}{8g^2} \left[ {\cal L}_{\mu\nu} {\cal L}^{\mu\nu}
 + {\cal R}_{\mu\nu} {\cal R}^{\mu\nu} \right]
\ ,
\end{eqnarray}
where we fixed $N_f=3$.
Comparing this with the ${\cal O}(p^4)$ terms of the ChPT Lagrangian
given in (\ref{ChPT-Op4}),
we obtain the contributions of vector mesons to the low-energy
parameters of the ChPT:
\begin{equation}
\label{GL-LEC-integrate-rho-out}
\begin{array}{ccc}
\displaystyle L_1^V = \frac{1}{32g^2} \ , \quad
& \displaystyle L_2^V = \frac{1}{16g^2} \ , \quad
& \displaystyle L_3^V = - \frac{3}{16g^2} \ ,
\\
& & \\
\displaystyle L_9^V = \frac{1}{4g^2} \ , \quad
& \displaystyle L_{10}^V = - \frac{1}{4g^2} \ . \quad
& 
\end{array}
\end{equation}
Using

\section{Inclusion of ${\cal O}(R^4)$ Corrections in the $D6$-Brane DBI Action}
\setcounter{equation}{0} \seceqdd

Inclusive of the ${\cal O}(R^4)$-corrections indicated by a $\tilde{}$ (e.g., the  ${\cal M}$-theory  metric: $\tilde{G}_{MN}^{\cal M}=G^{\rm MQGP}_{MN}\left(1+f_{MN}\right)$ \cite{OR4-Yadav+Misra}) in (\ref{V_12}), one sees that
\begin{eqnarray}
\label{V_12}
& & {\cal V}_1 = 2\sqrt{h} e^{-\tilde{\phi}_{\rm IIA}}\sqrt{-\tilde{g}_{7\times7}^{\rm IIA}}\tilde{g}^{ZZ}_{\rm IIA},\nonumber\\
& & {\cal V}_2 = h e^{-\tilde{\phi}_{\rm IIA}}\sqrt{-\tilde{g}^{\rm IIA}_{7\times7}},
\end{eqnarray}
wherein using (\ref{TypeIIA-from-M-theory-Witten-prescription}):
\begin{eqnarray}
\label{O4-corrections}
& & e^{-\tilde{\phi}_{\rm IIA}}\ ^{{\cal O}(R^4)} = \tilde{G}^{\cal M}_{x^{10}x^{10}}\ ^{-\frac{3}{4}}
= \left(G^{\cal M}_{x^{10}x^{10}} + {\cal F}_{x^{10}x^{10}}\right)^{-\frac{3}{4}} \equiv e^{-{\phi}_{\rm IIA}} + {\cal E}_{x^{10}x^{10}}{\cal F}_{x^{10}x^{10}},\nonumber\\
& & {\cal E}_{x^{10}x^{10}} = -\frac{3}{4}\frac{1}{\left(G^{\cal M}_{x^{10}x^{10}}\right)^{\frac{7}{4}}}\nonumber\\
& & \tilde{G}_{\cal M}^{ZZ} = \frac{e^{-2 Z}}{\sqrt{\tilde{G}^{\cal M}_{x^{10}x^{10}}} G_{\cal M}^{ZZ} {r_0}^2}-\frac{ e^{-2 Z} ({\cal F}_{x^{10}x^{10}}
  G_{\cal M}^{ZZ}+2 {\cal F}_{ZZ} \tilde{G}^{\cal M}_{x^{10}x^{10}})}{2 \tilde{G}^{\cal M}_{x^{10}x^{10}}\ ^{3/2} G_{\cal M}^{rr}\ ^2r_0^2},\nonumber\\
& & \sqrt{-\tilde{g}_{7\times7}} = \sqrt{-g_{7\times7}} + {\rm Tr}({\cal C}{\cal F}).
\end{eqnarray}
In (\ref{O4-corrections}),
{\footnotesize
\begin{eqnarray*}
& & \hskip -0.8in {\cal F}_{MN} \equiv f_{MN} G^{\cal M}_{MN};\nonumber\\
& & \hskip -0.8in {\cal C}_{MN} \equiv \delta^t_M\delta^t_N {\cal C}_{tt} + \delta^{x^1}_M\delta^{x^1}_N {\cal C}_{x^1x^1}
+ \delta^{x^2}_M\delta^{x^2}_N {\cal C}_{x^2x^2} + \delta^{x^3}_M\delta^{x^3}_N {\cal C}_{x^3x^3} \nonumber\\
& &  \hskip -0.8in + \delta^{r_0}_M\delta^{r_0}_N {\cal C}_{rr} + \delta^{\theta_2}_M\delta^{\theta_2}_N {\cal C}_{\theta_2\theta_2}
+ \delta^{\theta_2}_M\delta^{\phi_2}_N {\cal C}_{\theta_2y} + \delta^{\phi_2}_M\delta^{\phi_2}_N {\cal C}_{yy} + \delta^{x^{11}}_M\delta^{x^{11}}_N {\cal C}_{x^{10}x^{10}},\nonumber\\
& & \hskip -0.8in {\cal C}_{tt} =\nonumber\\
& & \hskip -0.8in  -\frac{\sqrt{-g} \left(-4 {\cal A}^{\cal M}_{\theta_2\theta_2} {G^{\cal M}_{x^1x^1}} {G^{\cal M}_{x^{10}x^{10}}}^3 {G^{\cal M}_{x^2x^2}} {G^{\cal M}_{x^3x^3}} {G^{\cal M}_{rr}}
   {G^{\cal M}_{yy}}-4 B_{r\theta_2}^2 {G^{\cal M}_{x^1x^1}} {G^{\cal M}_{x^{10}x^{10}}} {G^{\cal M}_{x^2x^2}} {G^{\cal M}_{x^3x^3}} {G^{\cal M}_{rr}}+4 {G^{\cal M}_{x^1x^1}}
   {G^{\cal M}_{x^{10}x^{10}}}^2 {G^{\cal M}_{x^2x^2}} {G^{\cal M}_{x^3x^3}} {G^{\cal M}_{rr}} {G^{\cal M}_{\theta_2y}}^2\right)}{\Delta}\nonumber\\
& & \hskip -0.8in = \frac{2 \sqrt[6]{\pi } {g_s}^{3/2} M N^{3/10} {N_f} \log \left({r_0} e^Z\right) \left(72 a^2 {r_0} e^Z \log
   \left({r_0} e^Z\right)-3 a^2+2 {r_0}^2 e^{2 Z}\right)}{3^{2/3} \log N  \left(1-e^{-4 Z}\right) \alpha _{\theta _1}
   \alpha _{\theta _2}^2 {\cal G}{}^{2/3}}\nonumber\\
& & \hskip -0.8in {\cal C}_{x^1x^1} = \nonumber\\
& & \hskip -0.8in -\frac{\sqrt{-g} \left(-4 {\cal A}^{\cal M}_{\theta_2\theta_2} {G^{\cal M}_{x^{10}x^{10}}}^3 {G^{\cal M}_{x^2x^2}} {G^{\cal M}_{x^3x^3}} {G^{\cal M}_{rr}} {G^{\cal M}_{tt}}
   {G^{\cal M}_{yy}}-4 B_{r\theta_2}^2 {G^{\cal M}_{x^{10}x^{10}}} {G^{\cal M}_{x^2x^2}} {G^{\cal M}_{x^3x^3}} {G^{\cal M}_{rr}} {G^{\cal M}_{tt}}+4 {G^{\cal M}_{x^{10}x^{10}}}^2
   {G^{\cal M}_{x^2x^2}} {G^{\cal M}_{x^3x^3}} {G^{\cal M}_{rr}} {G^{\cal M}_{\theta_2y}}^2 {G^{\cal M}_{tt}}\right)}{\Delta}\nonumber\\
& & \hskip -0.8in= \frac{2 \sqrt[6]{\pi } {g_s}^{3/2} M N^{3/10} {N_f} \log \left({r_0} e^Z\right) \left(72 a^2 {r_0} e^Z \log
   \left({r_0} e^Z\right)-3 a^2+2 {r_0}^2 e^{2 Z}\right)}{3^{2/3} \log N  \alpha _{\theta _1} \alpha _{\theta _2}^2
  {\cal G}{}^{2/3}},\nonumber\\
& & \hskip -0.8in {\cal C}_{x^2x^2} = {\cal C}_{x^3x^3} =\nonumber\\
& & \hskip -0.8in -\frac{\sqrt{-g} \left(-4 {\cal A}^{\cal M}_{\theta_2\theta_2} {G^{\cal M}_{x^1x^1}} {G^{\cal M}_{x^{10}x^{10}}}^3 {G^{\cal M}_{x^3x^3}} {G^{\cal M}_{rr}} {G^{\cal M}_{tt}}
   {G^{\cal M}_{yy}}-4 B_{r\theta_2}^2 {G^{\cal M}_{x^1x^1}} {G^{\cal M}_{x^{10}x^{10}}} {G^{\cal M}_{x^3x^3}} {G^{\cal M}_{rr}} {G^{\cal M}_{tt}}+4 {G^{\cal M}_{x^1x^1}}
   {G^{\cal M}_{x^{10}x^{10}}}^2 {G^{\cal M}_{x^3x^3}} {G^{\cal M}_{rr}} {G^{\cal M}_{\theta_2y}}^2 {G^{\cal M}_{tt}}\right)}{\Delta}\nonumber\\
& & \hskip -0.8in = \frac{2 \sqrt[6]{\pi } {g_s}^{3/2} M N^{3/10} {N_f} \log \left({r_0} e^Z\right) \left(72 a^2 {r_0} e^Z \log
   \left({r_0} e^Z\right)-3 a^2+2 {r_0}^2 e^{2 Z}\right)}{3^{2/3} \log N  \alpha _{\theta _1} \alpha _{\theta _2}^2
  {\cal G}{}^{2/3}},\nonumber\\
\end{eqnarray*}
\begin{eqnarray*}
& & \hskip -0.8in {\cal C}_{rr} =\nonumber\\
& & \hskip -0.8in  -\frac{\sqrt{-g} \left(-4 {\cal A}^{\cal M}_{\theta_2\theta_2} {G^{\cal M}_{x^1x^1}} {G^{\cal M}_{x^{10}x^{10}}}^3 {G^{\cal M}_{x^2x^2}} {G^{\cal M}_{x^3x^3}} {G^{\cal M}_{tt}}
   {G^{\cal M}_{yy}}-4 B_{r\theta_2}^2 {G^{\cal M}_{x^1x^1}} {G^{\cal M}_{x^{10}x^{10}}} {G^{\cal M}_{x^2x^2}} {G^{\cal M}_{x^3x^3}} {G^{\cal M}_{tt}}+4 {G^{\cal M}_{x^1x^1}}
   {G^{\cal M}_{x^{10}x^{10}}}^2 {G^{\cal M}_{x^2x^2}} {G^{\cal M}_{x^3x^3}} {G^{\cal M}_{\theta_2y}}^2 {G^{\cal M}_{tt}}\right)}{\Delta}\nonumber\\
& & \hskip -0.8in= \frac{\sqrt{{g_s}} M \left(\frac{1}{N}\right)^{7/10} {N_f} {r_0}^4 \left(e^{4 Z}-1\right) \left(9 a^2+{r_0}^2 e^{2
   Z}\right) \log \left({r_0} e^Z\right) \left(72 a^2 {r_0} e^Z \log \left({r_0} e^Z\right)-3 a^2+2 {r_0}^2 e^{2
   Z}\right)}{2\ 3^{2/3} \pi ^{5/6} \log N  \alpha _{\theta _1} \alpha _{\theta _2}^2 \left(6 a^2+{r_0}^2 e^{2 Z}\right)
  {\cal G}{}^{2/3}},\nonumber\\
& &\hskip -0.8in {\cal C}_{\theta_2\theta_2} = \frac{{G^{\cal M}_{x^{10}x^{10}}} {G^{\cal M}_{yy}} \sqrt{-g}}{2 \left({\cal A}^{\cal M}_{\theta_2\theta_2} {G^{\cal M}_{x^{10}x^{10}}}^2
   {G^{\cal M}_{yy}}+B_{r\theta_2}^2-{G^{\cal M}_{x^{10}x^{10}}} {G^{\cal M}_{\theta_2y}}^2\right)} = \frac{\sqrt[6]{\pi } \left(\frac{1}{N}\right)^{13/10} {r_0}^4 e^{4 Z} \alpha _{\theta _1} \alpha _{\theta _2}^2 \left(72 a^2
   {r_0} e^Z \log \left({r_0} e^Z\right)-3 a^2+2 {r_0}^2 e^{2 Z}\right){\cal G}{}^{4/3}}{32\ 3^{2/3} {g_s}^{5/2}
   \log N  M {N_f}^3 \log \left({r_0} e^Z\right) \left(36 a^2 \log \left({r_0} e^Z\right)+{r_0} e^Z\right)^2}\nonumber\\
& &\hskip -0.8in {\cal C}_{\theta_2y} = -\frac{{G^{\cal M}_{x^{10}x^{10}}} {G^{\cal M}_{\theta_2y}} \sqrt{-g}}{{\cal A}^{\cal M}_{\theta_2\theta_2} {G^{\cal M}_{x^{10}x^{10}}}^2 {G^{\cal M}_{yy}}+B_{r\theta_2}^2-{G^{\cal M}_{x^{10}x^{10}}}
   {G^{\cal M}_{\theta_2y}}^2} = -\frac{\pi ^{5/12} \left(\frac{1}{N}\right)^{19/20} {r_0}^4 e^{4 Z} \alpha _{\theta _2}^3 \left(72 a^2 {r_0} e^Z \log
   \left({r_0} e^Z\right)-3 a^2+2 {r_0}^2 e^{2 Z}\right){\cal G}{}^{4/3}}{72 \sqrt{2} 3^{2/3} {g_s}^{9/4} \log N  M
   {N_f}^3 \alpha _{\theta _1} \log \left({r_0} e^Z\right) \left(36 a^2 \log \left({r_0} e^Z\right)+{r_0}
   e^Z\right)^2}\nonumber\\
\label{C_{MN}}
& &\hskip -0.8in {\cal C}_{yy} = \frac{{\cal A}^{\cal M}_{\theta_2\theta_2} {G^{\cal M}_{x^{10}x^{10}}}^2 \sqrt{-g}}{2 \left({\cal A}^{\cal M}_{\theta_2\theta_2} {G^{\cal M}_{x^{10}x^{10}}}^2
   {G^{\cal M}_{yy}}+B_{r\theta_2}^2-{G^{\cal M}_{x^{10}x^{10}}} {G^{\cal M}_{\theta_2y}}^2\right)}\nonumber\\
& & \hskip -0.8in= \frac{{g_s} M \sqrt[5]{\frac{1}{N}} {N_f} {r_0}^2 e^{2 Z} \log \left({r_0} e^Z\right) \left(72 a^2 {r_0} e^Z
   \log \left({r_0} e^Z\right)-3 a^2+2 {r_0}^2 e^{2 Z}\right)}{3^{2/3} \sqrt[3]{\pi } \log N  \alpha _{\theta _1}
   \alpha _{\theta _2}^2 {\cal G}{}^{2/3}},\nonumber\\
& &\hskip -0.8in {\cal C}_{x^{10}x^{10}} = \nonumber\\
& & \hskip -0.8in  -\frac{\sqrt{-g} \left(-18 {\cal A}^{\cal M}_{\theta_2\theta_2} {G^{\cal M}_{x^1x^1}} {G^{\cal M}_{x^{10}x^{10}}}^2 {G^{\cal M}_{x^2x^2}} {G^{\cal M}_{x^3x^3}} {G^{\cal M}_{rr}} {G^{\cal M}_{tt}}
   {G^{\cal M}_{yy}}-10 B_{r\theta_2}^2 {G^{\cal M}_{x^1x^1}} {G^{\cal M}_{x^2x^2}} {G^{\cal M}_{x^3x^3}} {G^{\cal M}_{rr}} {G^{\cal M}_{tt}}+14 {G^{\cal M}_{x^1x^1}} {G^{\cal M}_{x^{10}x^{10}}} {G^{\cal M}_{x^2x^2}}
   {G^{\cal M}_{x^3x^3}} {G^{\cal M}_{rr}} {G^{\cal M}_{\theta_2y}}^2 {G^{\cal M}_{tt}}\right)}{\Delta}\nonumber\\
& & \hskip -0.8in = \frac{27 \sqrt[3]{3} {g_s} M \sqrt[5]{\frac{1}{N}} {N_f} {r_0}^2 e^{2 Z} \log \left({r_0} e^Z\right) \left(72 a^2
   {r_0} e^Z \log \left({r_0} e^Z\right)-3 a^2+2 {r_0}^2 e^{2 Z}\right) {\cal G}{}^{4/3}}{128 \pi ^{7/3} \log N
   \alpha _{\theta _1} \alpha _{\theta _2}^2};\nonumber\\
& & \hskip -0.8in \Delta = 8 {G^{\cal M}_{x^1x^1}} {G^{\cal M}_{x^{10}x^{10}}}
   {G^{\cal M}_{x^2x^2}} {G^{\cal M}_{x^3x^3}} {G^{\cal M}_{rr}} {G^{\cal M}_{tt}} \left({\cal A}^{\cal M}_{\theta_2\theta_2} {G^{\cal M}_{x^{10}x^{10}}}^2
   {G^{\cal M}_{yy}}+B_{r\theta_2}^2-{G^{\cal M}_{x^{10}x^{10}}} {G^{\cal M}_{\theta_2y}}^2\right);
\nonumber\\
\end{eqnarray*}
\begin{eqnarray}
\label{CMN}
& & \hskip -0.8in {\cal A}^{\cal M}_{\theta_2\theta_2} = \left(x \left(\widetilde{\widetilde{\widetilde{F_3^{IIB}}}}\right)_{x\theta_2} + y \left(\widetilde{\widetilde{\widetilde{F_3^{IIB}}}}\right)_{y\theta_2} + z \left(\widetilde{\widetilde{\widetilde{F_3^{IIB}}}}\right)_{z\theta_2} + r \left(\widetilde{\widetilde{\widetilde{*F_5^{IIB}}}}\right)_{r\theta_2}\right)^2 \sim \left(z \left(\widetilde{\widetilde{\widetilde{F_3^{IIB}}}}\right)_{z\theta_2} + r \left(\widetilde{\widetilde{\widetilde{*F_5^{IIB}}}}\right)_{r\theta_2}\right)^2\nonumber\\
& & \hskip -0.8in = \frac{9 {g_s}^{7/2} M^2 N^{11/10} {N_f}^4 e^{-2 Z} \log ^2\left({r_0} e^Z\right) \left(36 a^2 \log \left({r_0}
   e^Z\right)+{r_0} e^Z\right)^2}{2 \pi ^{5/2} {r_0}^2 \alpha _{\theta _1}^2 \alpha _{\theta _2}^4},
\end{eqnarray}
}
where in the last line use has been made of that in the IR:
\begin{eqnarray}
& & x \left(\widetilde{\widetilde{\widetilde{F_3^{IIB}}}}\right)_{x\theta_2}\sim N^{\frac{3}{20}}\log r,\  y\left(\widetilde{\widetilde{\widetilde{F_3^{IIB}}}}\right)_{y\theta_2}\sim N^{\frac{1}{4}}\log r,\nonumber\\
& & \   z \left(\widetilde{\widetilde{\widetilde{F_3^{IIB}}}}\right)_{z\theta_2}\sim N^{\frac{11}{20}}\log r,\ \left(\widetilde{\widetilde{\widetilde{*F_5^{IIB}}}}\right)_{r\theta_2}\sim N^{\frac{1}{20}}\left(\frac{\log r}{r_0}\right)^2 ;
\end{eqnarray}
 the triple tildes $\hskip 0.1in\widetilde{\widetilde{\widetilde{} }}\hskip 0.1in$ imply a tripe T-dual of the type IIB background of \cite{metrics}; $x, y, z$ are the delocalized $T^3$ coordinates using for effecting SYZ mirror symmetry via a triple T dual in \cite{MQGP}. In (\ref{C_{MN}}), ${\cal G} \equiv  8\pi e^{-\phi^{\rm IIB}}$.

One hence obtains:
\begin{eqnarray}
\label{E1010F1010}
& & \hskip -0.5in {\cal E}_{x^{10}x^{10}}{\cal F}_{x^{10}x^{10}} \nonumber\\
& & \hskip -0.5in  = \frac{243 b^{10} \left(9 b^2+1\right)^4 \beta  M \left(\frac{1}{N}\right)^{5/4} e^Z \left(e^Z-2\right) \left(-19683 \sqrt{6} \alpha _{\theta _1}^6-6642 \alpha _{\theta _2}^2
   \alpha _{\theta _1}^3+40 \sqrt{6} \alpha _{\theta _2}^4\right) \left(r_0 ^2-3 a^2\right) \log ^3(r_0 )}{16 \pi ^2 \left(3 b^2-1\right)^5  \left(\log N\right) ^3 \alpha
   _{\theta _2}^3 \left(6 b^2 r_0 +r_0 \right)^4},\nonumber\\
& & \nonumber\\
\end{eqnarray}
relevant to obtaining the ${\cal O}(R^4)$-corrected type IIA dilaton via (\ref{O4-corrections}).

\end{document}